\documentclass[aps,prc,twocolumn,floatfix,superscriptaddress]{revtex4}
\usepackage{epsfig}
\usepackage{amsmath}
\usepackage{color}
\usepackage{graphicx} 
\usepackage[utf8]{inputenc}

\definecolor{blue}{rgb}{0.05, 0.05, 0.5}

\begin{document}

\title{Production of Charm Quarks in a Parton Cascade Model for
Relativistic Heavy Ion  Collisions at $\sqrt{s_{\textrm {NN}}}$= 200 GeV}
\author{Dinesh K. Srivastava}
\email{dinesh@vecc.gov.in}
\affiliation{Variable Energy Cyclotron Centre, HBNI, 1/AF, Bidhan Nagar, Kolkata 700064, India}
\affiliation{Institut für Theoretische Physik, Johann Wolfgang Goethe-Universität, Max-von-Laue-Str. 1, D-60438 Frankfurt am Main, Germany}
\affiliation{ExtreMe Matter Institute EMMI, GSI Helmholtzzentrum für Schwerionenforschung, Planckstrasse 1, 64291 Darmstadt, Germany}
\author{Steffen A. Bass}
\email{bass@phy.duke.edu}
\affiliation{Duke University, Dept. of Physics, 139 Science Drive,
Box 90305, Durham NC 27708, USA}
\author{Rupa Chatterjee}
\email{rupa@vecc.gov.in}
\affiliation{Variable Energy Cyclotron Centre, HBNI, 1/AF, Bidhan Nagar, Kolkata 700064, India}

\begin{abstract}
We study the production and dynamics of heavy quarks in the parton cascade model for relativistic
heavy ion collisions. The model  is motivated by the QCD parton picture and describes the dynamics of an ultra-relativistic heavy-ion collision in terms of
 cascading partons which undergo scattering and multiplication while propagating.
We focus on the dynamics of charm quark production and evolution in p+p and Au+Au collisions for several different interaction scenarios, {\it {viz.}}, collisions
only between primary partons
without radiation of gluons, multiple collisions without
radiation of gluons and multiple collisions with radiation of gluons, allowing us to isolate the contributions of parton rescattering and radiation to charm production.
We also discuss results of an eikonal approximation of the collision which 
provides a valuable comparison with mini-jet calculations 
and clearly brings out the importance of multiple collisions. 
\end{abstract}


\maketitle

\maketitle
\section{Introduction} 

The quark-gluon plasma (QGP), a deconfined strongly interacting matter, which filled
the nascent universe a few micro-seconds after the Big Bang, is now being routinely
produced and studied in relativistic heavy ion collisions at RHIC and LHC.  Its presence has been confirmed through a host of previously predicted phenomena~\cite{Gyulassy:2004zy,Muller:2006ee,Muller:2012zq,Wang:2016opj}, as well as through novel and unexpected discoveries, such as strong elliptic flow \cite{Kolb:2003dz,Song:2010mg}, initial state fluctuations leading to higher order flow~\cite{Alver:2007qw,Alver:2008zza, Alver:2010gr} of hadrons as well as phenomena related to parton recombination~\cite{Fries:2003vb,Greco:2003xt}.

Heavy quarks serve as excellent probes of the QGP fireball as they are primarily produced by early-state hard scatterings and thus have the potential to probe the whole space-time history of the transient matter. 
It was generally expected that heavy quarks do not lose as much energy as light partons while traversing the
quark gluon plasma due to their large mass. However, early estimates of
drag and diffusion coefficients~\cite{Svetitsky:1987gq, GolamMustafa:1997id} as well as an early estimate of heavy-quark energy loss due to radiation of gluons~\cite{Mustafa:1997pm} suggested that heavy quarks
could possibly lose as much energy as light quarks and gluons during their passage through the
QGP. These predictions have been subsequently confirmed by multiple other calculations~\cite{GolamMustafa:1997id,Gyulassy:2000fs,Djordjevic:2003zk,Moore:2004tg,Peigne:2008nd,Das:2010tj,Gossiaux:2010yx,Abir:2012pu,Das:2015ana}, with tools ranging from
simple phenomenological models used to estimate  the medium modification of charm production for heavy ion 
collisions~\cite{Younus:2010sx}, to more detailed comparisons with models of
energy loss embedded in a hydrodynamic evolution of the plasma~\cite{Cao:2013ita}.
Studies including the later hadronic stages of the evolution have also indicated that charm-mesons may lose energy due to (resonant)
scatterings with hadrons~\cite{vanHees:2007me, Rapp:2009my}. 
The experimental observation of large elliptic flow for D and B mesons as well as the small value of $R_{AA}$, the nuclear modification factor, have added experimental evidence to these findings \cite{Andronic:2015wma,Xie:2016iwq}.
Several of these aspects have been studied in great detail using the Parton Hadron String Dynamics Model at energies reached at RHIC and 
LHC~\cite{Cassing:2000vx,Bratkovskaya:2003ux,Bratkovskaya:2004ec,Linnyk:2008uf,Linnyk:2008hp,Song:2015sfa,Song:2015ykw}.

In the high momentum domain the production of heavy quarks via the interaction between (mini)-jets following
the initial scattering as well as due to interaction of thermal partons at 
high incident energies~\cite{Levai:1994dx,Lin:1994xma, Younus:2010sx} has been studied. 
An estimate at next to
leading order using the formalism developed by Mangano et al.~\cite{Mangano:1991jk, Frixione:1997ma} was
also used to suggest~\cite{Younus:2011mn} that angular correlations of charm-anticharm quarks
in p+p collisions would be drastically different  for various  production processes, e.g., gluon fusion or quark-antiquark annihilation, gluon splitting
and when a gluon is radiated off one of the legs of the scattering diagram.
This angular correlation is expected to get
even more interesting for nucleus-nucleus collisions because of the interaction of the
heavy quarks with the medium.

State of the art approaches for the investigation of heavy quarks in a hot and dense QCD medium need to incorporate a dynamical treatment for both, the heavy quarks as well as the QCD medium. This can be either accomplished via hybrid approaches that utilize a hydrodynamic evolution for the medium in concert with a dynamical interaction model for the heavy quarks~\cite{Moore:2004tg,He:2011qa,Gossiaux:2011ea,Cao:2013ita} or in the context of a purely microscopic transport for both, the heavy quarks and the medium~\cite{Uphoff:2012gb,Scardina:2014lda,Uphoff:2014hza}. Parton Cascade Models~\cite{Geiger:1991nj, Geiger:1992si, Geiger:1992ac, Geiger:1993py, Geiger:1993ix, Geiger:1994he,Bass:2002fh} as used in this study, fall into the latter category. Over the past decade, PCMs have been used for multiple studies of heavy quark dynamics in a QCD medium, ranging from the study of heavy quark energy loss in infinite matter~\cite{Younus:2013rja}, to studies of equilibration and elliptic flow built-up at RHIC and LHC energies~\cite{Uphoff:2012gb,Uphoff:2014hza}. However, most of these studies focused on the low to intermediate momentum regime (mostly due to the availability of data in that regime). 

As more heavy quark data at high transverse momenta become available, it is now the right time to ask whether approaches for the dynamics of heavy quarks based purely on perturbative QCD are able to describe the experimental observations in that domain. In addition, even if pQCD approaches cannot fully account for the observed features, it remains an open question to what extent the observed features of final state observables are already imprinted on the heavy quarks prior to the formation of the QGP, i.e. via initial state effects as well as contributions from early non-equilibrium evolution.
To address these questions we use a Monte Carlo implementation of the PCM, {\tt VNI/BMS} \cite{Bass:2002fh}, that is based on a Boltzmann transport description with pQCD matrix elements for parton-parton interactions. This particular PCM implementation also contains parton shower emission.  In the present work we discuss the implementation of heavy quarks into the PCM and calculate 
charm production in relativistic collisions of gold nuclei at 
$\sqrt{s_{\textrm NN}}$ = 200 GeV. We consider several different
interaction scenarios for pp and nucleus-nucleus systems,
namely,
an implementation of the eikonal approximation  for a direct comparison with 
minijet calculations and to clearly demonstrate the
dynamics which emerges as the partons change their momenta
after collisions (which is neglected in the eikonal approximations),
a scenario which involves only collisions 
between primary partons to clearly bring out the consequences of
multiple scattering, and finally a scenario which additionally
includes gluonic multiplication by radiation of gluons 
following scattering. The last scenario  considerably increases
the number of scatterings and partons (including the heavy quarks) are capable of losing
energy by radiation of gluons.

\section{Formulation} 

The details of the Monte Carlo implementation of the parton cascade model
have been discussed in ~\cite{Bass:2002fh}. However, given how our understanding of the hot and dense QCD matter created in relativistic heavy-ion collisions has changed since the inception of the PCM, a brief discussion of its past and future uses and relevancy is in order: at the time when the first Parton Cascade Models were developed, the notion of a Quark-Gluon-Plasma was still in its infancy and it was thought of as a weakly interacting gas of quark and gluons that could be described using perturbative QCD. Having a space-time evolution based on the Boltzmann equation with quark and gluon degrees of freedom and pQCD cross sections was thought to provide a full picture of the dynamics of the deconfined system up to hadronization. With the discovery of the near perfect fluidity of the QGP at RHIC and LHC, this picture has been found to be inadequate and most PCM implementations have failed to generate the observed amount of collective flow, while utilizing pQCD based interactions for quark and gluon degrees of freedom \footnote{the one exception is the BAMPS implementation \cite{}}. 

However, the production of hard probes -- jets, photons and heavy quarks -- can be perfectly well understood in terms of perturbative QCD production cross sections and their dynamical evolution in the early reactions stages prior to the formation of the QGP at $\tau_0 \approx 1$~fm/c is dominated by interactions with fairly large  momentum transfers. Parton Cascade Models are thus ideally suited to describe the early out-of-equilibrium evolution of hard probes prior to the formation of a thermalized Quark-Gluon-Plasma. Their ability to describe the dynamical evolution of hard probes during the early times of the collision evolution, including multiple interactions (rescattering and gluon splitting), provides significant added insight beyond what can be gained from initial state production Monte-Carlo codes such as PYTHIA or Herwig \footnote{historically, VNI/BMS is based on {\tt PYTHIA 6} with the addition of a Boltzmann equation solver for the space-time evolution of all partons}. 

Thus, we shall utilize the PCM for the study of heavy quarks during early times of Au+Au collisions at RHIC. In the following we shall first describe the implementation
of heavy quark (Q) production and interactions before moving on to our analysis.

\subsection{Parton cascade model involving light quarks and gluons}

The $2\rightarrow 2$ scatterings included in the version of the parton cascade model utilized for our work, {\tt VNI/BMS}, are:
\begin{eqnarray} 
q_i q_j &\rightarrow& q_i q_j , \, q_i \bar{q}_i \rightarrow q_j \bar{q}_j \, , \nonumber\\
q_i\bar{q}_i &\rightarrow & gg , \, q_i \bar{q}_i \rightarrow g \gamma \, ,\nonumber\\
q \bar{q}_i &\rightarrow & \gamma \gamma , \, q_i g \rightarrow q_i g \, , \nonumber\\
q_i g & \rightarrow & q_i \gamma , \, gg \rightarrow q_i \bar{q}_i , \, \nonumber\\
gg &\rightarrow & gg
\end{eqnarray}
The $2\rightarrow 3$ reactions are included via time-like branchings of the final-state
partons:
\begin{eqnarray}
g^{*} &\rightarrow&  q_i \bar{q}_i  \, , \, {q_i}^* \rightarrow q_i g \, , \nonumber\\
g^{*} &\rightarrow& gg \, , \, {q_i}^* \rightarrow q_i \gamma
\end{eqnarray}
For details we refer the reader to Ref~\cite{Bass:2002fh}.
We add that the gluon splitting in the above, includes the splitting $g^{*} \rightarrow Q \bar{Q}$ where $Q$ stands for heavy quarks (charm or bottom) if the virtuality is large enough to admit this process. We shall see later that this process plays an important role, which is likely to increase with the increasing centre of mass energy of the collision.
 Due to the specifics of the implementation of the collision term in {\tt VNI/BMS} the reverse process, parton fusion, is not included, since it would lead to the propagation of partons off-shell, which is not included in the {\tt VNI/BMS} formulation. Technically, parton fusion with on-shell partons can be formulated via a $3\rightarrow 2$ process that can be implemented using collision rates in the collision term. This has been successfully done in the {\tt BAMPS} model \cite{Xu:2004mz}. The main drawback of the lack of parton fusion in {\tt VNI/BMS} is that it violates detailed balance in the collision term and thus leads to an incorrect equilibrium limit for the bulk in the infinite time and size limit. While this is a valid concern if we were to study the equilibration of bulk QCD matter in our approach, our focus on high-momentum heavy quarks that are not thought to thermalize fully \cite{Cao:2011et} reduces the systematic uncertainty induced by the omission of these processes.

\subsection{Heavy quark production by gluon fusion and annihilation of light quarks}

 At lowest order pQCD, heavy quarks are produced from fusion of gluons ($gg \rightarrow Q\bar{Q}$)
and annihilation of light quarks ($q\bar{q} \rightarrow Q\bar{Q}$).

The partial differential cross-section can be written as
\begin{equation}
\frac{d\hat{\sigma}}{d\hat{t}}= \frac{1}{16\pi \hat{s}^2}\sum |{\cal M} |^2\, ,
\end{equation}
where
the summed spin and coloured averaged squared matrix element for the process
$gg \rightarrow Q\bar{Q}$ is given by

\begin{equation}
\sum |{\cal M}|^2= \pi^2 \alpha_s^2(Q^2)\left[a_1+a_2+a_3+a_4+a_5+a_6 \right]
\end{equation}
where
\begin{eqnarray}
a_1 & =& \frac{12}{\hat{s}^2}(M^2-\hat{t})(M^2-\hat{u})\nonumber\\
a_2 & = & \frac{8}{3}\frac{(M^2-\hat{t})(M^2-\hat{u})-2M^2(M^2+\hat{t})}{(M^2-\hat{t})^2}\nonumber\\
a_3 & = & \frac{8}{3}\frac{(M^2-\hat{t})(M^2-\hat{u}-2M^2(M^2+\hat{u})}{(M^2-\hat{u})^2}\nonumber\\
a_4 & = & -\frac{2M^2(\hat{s}-4M^2)}{3(M^2-\hat{t})(M^2-\hat{u})}\nonumber\\
a_5 & = & -6\frac{(M^2-\hat{t})(M^2-\hat{u})+M^2(\hat{u}-\hat{t})}{\hat{s}(M^2-\hat{t})}\nonumber\\
a_6 & = & -6 \frac{(M^2-\hat{t})(M^2-\hat{u})+M^2(\hat{t}-\hat{u})}{\hat{s}(M^2-\hat{u})}
\end{eqnarray} 

The total cross-section $\hat{\sigma}(\hat{s})$ is obtained from the above
by integrating over $\hat{t}$ :
\begin{equation}
\hat{\sigma}(\hat{s})=\frac{1}{16 \pi {\hat{s}}^2} 
\int_{M^2-\hat{s}(1+\chi)/2}^{M^2-\hat{s}(1-\chi)/2} \, d\hat{t} \sum 
\left | \mathcal{M} \right |^2 \, ,
\end{equation}
where
\begin{equation}
\chi=\sqrt{1-\frac{4M^2}{\hat{s}}} \, .
\end{equation}

We note that due to the mass of the heavy quark $M$ this cross-section remains finite
and the total cross-section reduces to
\begin{eqnarray}
\hat{\sigma}_{\mathrm {gg \rightarrow Q\bar{Q}}}(\hat{s})& =& \frac{\pi \alpha_s^2(Q^2)}{3\hat{s}} 
                  \left [-\left (7+\frac{31M^2}{\hat{s}} \right )\frac{1}{4}\chi \right. \nonumber\\
                && +\left( 1+ \frac{4M^2}{\hat{s}}+\frac{M^4}{\hat{s}^2}\right )
                    \left.  \log\frac{1+\chi}{1-\chi} \right ] \,.
\end{eqnarray}

Similarly, the summed spin and coloured averaged squared matrix element for the process
$q\bar{q} \rightarrow Q\bar{Q}$ is given by:
\begin{eqnarray}
\sum |{\cal M}|^2 &=& \frac{64}{9}\pi^2 \alpha_s^2(Q^2)\times\nonumber\\
& &\left[ \frac{(M^2-\hat{t})^2+(M^2-\hat{u})^2+2M^2\hat{s}}{\hat{s}^2}\right],
\end{eqnarray}
so that the total cross-section becomes.
\begin{equation}
\hat{\sigma}_{\mathrm {q\bar{q} \rightarrow Q\bar{Q}}}(\hat{s}) = \frac{8\pi \alpha_s^2(Q^2)}{27\hat{s}^2} 
(\hat{s}+2M^2)\chi~.
\end{equation}
Once again, these cross-sections remain finite due to the mass of the heavy quarks.


\subsection{Heavy quark production due to flavour excitation}

The implementation of heavy quark production due to flavour excitation has been subject of significant debate in the community. In principle, the (off-shell) heavy sea-quarks in the target (projectile)
can be brought on-shell by scattering
with other off-shell partons from the projectile (target) (or by scattering with
on-shell partons that are produced during an earlier partonic interaction). This contribution
was referred to as flavour excitation in the original paper of Combridge~\cite{Combridge:1978kx}.
Its inclusion in the parton cascade model, where all parton scatter are 
treated at lowest order, has been questioned~\cite{Lin:1994xma}. It  is also seen to
give rise to a substantial production of charm and bottom quarks~\cite{Geiger:1993py},
as was also pointed out in the original work of
Combridge. It has been argued that this contribution  should be strongly suppressed if higher order 
coherent interference terms are 
accounted for (see Ref.~\cite{Collins:1985gm} and references there in). 

In our calculation, we have excluded this process for the production
of charm or bottom quarks. This is done in two steps. Firstly,
we initialize the quark and gluon distribution functions at  a rather low
 $Q_0^2$ of 0.589 GeV$^2$ (see \cite{Bass:2002fh}) which effectively eliminates the presence of heavy sea-quarks. 
In addition, we disallow any partonic interaction if one of the
partons is a heavy (charm or bottom) quark from the sea.

The flavour excitation process as it is understood now is included in NLO calculations,
e.g., Ref.~\cite{Mangano:1991jk, Frixione:1997ma}, and proceeds via a splitting of a gluon into a $Q \bar{Q}$ 
pair followed by scattering of one of the heavy quarks with another parton in the system. The contribution from this process to heavy quark production 
 is not large.
We re-emphasize that as mentioned earlier, the splitting  of the final state gluon into a charm- anti charm pair (or for that matter bottom-anti bottom pair as well) is included in our calculations. We also add that Soft Collinear Effective Theory has
been used recently to estimate the gluon splitting into heavy quarks~\cite{Kang:2016ofv}, which suggests that this contribution
can be large at LHC energies.
Once produced,
the heavy quarks are allowed to scatter with quarks and gluons. We shall see
later that this plays an important role in their dynamics.

\subsection{Scattering of light quarks and gluons with heavy quarks}
We consider elastic scattering of heavy quarks with
light quarks and gluons, i.e, 
\begin{equation}
  Qq(\bar{q})\rightarrow Qq(\bar{q})\, \, \&
\, \,  Qg\rightarrow Qg \, .
\end{equation}
The differential scattering cross section  is defined as: 
\begin{equation}
 \frac{d\hat{\sigma}}{d\hat{t}}=\frac{1}{16\pi(\hat{s}-M^2)^2}\sum{|\mathcal{M}|^2}.
\end{equation}

The summed spin and coloured averaged squared matrix element for the process
$Qq(\bar{q}) \rightarrow Qq(\bar{q})$ have been derived by Combridge~\cite{Combridge:1978kx}:
\begin{eqnarray}
\sum{|\mathcal{M}|^2}&=&\frac{64}{9}\pi^2\alpha_{s}^2(Q^2)\times  \nonumber\\
 & &\left[\frac{(M^2-\hat{u})^2+
(\hat{s}-M^2)^2+2M^2\hat{t}}{\hat{t}^2}\right]\,.
\end{eqnarray}

The corresponding matrix element for the process $Qg \rightarrow Qg$ is given by:
\begin{eqnarray}
\sum{|\mathcal{M}|^2}&=&\pi^2\alpha_{s}^2(Q^2)\times \nonumber\\
& &\left[b_1+b_2+b_3+b_4+b_5+b_6\right]\,,
\end{eqnarray}
where
\begin{align}
b_1&=32\frac{(\hat{s}-M^2)(M^2-\hat{u})}{\hat{t}^2}\,,\nonumber\\
b_2&=\frac{64}{9}\frac{(\hat{s}-M^2)(M^2-\hat{u})+2M^2(\hat{s}+M^2)}
{(\hat{s}-M^2)^2}\,,\nonumber\\
b_3&=\frac{64}{9}\frac{(\hat{s}-M^2)(M-\hat{u})
+2M^2(M^2+\hat{u})}{(M^2-\hat{u})^2}\,,\nonumber\\
b_4&=\frac{16}{9}\frac{M^2(4M^2-\hat{t})}{(\hat{s}-M^2)(M^2-\hat{u})}\,,\nonumber\\
b_5&=16\frac{(\hat{s}-M^2)(M^2-\hat{u})+M^2(\hat{s}-\hat{u})}
{\hat{t}(\hat{s}-M^2)}\,,\nonumber\\
b_6&=-16\frac{(\hat{s}-M^2)(M^2-\hat{u})-M^2(\hat{s}-\hat{u})}
{\hat{t}(M^2-\hat{u})}\,.
\end{align}

The total cross section is formally defined as:
\begin{equation}
\hat{\sigma}_{tot}=\frac{1}{16 \pi (\hat{s}-M^2)^2}\int_{-(\hat{s}-M^2)^2/\hat{s}}^{0}
d\hat{t} \sum |\mathcal{M}|^2\,.
\end{equation}
We immediately note that the total cross-sections for both the processes diverge
because of the pole at $\hat{t}$ equal to zero. We reduce the upper limit of
integration to $\hat{t}=-Q_0^2$, where $Q_0^2=p_0^2$ and $p_0$ is the cut-off 
used in the parton cascade model
to regularize the $2 \rightarrow 2$ pQCD matrix element, as in {\tt PYTHIA}.

\subsection{Radiation of gluons by heavy quarks}
The parton cascade model includes final state radiation of gluons 
or photons following a hard ($2 \rightarrow 2$) scattering using the well
established approach of parton showers~\cite{Sjostrand:1985xi} as implemented 
in {\tt PYTHIA}, in a leading logarithmic approximation.
It incorporates a sequence of nearly collinear splittings, $a \rightarrow bc$,
where the initial parton $a$ is called mother parton and $b$ and $c$
are called daughters, which can split further and populate a tree-like structure.
Strong and electromagnetic interactions allow for several different possibilities 
of splittings, which are all included in the PCM. The differential
probability for a  splitting to occur is given by:

\begin{equation}
\mathcal{P}_a=\sum_{b,c}\frac{\alpha_{abc}}{2\pi}P_{a\rightarrow bc}\frac{dQ^2}{Q^2} dz
\end{equation}
where the sum runs over all the allowed splittings, $\alpha_{abc}$ is equal to $\alpha_{em}$
for emission of photons and for QCD splittings it is $\alpha_{s}$. Further, $Q^2$ is the 
momentum scale of the splitting, $z$ gives fraction of energy carried by the daughter $b$ 
and $1-z$ is the remaining fraction, carried by the daughter $c$. The kernels describing the
splitting $P_{a\rightarrow bc}$ are taken from Altarelli and Parisi~\cite{Altarelli:1977zs}.

The parton cascade model uses a cut-off $\mu_0$, taken as $\geq$ 1 GeV, 
to regulate the collinear
singularities, by terminating the splitting once the virtuality of the time-like parton drops
to $\mu_0$. The soft gluon interference is accounted for by implementing an angular ordering
of the radiated gluons. The mass of the partons is accounted for by modifying 
the cut-off $\mu_0$ such that $\mu_0^2$= 1 GeV$^2$ for gluons and $\mu_0^2=m_q^2+1$ (GeV$^2$)
for quarks~\cite{Geiger:1991nj} and thus radiation of gluons from heavy quarks is
at the same footing as for that from light quarks.  
This value of $\mu_0$ is same as that used in {\tt PYTHIA} for regularizing the final (and initial) state
radiations. Increasing this will lead to a hardening of the transverse momentum spectra as well as
a reduction in the number of collisions due to a reduction in number of gluons, while taking a smaller
value would have an opposite effect. We shall see later that we get a reasonable description of the
spectra for charm quarks both for $pp$ and $AA$ collisions and thus have kept it fixed, though exploratory
calculations were done to verify the comments made above.
\begin{figure}
\centerline{\includegraphics*[width=8.0 cm]{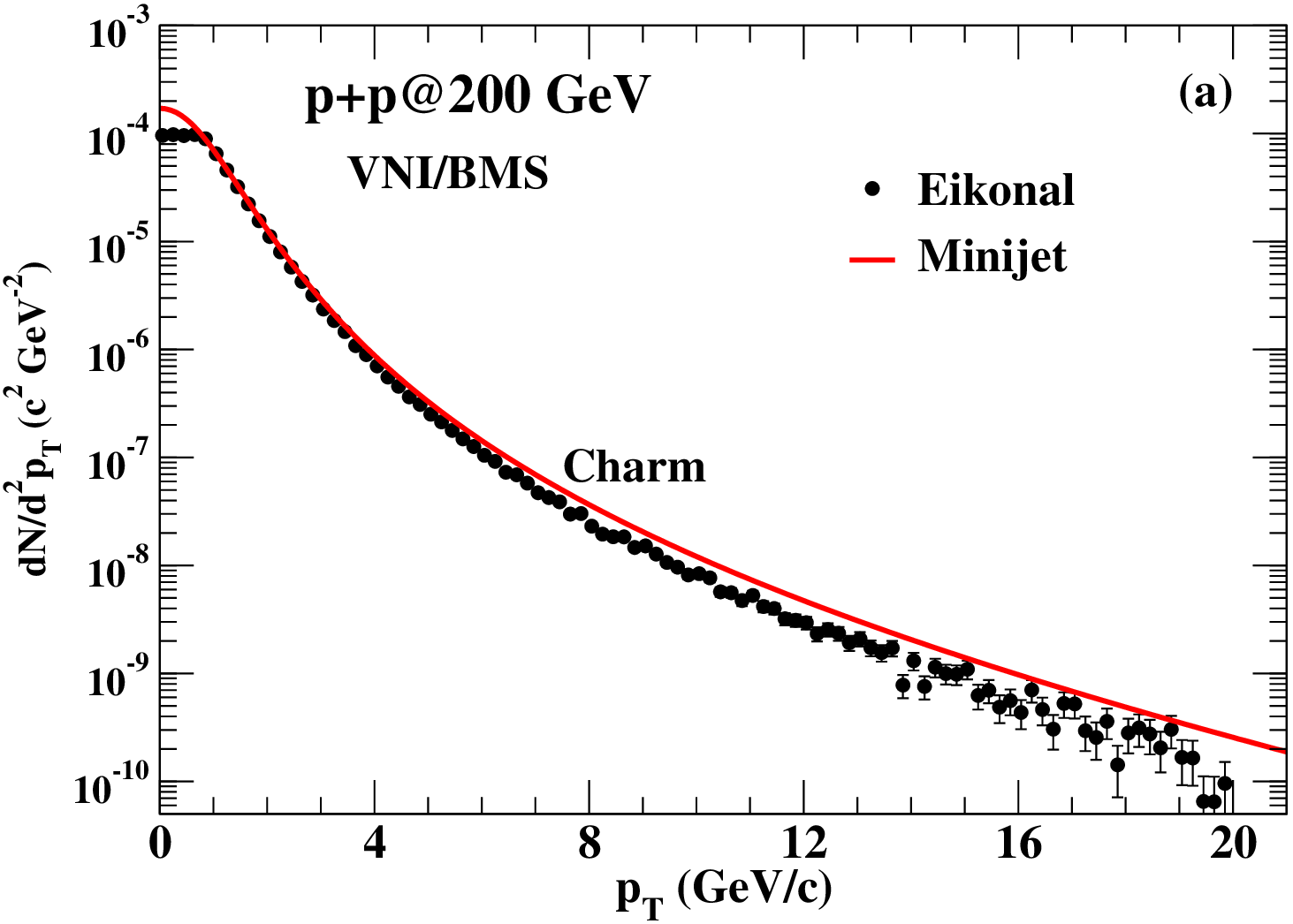}}
\centerline{\includegraphics*[width=8.0 cm]{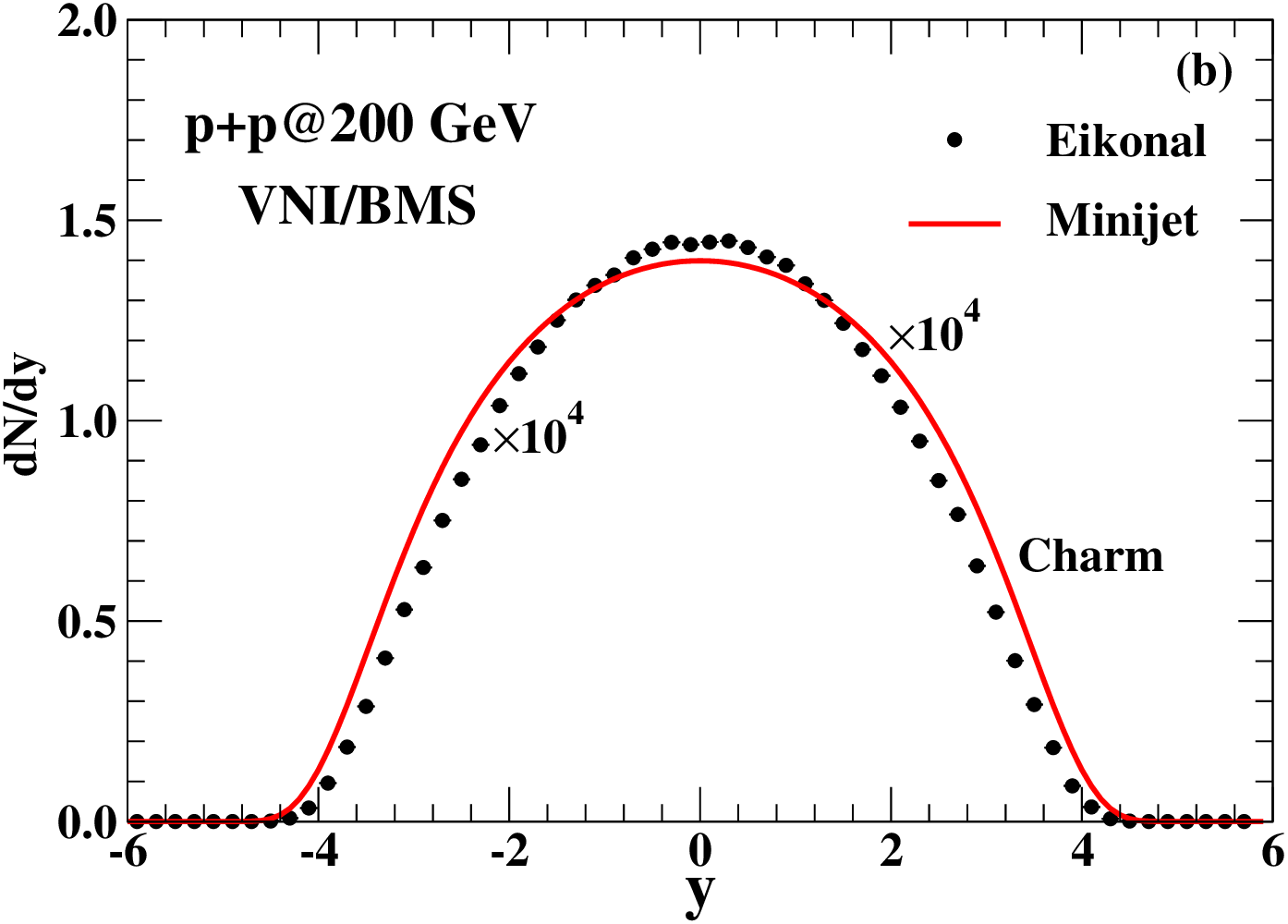}}
\caption{(colour online) (a) The rapidity ($y$) integrated $p_T$ spectra of charm quarks
using mini-jet calculations (solid curves) and eikonal approximation. 
 The lower panel (b) gives the $p_T$ integrated rapidity
spectra.}
\label{pp_eik_min}
\end{figure}
\begin{figure}
\centerline{\includegraphics*[width=8.0 cm]{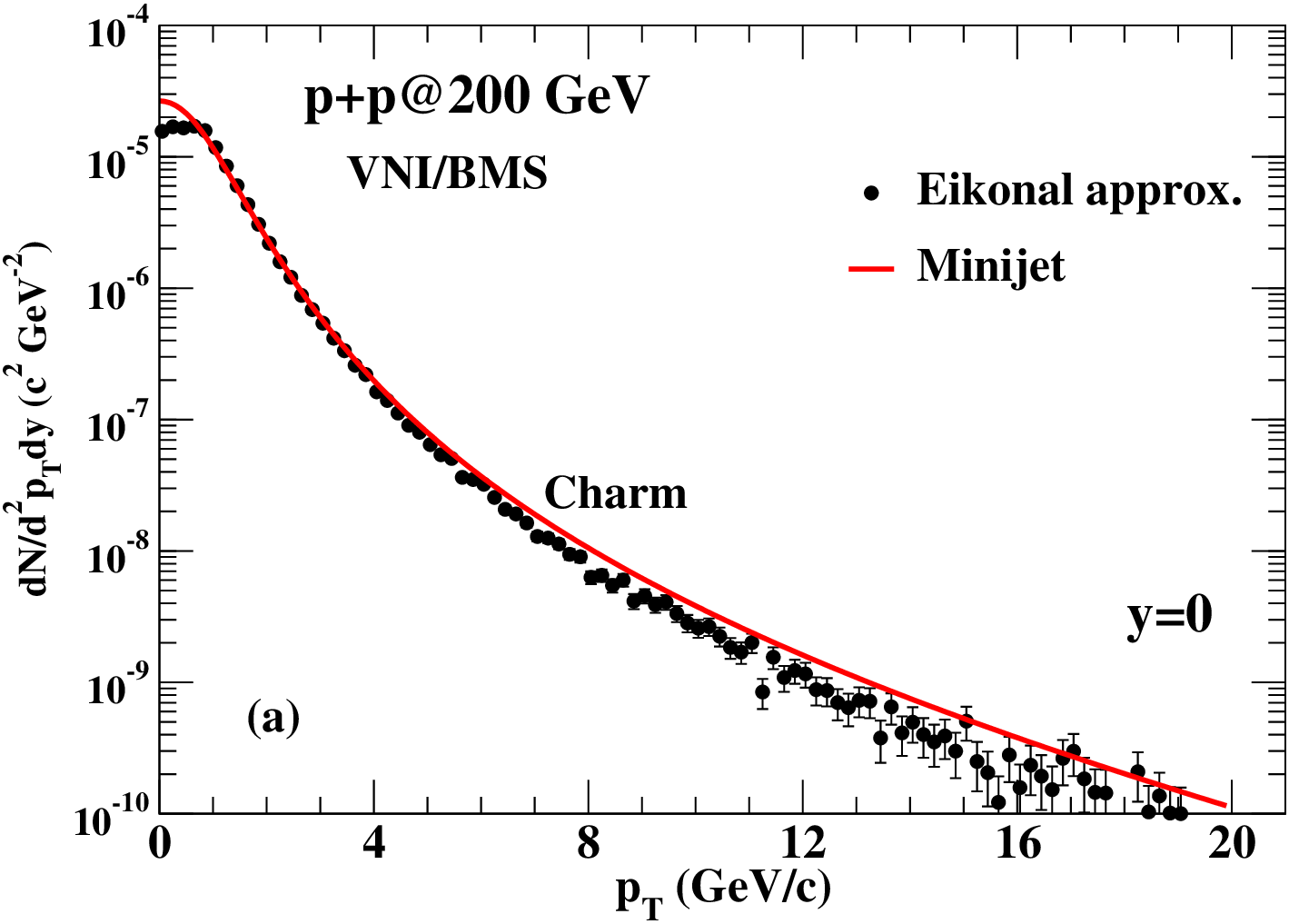}}
\centerline{\includegraphics*[width=8.0 cm]{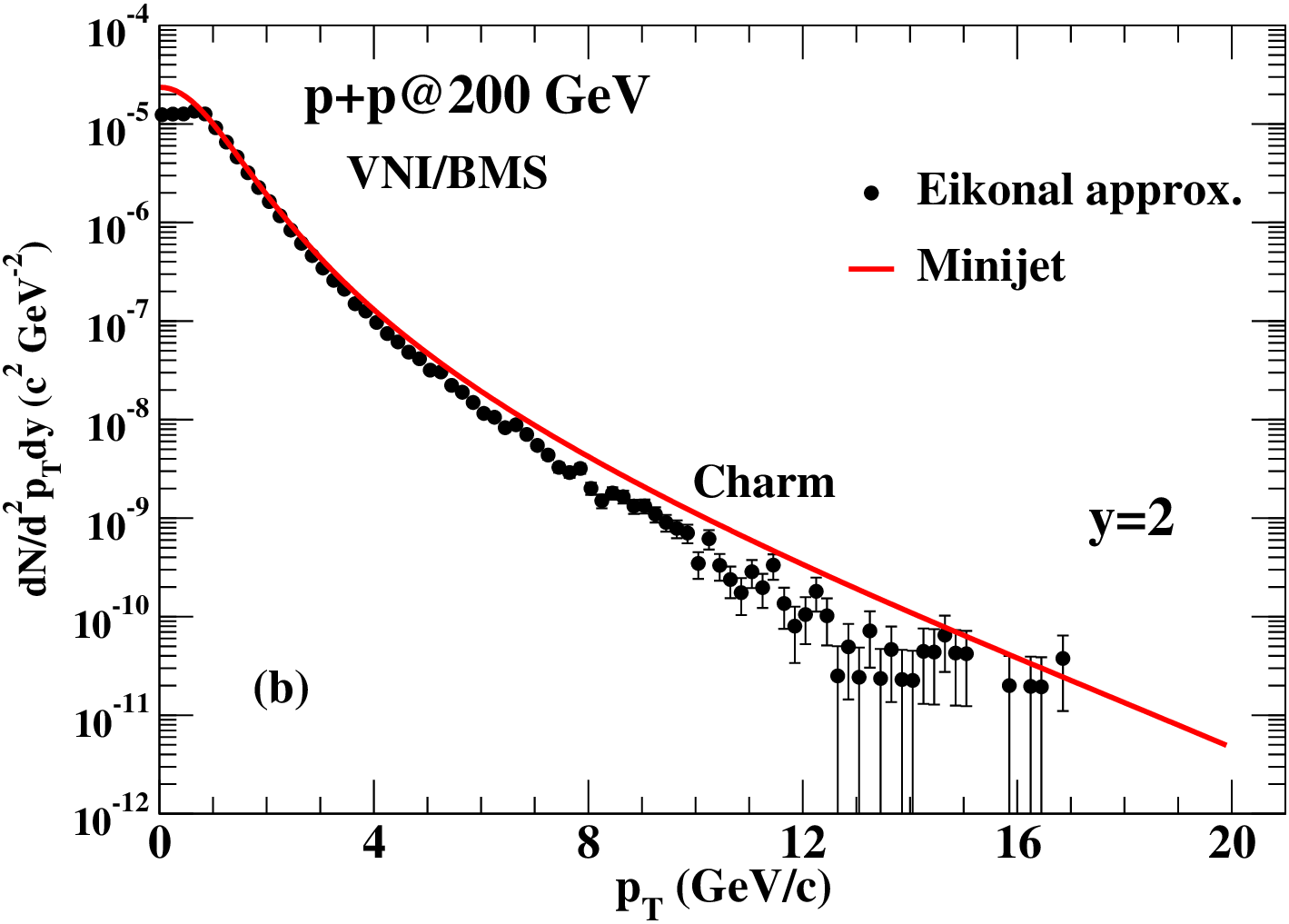}}
\centerline{\includegraphics*[width=8.0 cm]{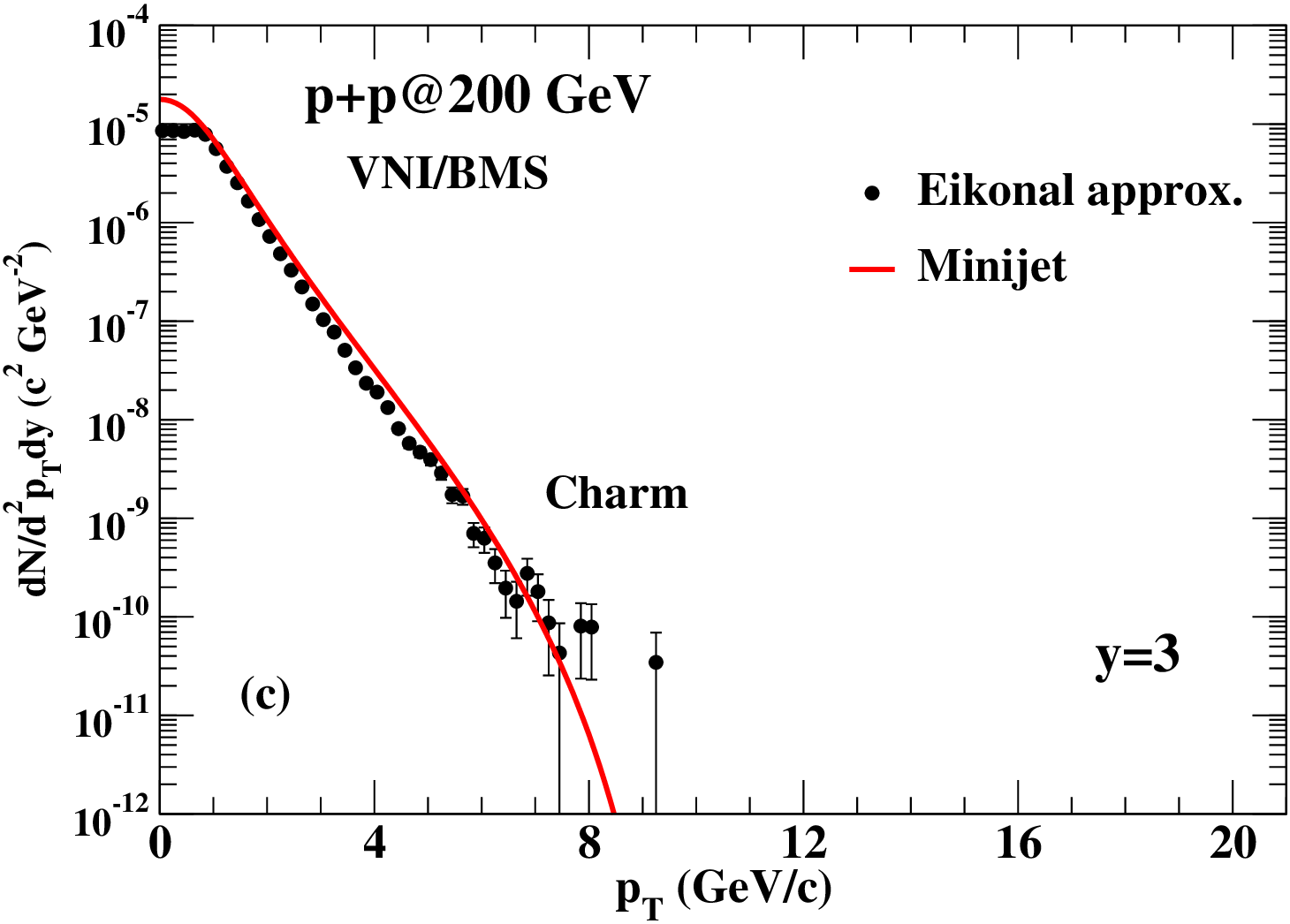}}
\caption{(colour online) The $p_T$ spectra for charm quarks
using mini-jet calculations (solid curves) and eikonal approximation at $y$ =0 (a),
$y=2$ (b), and $y=3$ (c).} 
\label{pp_eik_min_y}
\end{figure}
\subsection{Production of heavy quarks using the minijet formalism}

In order to validate the Monte-Carlo implementation of the PCM that we are using,  we perform a separate calculation of heavy quark production using the minijet formalism~\cite{Eichten:1984eu}. The cross-section for the production of a heavy-quark pair ($Q\bar{Q}$) is written in
terms of the rapidities of the two quarks $y_1$ and $y_2$ and their transverse momentum
$p_T$ as
\begin{eqnarray}
\frac{d\sigma}{dp_T^2 dy_1 dy_2}& = &\sum_{ij}  \frac{1}{1+\delta_{ij}}\times \nonumber\\
 & &\left [ x_a f_i^{(a)}(x_a,Q^2)  x_b f_j^{(b)}(x_b,Q^2)\frac{d\hat{\sigma}_{ij}}{d\hat{t}}(\hat{s},\hat{t},\hat{u})\right. \nonumber\\
 & & \left. x_a f_j^{(a)}(x_a,Q^2) x_b f_i^{(b)}(x_b,Q^2)\frac{d\hat{\sigma}_{ij}}{d\hat{t}}(\hat{s},\hat{u},\hat{t}) \right] \, . \nonumber\\
\end{eqnarray}
In the above $f_i^{(a)}$ denotes structure function of the parton $i$ for the nucleon $a$, etc., and $x_a$ is the fraction of linear-momentum of
the nucleon $a$ carried by the parton, $Q^2$ is the momentum scale, in the standard notation. However, in order to be able to directly compare to 
the parton cascade model, we fix the fragmentation scale at the momentum cut-off $p_0^2$ used in the calculation for the momentum cut-off
and the coupling constant is kept at a fixed value of $\alpha_s$.

\section{Results}

We shall first discuss our results for $pp$ collisions
at $\sqrt{s_{NN}}$ = 200 GeV for several different parton interaction scenarios. Next we give our 
findings for $Au+Au$ collisions at the same centre-of mass energy for central
collisions. Finally we compare the proton-proton results to the Au+Au results to get an idea about the emerging dynamics of the
propagation of charm quarks, subject to semi-hard scatterings and radiation of gluons in the QCD medium produced by
relativistic heavy ion collisions.

\subsection{$pp$ collisions at $\sqrt{s_{NN}}=$ 200 GeV}

\begin{figure}
\centerline{\includegraphics*[width=8.0 cm]{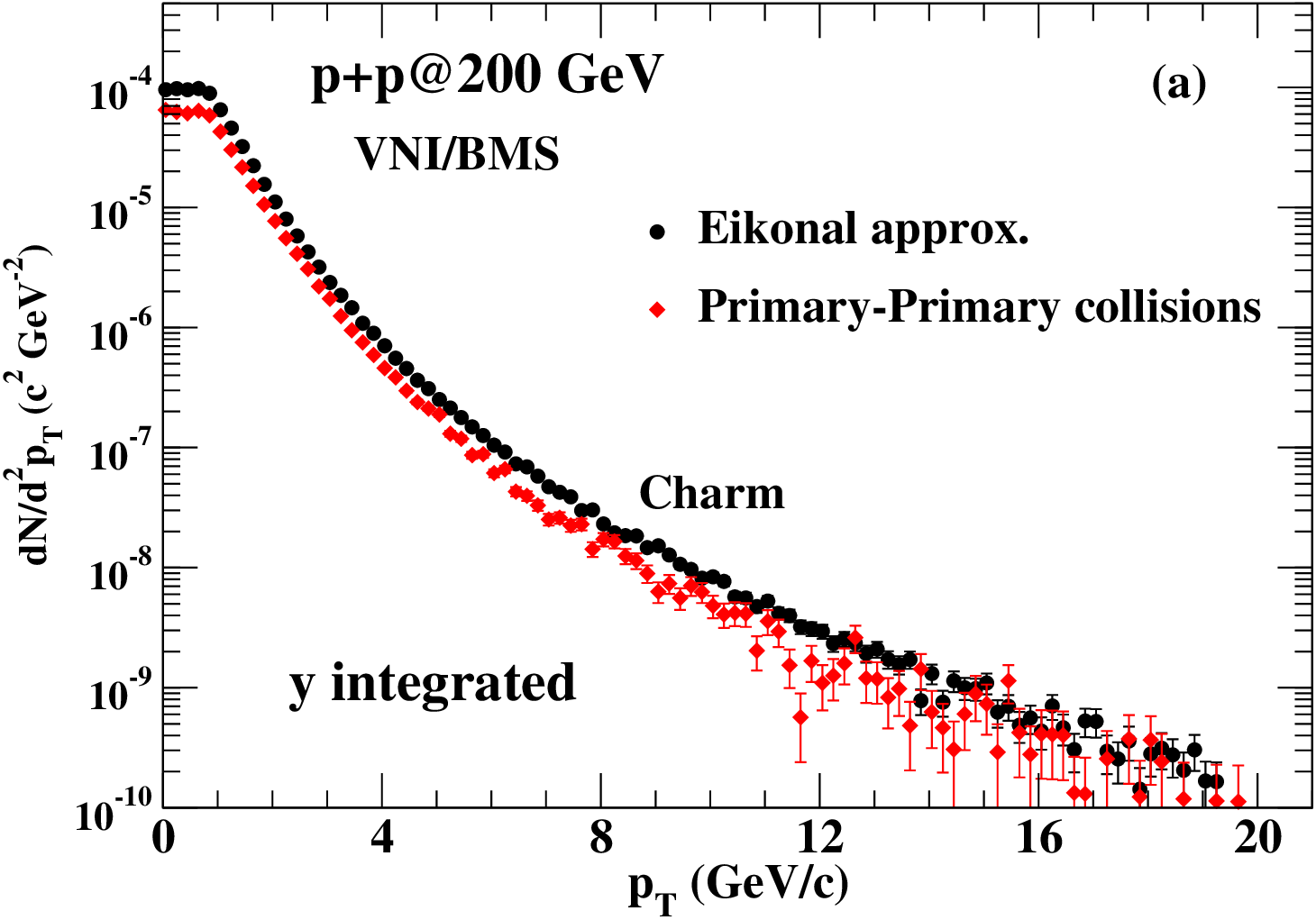}}
\centerline{\includegraphics*[width=8.0 cm]{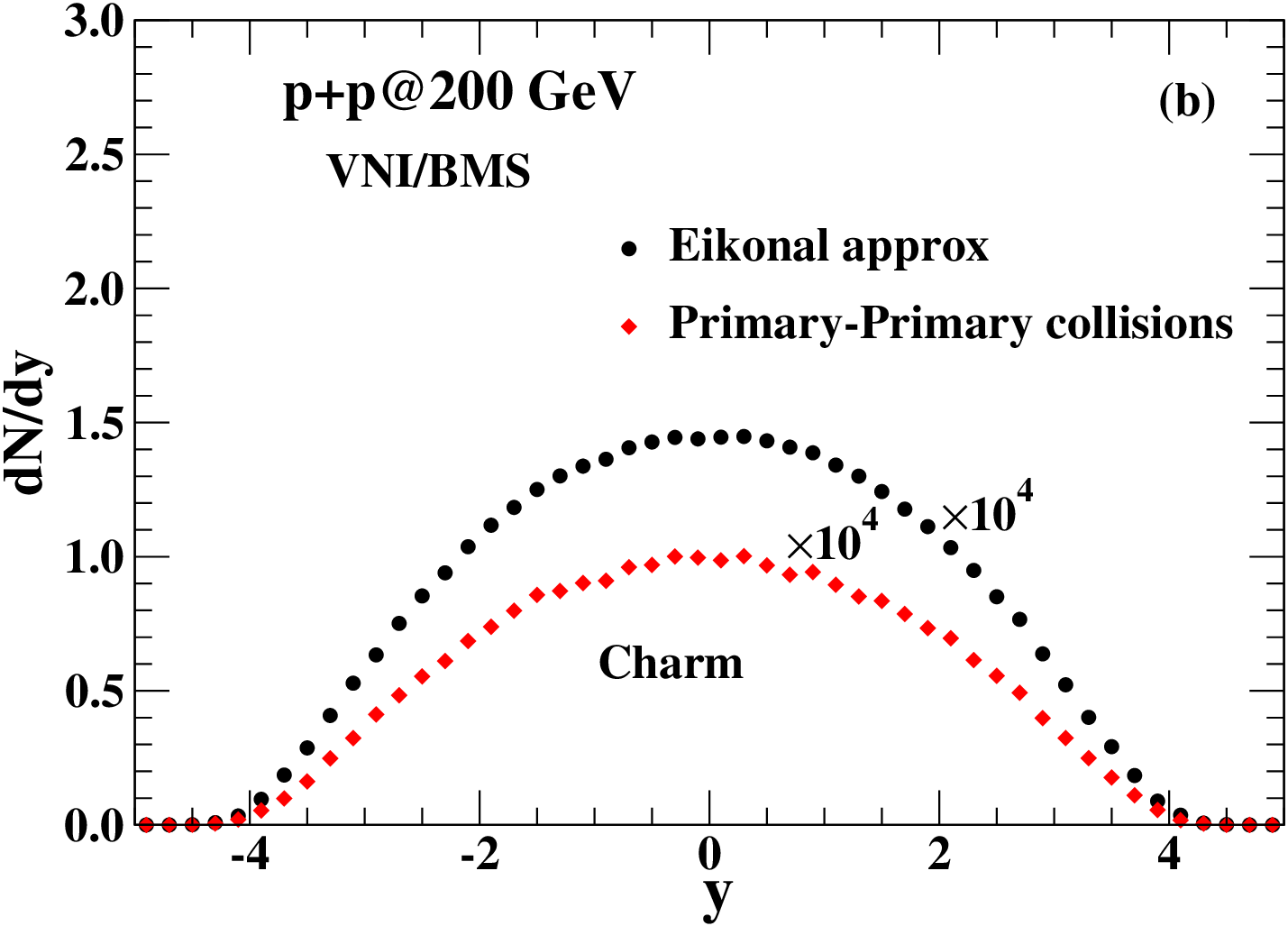}}
\caption{(colour online) (a) The rapidity integrated $p_T$ spectra and (b) $p_T$
integrated rapidity spectra for charm quarks using eikonal approximation
and primary-primary collisions for $pp$ system at $\sqrt{s_\textrm{NN}}$ = 200 GeV.}
\label{pp_eik_prim}
\end{figure}

\begin{figure}
\centerline{\includegraphics*[width=8.0 cm]{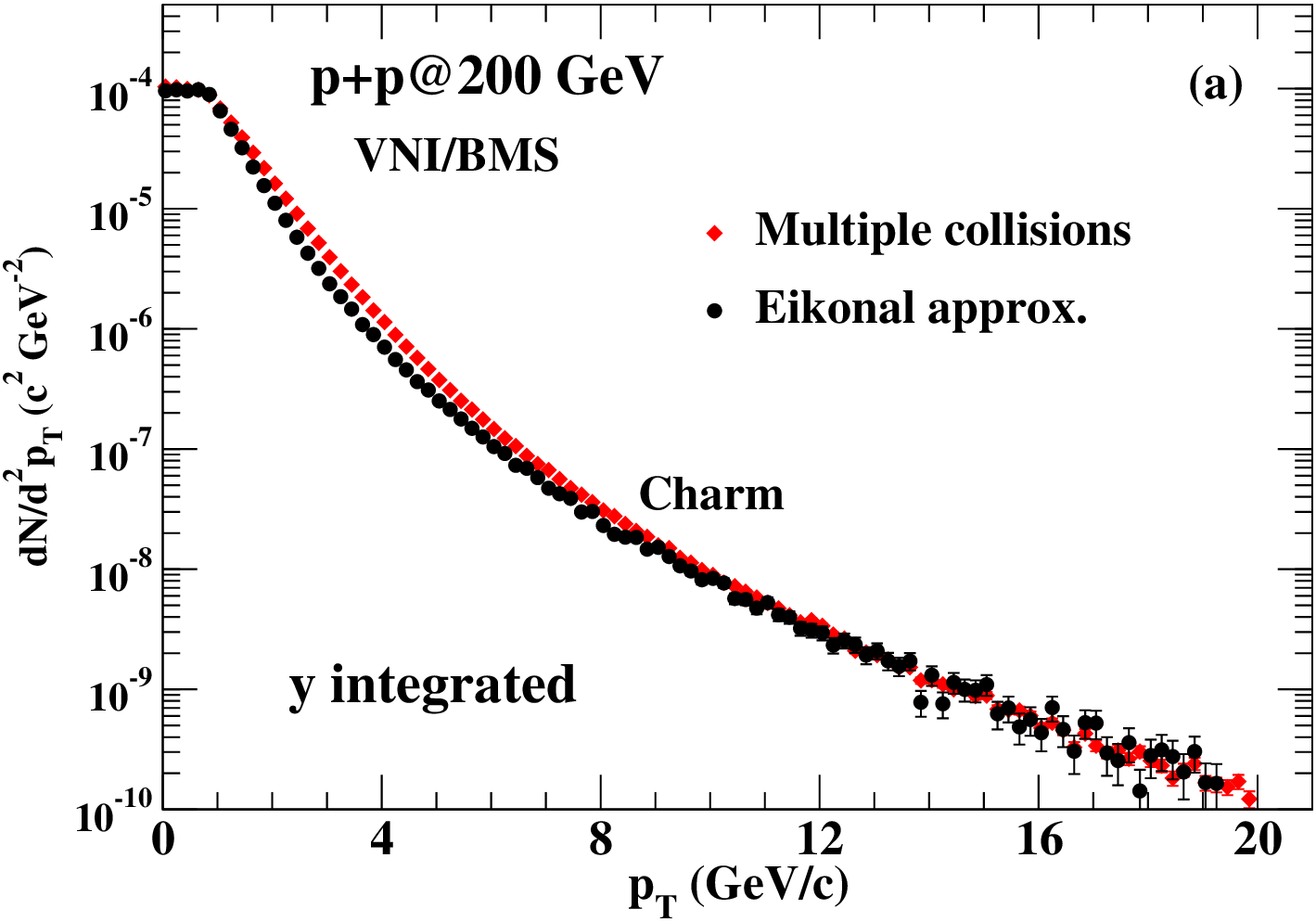}}
\centerline{\includegraphics*[width=8.0 cm]{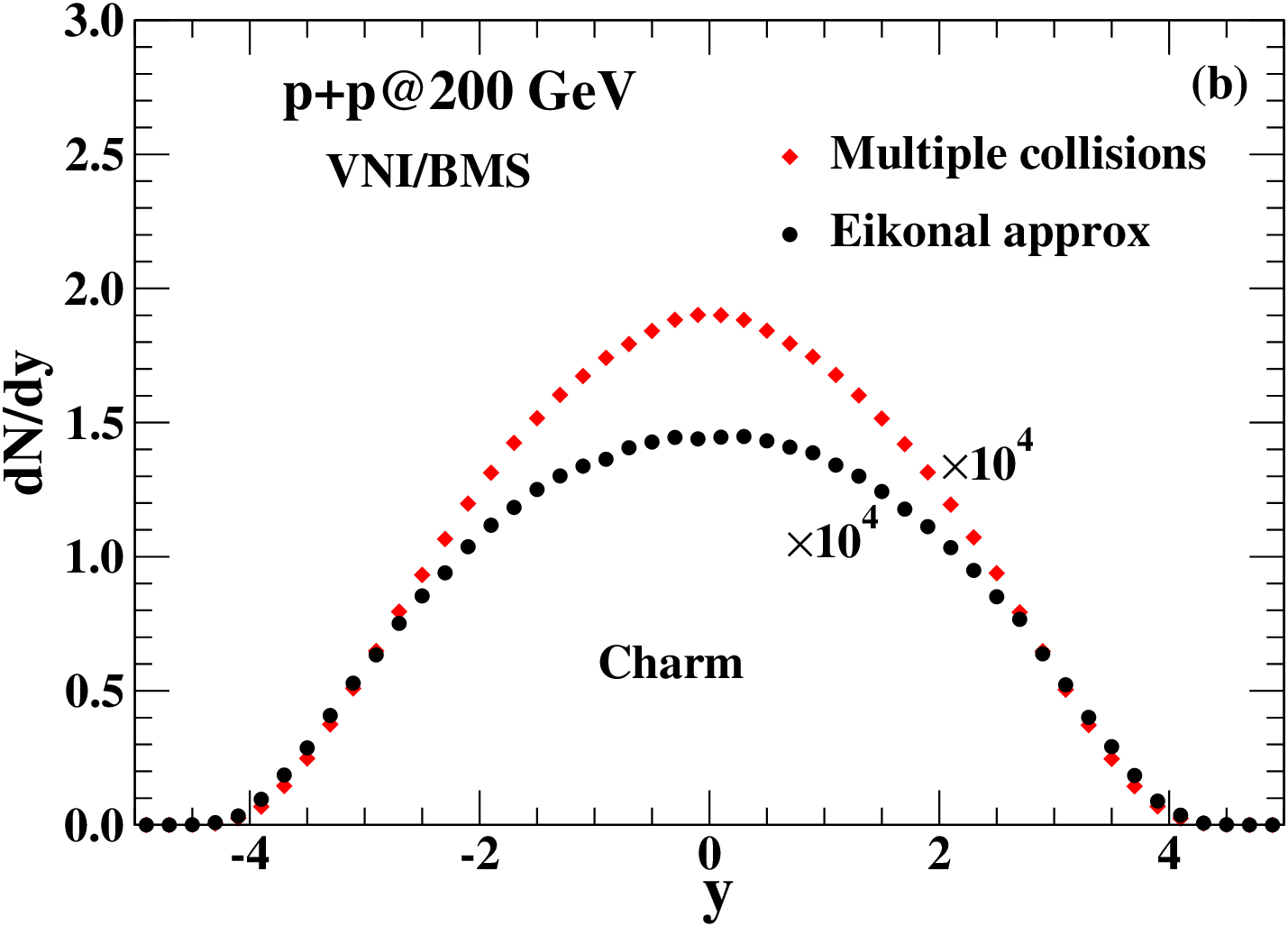}}
\caption{(colour online) (a) The rapidity integrated $p_T$ spectra and  (b) $p_T$
integrated rapidity spectra for charm quarks using eikonal approximation
and multiple collisions for $pp$ system at $\sqrt{s_\textrm{NN}}$ = 200 GeV.}
\label{pp_eik_mult}
\end{figure}

In order to test the accuracy of our implementation  of heavy quark production
in {\tt VNI/BMS} we first study collisions of protons at $\sqrt{s_{NN}}=$ 200 GeV.

We calculate the production of charm quarks due to
$gg\rightarrow Q\bar{Q}$ and $q\bar{q} \rightarrow Q\bar{Q}$ processes using a mini-jet calculation 
with a GRV-HO parametrization of the parton distribution function. In order
to compare it with our implementation of the PCM we fix $\alpha_s$ as 0.3 and
keep $Q_0^2$ fixed at 0.589 GeV$^2$ for the renormalization and factorization scales. Note that for a pure theory to theory comparison the specifics of the parton distribution function are not of particular importance, which is why we choose the well-established (even though somewhat dated) GRV-HO parametrization.

The minijet calculation is compared to a {\tt VNI/BMS} calculation using an eikonal approximation, where  the
final momenta and flavours of the particles produced in a hard scattering are  replaced by 
the momenta and flavours of the corresponding initial state partons. This procedure thus
mimics the eikonal approximation. A comparison of these two calculations can be found in figures~\ref{pp_eik_min} and~\ref{pp_eik_min_y}:
the rapidity integrated transverse momentum spectrum and the $p_T$ integrated rapidity spectrum of the produced charm (or anti-charm)  are given in Fig~\ref{pp_eik_min} and the $p_T$ spectra at several rapidities are given in Fig.~\ref{pp_eik_min_y}.
We observe satisfactory agreement between the {\tt VNI/BMS} calculation and the minijet reference. The minor differences at very low and very
high $p_T$ arise due to the $p_T$-cut-off in the parton cascade model, which is also  reflected in $x_\textrm{min}$ which enters in the parton cascade model. This difference is again reflected in the rapidity distribution as well.

Having established reasonable  quantitative agreement between the two independent calculations
and thus confirming that  heavy quark interactions are correctly implemented in the Monte Carlo scheme of {\tt VNI/BMS},
we proceed to study the dynamics of the cascade evolution.
As a first step we consider a calculation in which only primary-primary parton interactions are
permitted, i.e. each parton is  allowed to interact only once. A comparison of these calculations with the results of the eikonal approximation 
then immediately highlights the effect of multiple parton interactions in the eikonal approximation, already in $pp$ collisions. Again we show the rapidity integrated transverse momentum spectrum and the $p_T$ integrated
rapidity spectrum of the produced charm (or anti-charm) in Fig~\ref{pp_eik_prim}. 
 The spectra are quite similar
in shape, but differ by the factor which accounts for the increased number of collisions in
the eikonal approximation.
Details of the
evolution of the cascade are  quantified in Table~\ref{table1}.
We see from Table~\ref{table1} that the leading process which contributes to the production of charm 
quarks, i.e.,  $gg \rightarrow c\overline{c}$, is about 50\% more likely in the eikonal approximation 
compared to the case when only primary-primary collisions are allowed. 

\begin{figure}
\centerline{\includegraphics*[width=8.0 cm]{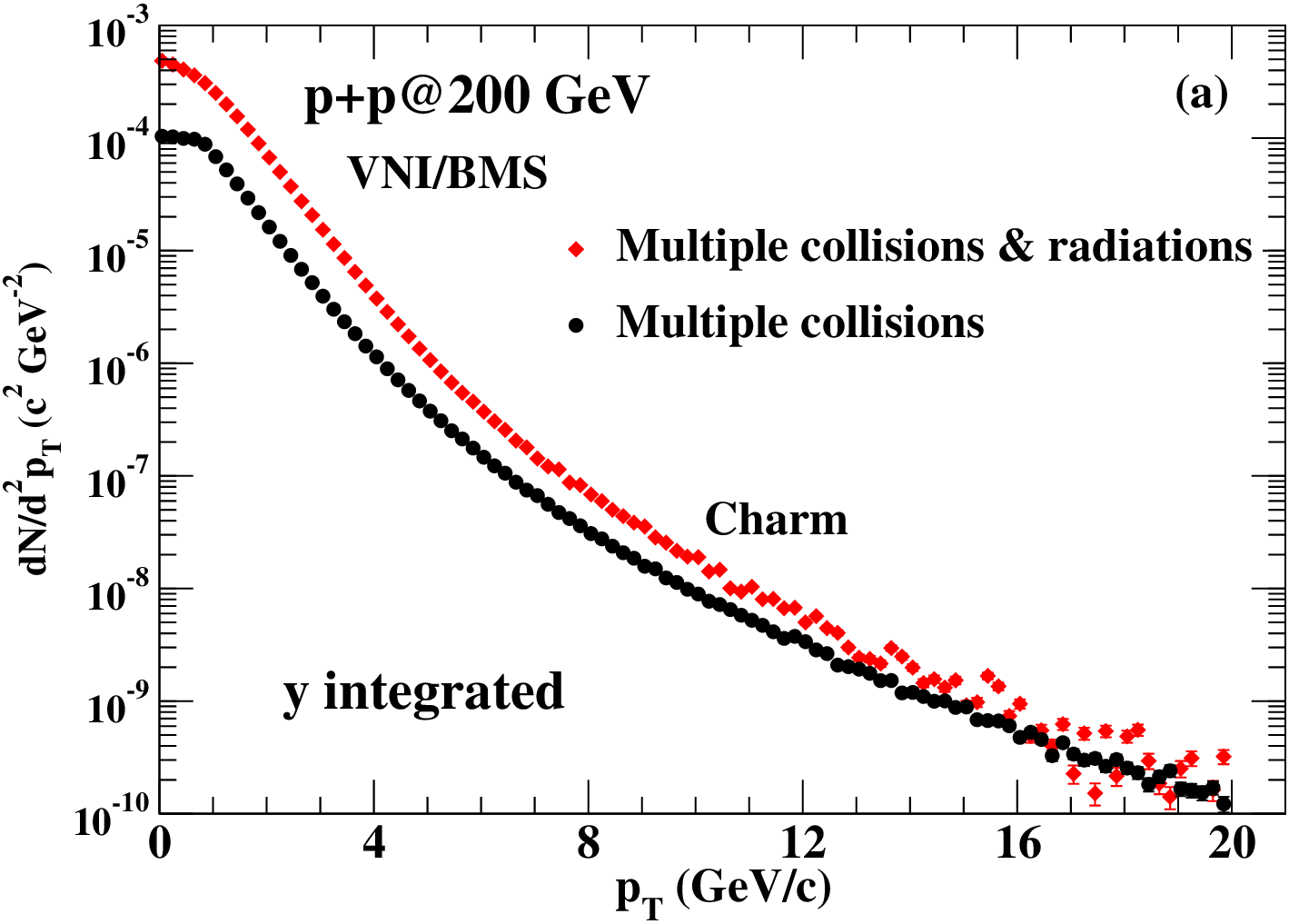}}
\centerline{\includegraphics*[width=8.0 cm]{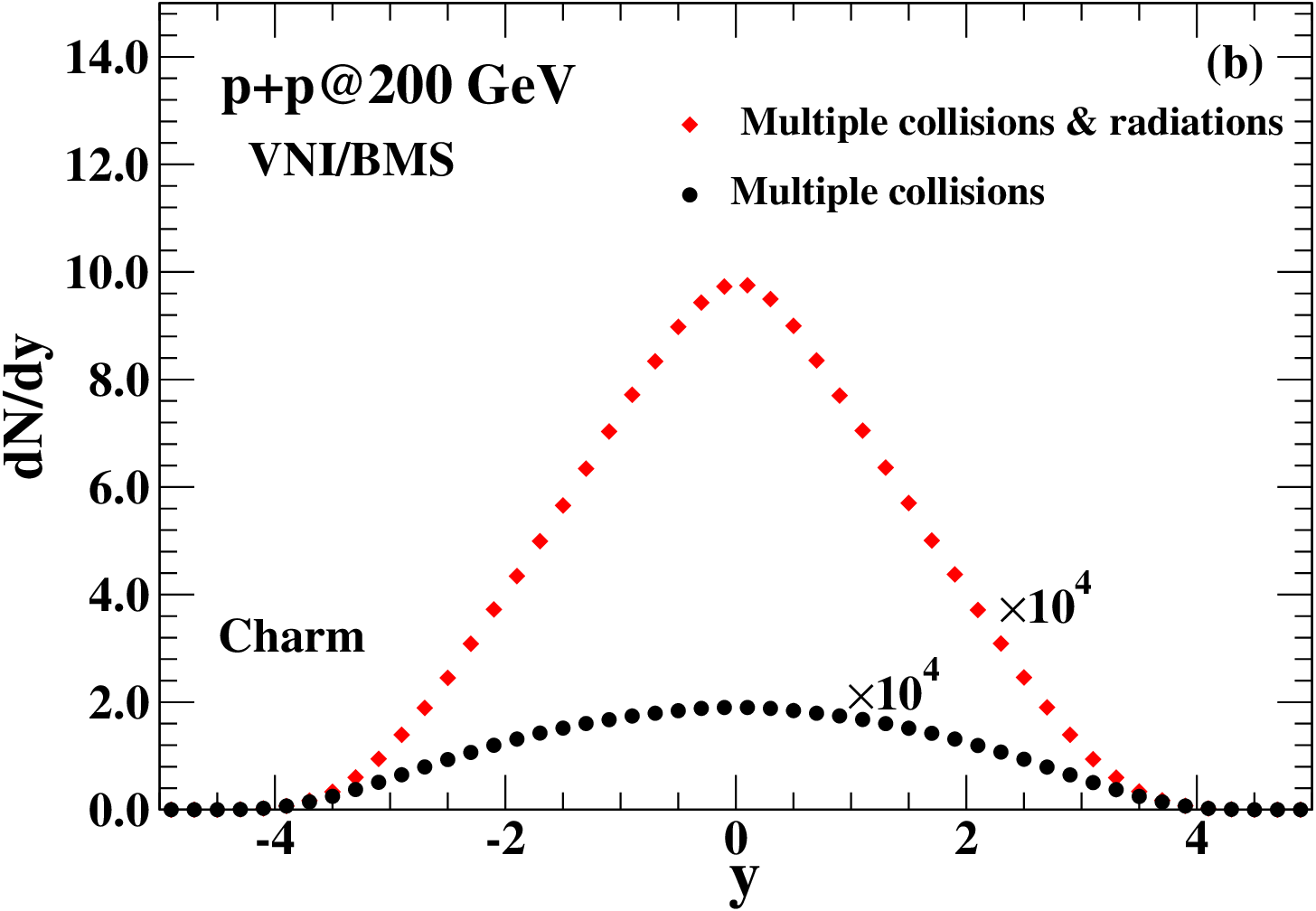}}
\caption{(colour online) (a) The rapidity integrated $p_T$ spectra and (b) $p_T$
integrated rapidity spectra for charm quarks using multiple collisions only
and multiple collisions as well as fragmentation of partons scattered off the final state partons,
following a semi-hard scattering in evolution of the cascade for
 $pp$ system at $\sqrt{s_\textrm{NN}}$ = 200 GeV.}
\label{pp_full_mult}
\end{figure}

\begin{figure}
\centerline{\includegraphics*[width=8.0 cm]{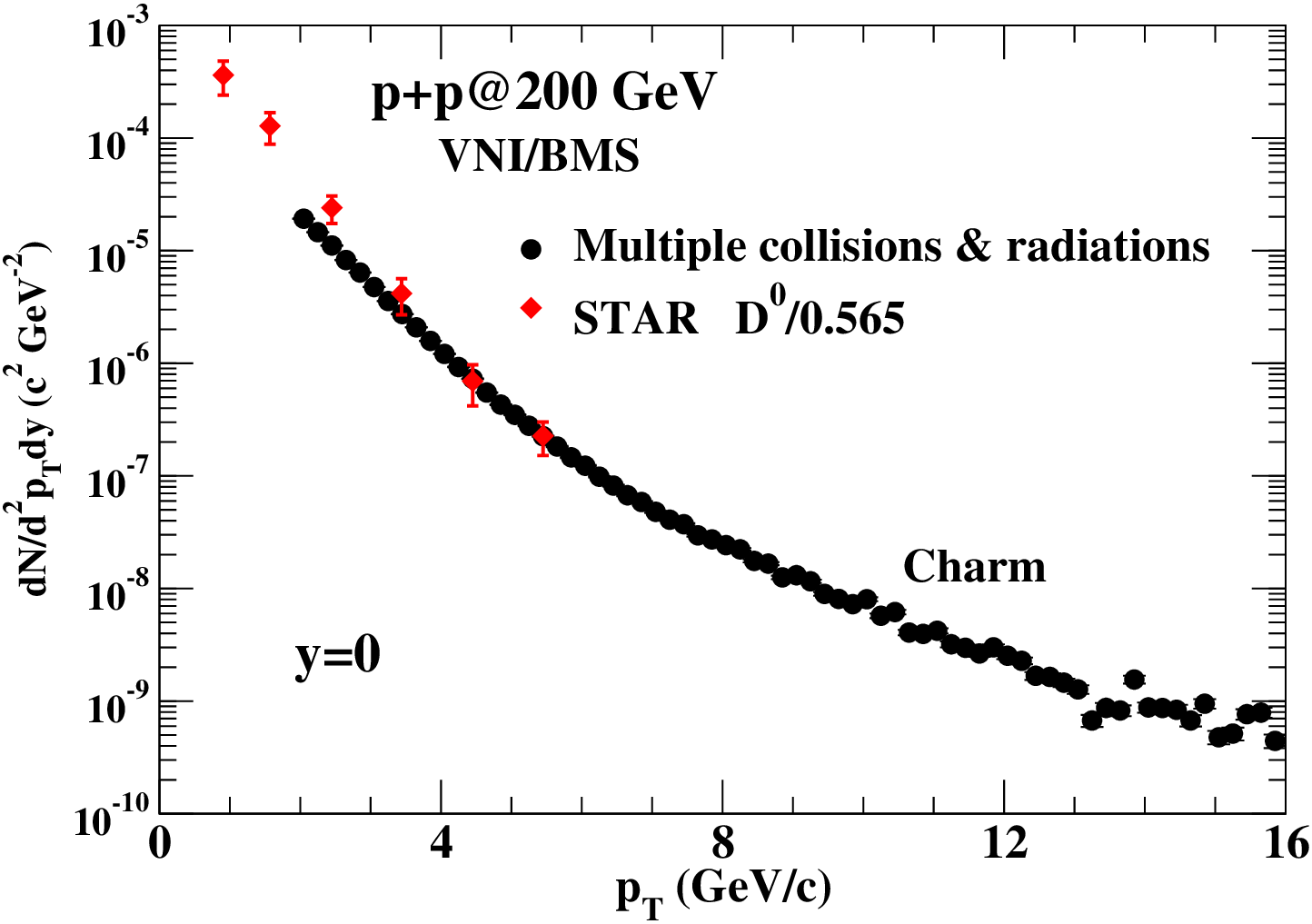}}
\caption{(colour online) The $p_T$ spectra
for charm quarks using multiple collisions as well as fragmentation of partons scattered off the final state partons,
following a semi-hard scattering in evolution of the cascade for
 $pp$ system at $\sqrt{s_\textrm{NN}}$ = 200 GeV. The data are from the STAR experiment\cite{Ye:2014ska}}.
\label{pp_data}
\end{figure}

\begin{figure}
\centerline{\includegraphics*[width=8.0 cm]{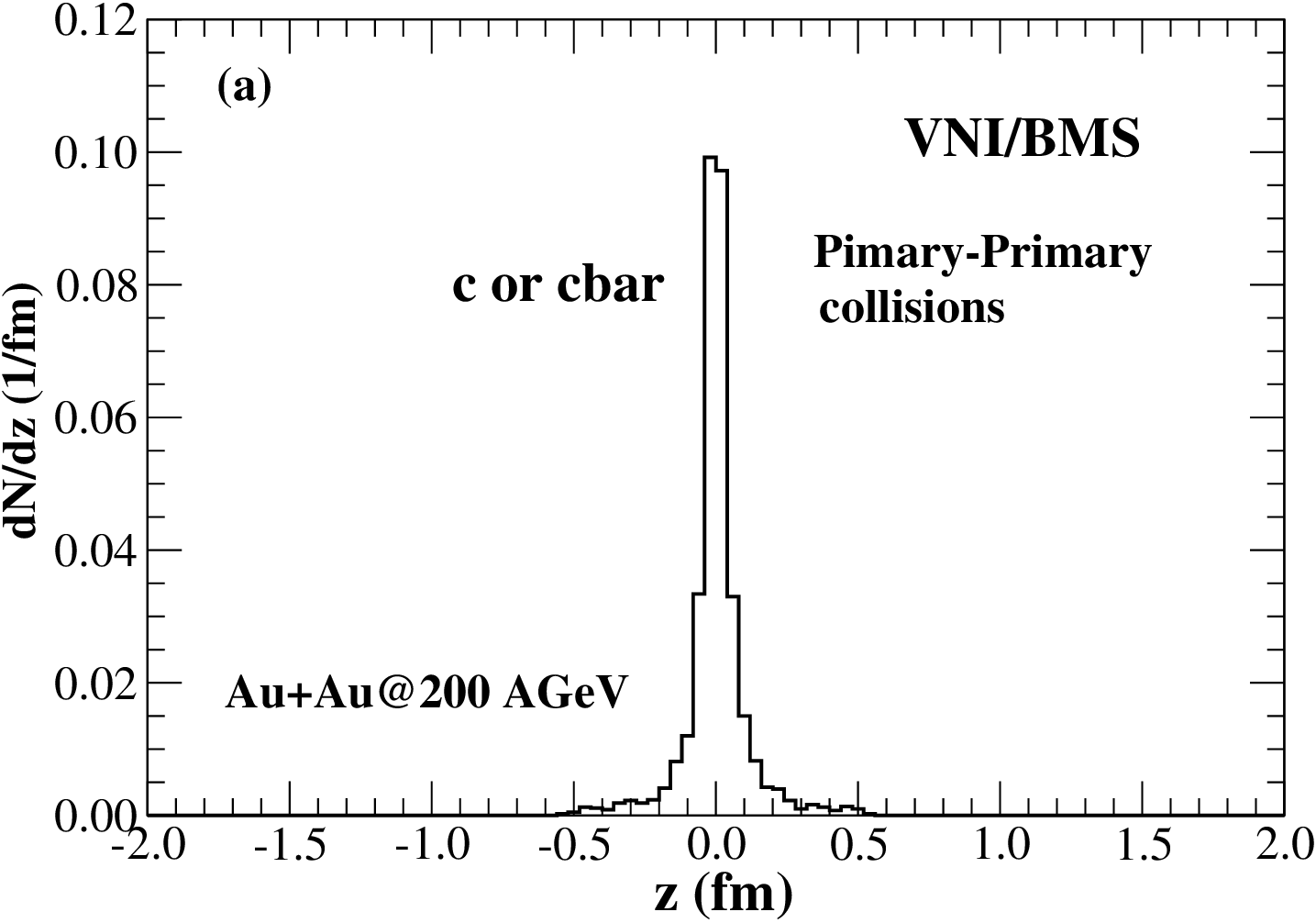}}
\centerline{\includegraphics*[width=8.0 cm]{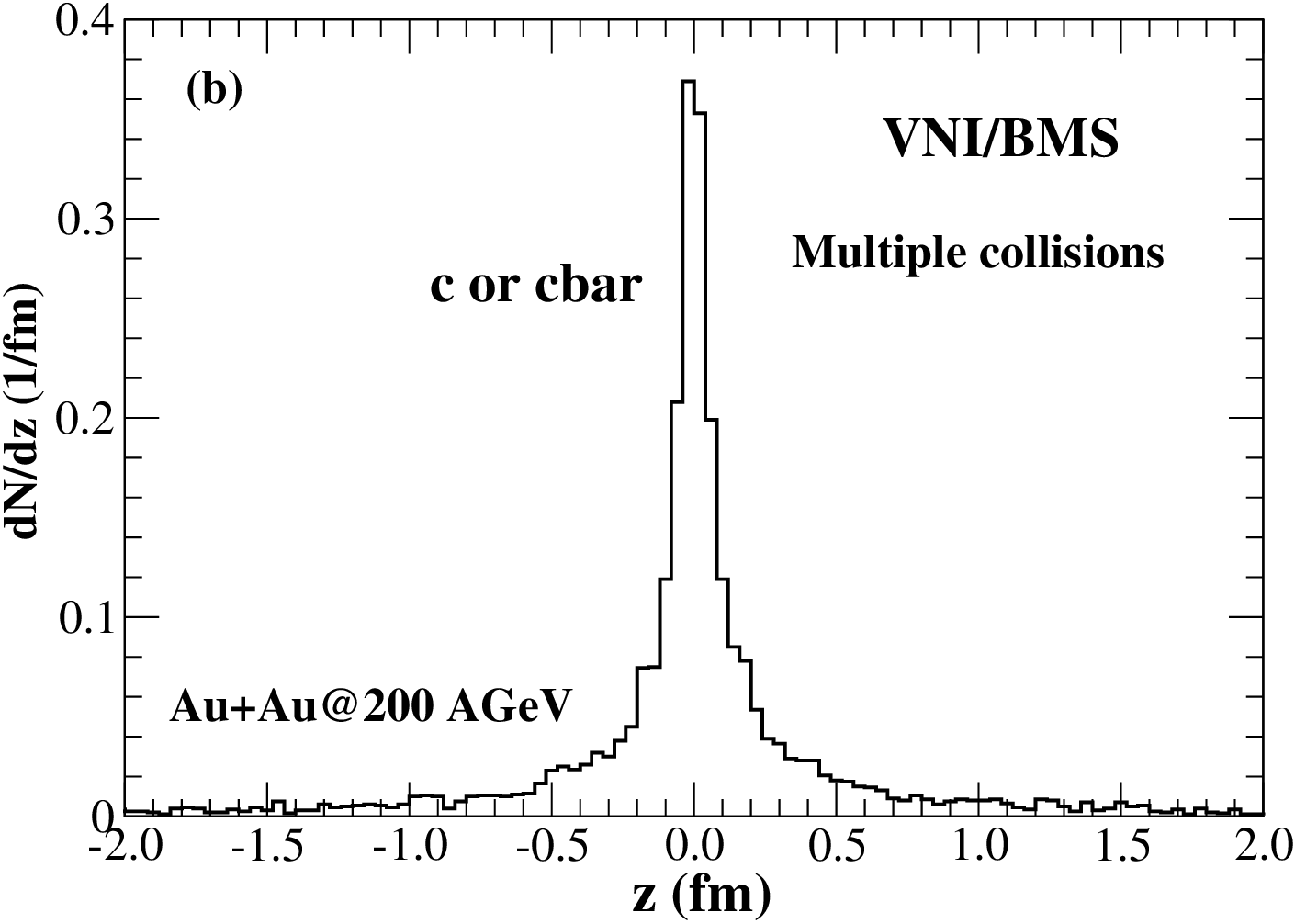}}
\centerline{\includegraphics*[width=8.0 cm]{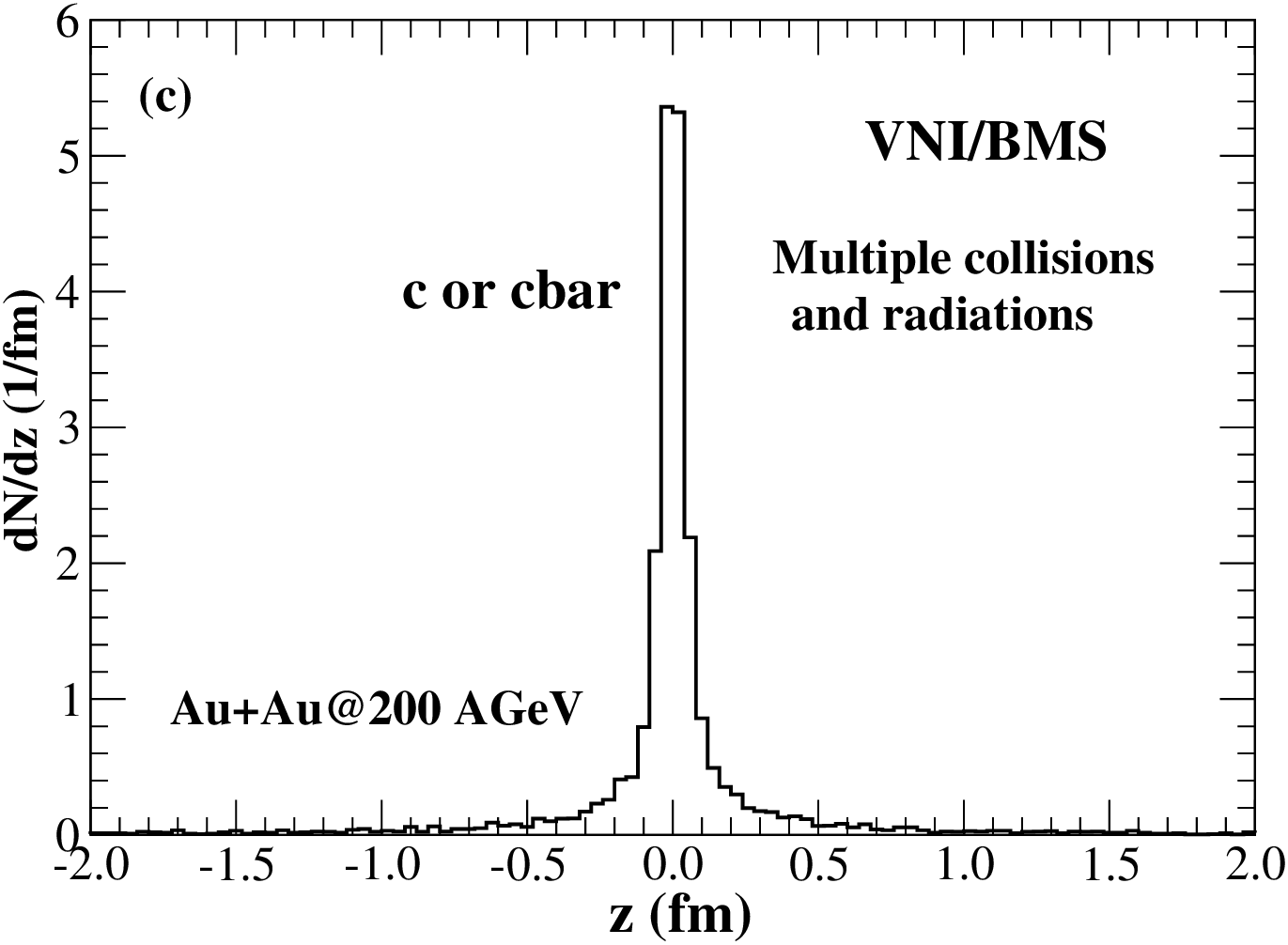}}
\caption{The distribution of $z$ co-ordinates of production vertices of
charm (or anti-charm) quarks for the cases involving only primary-primary collisions (a), only multiple collisions (b) and multiple collisions and radiations of gluons (c).}
\label{dndz}
\end{figure}

\begin{figure}
\centerline{\includegraphics*[width=8.0 cm]{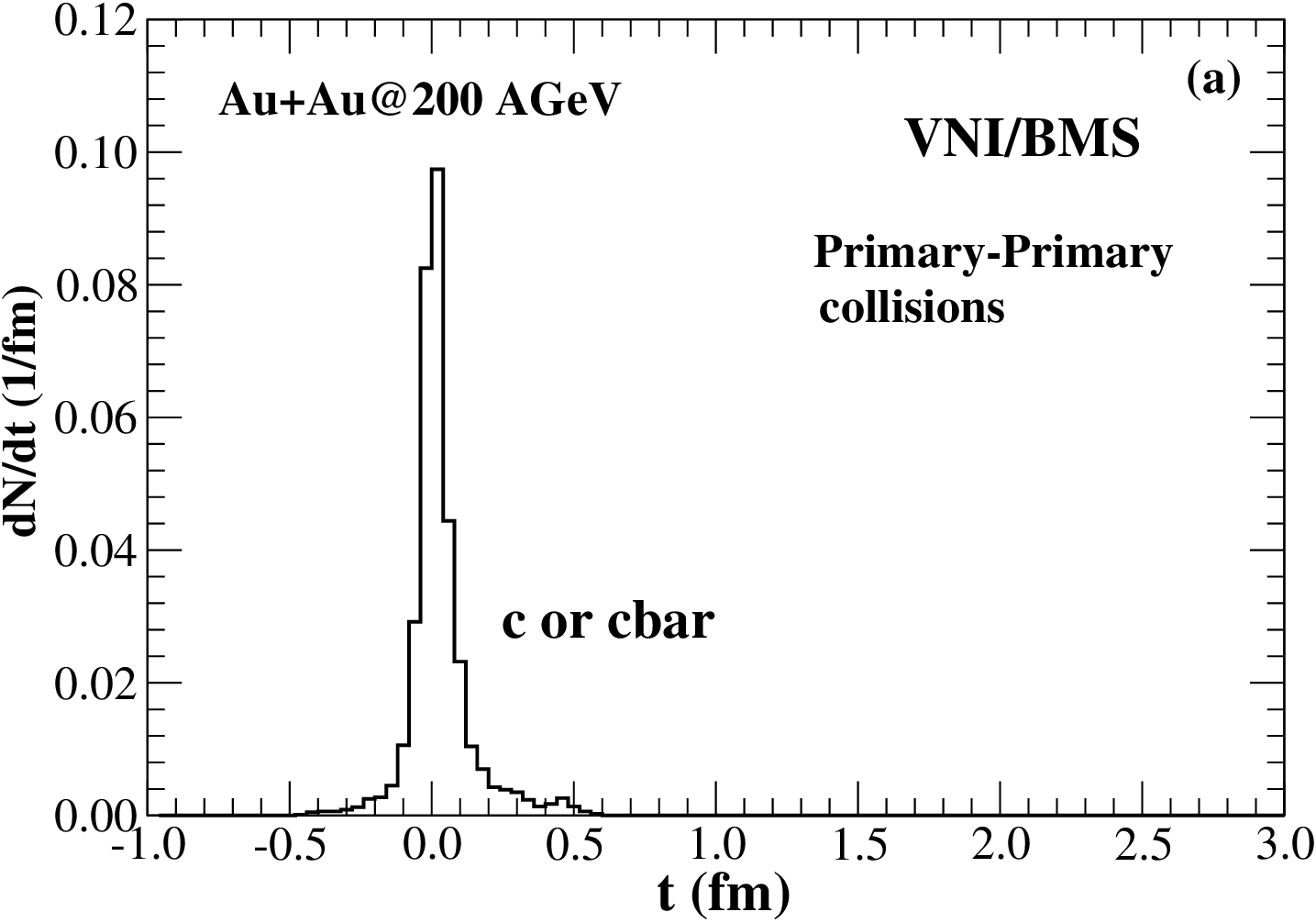}}
\centerline{\includegraphics*[width=8.0 cm]{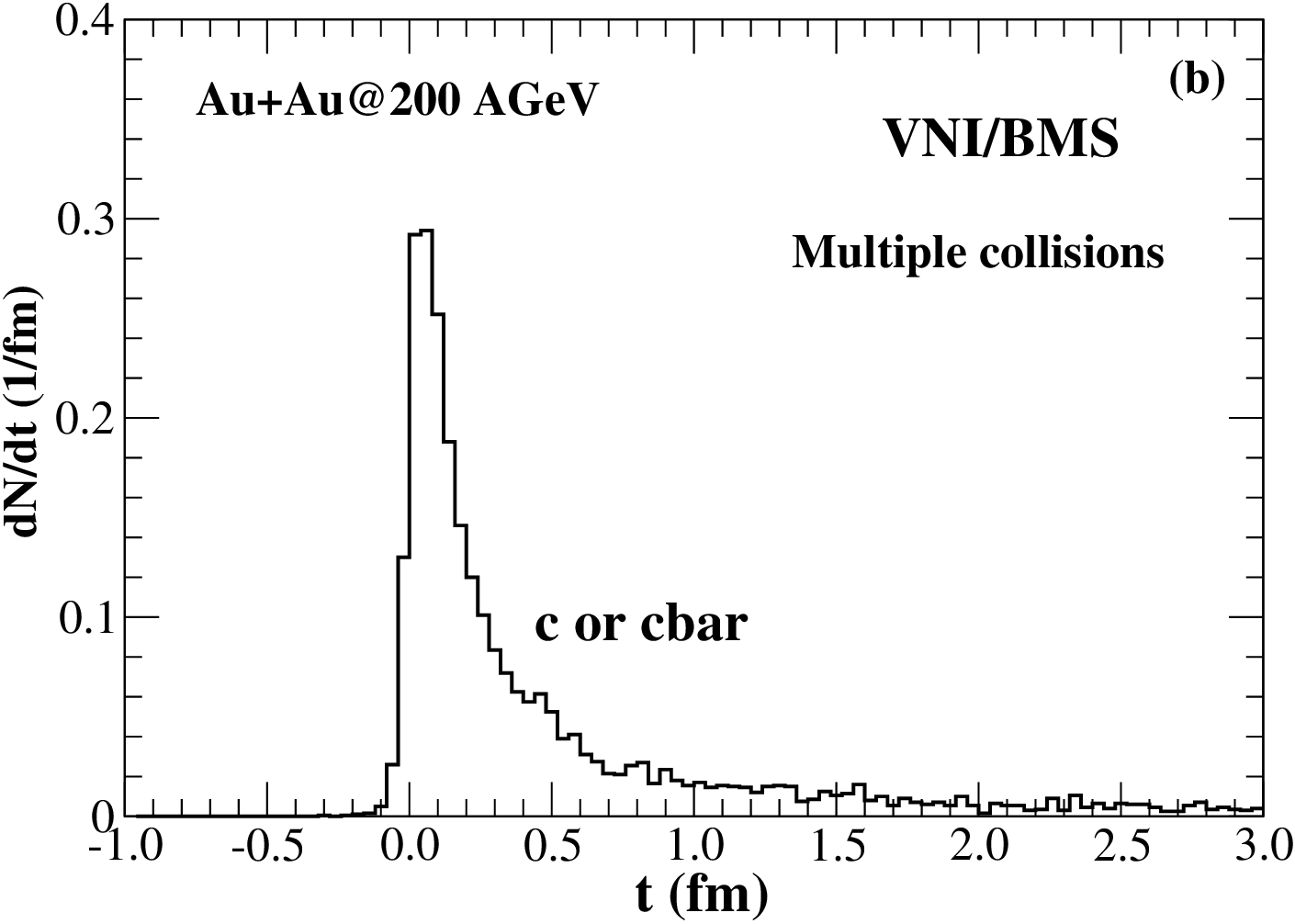}}
\centerline{\includegraphics*[width=8.0 cm]{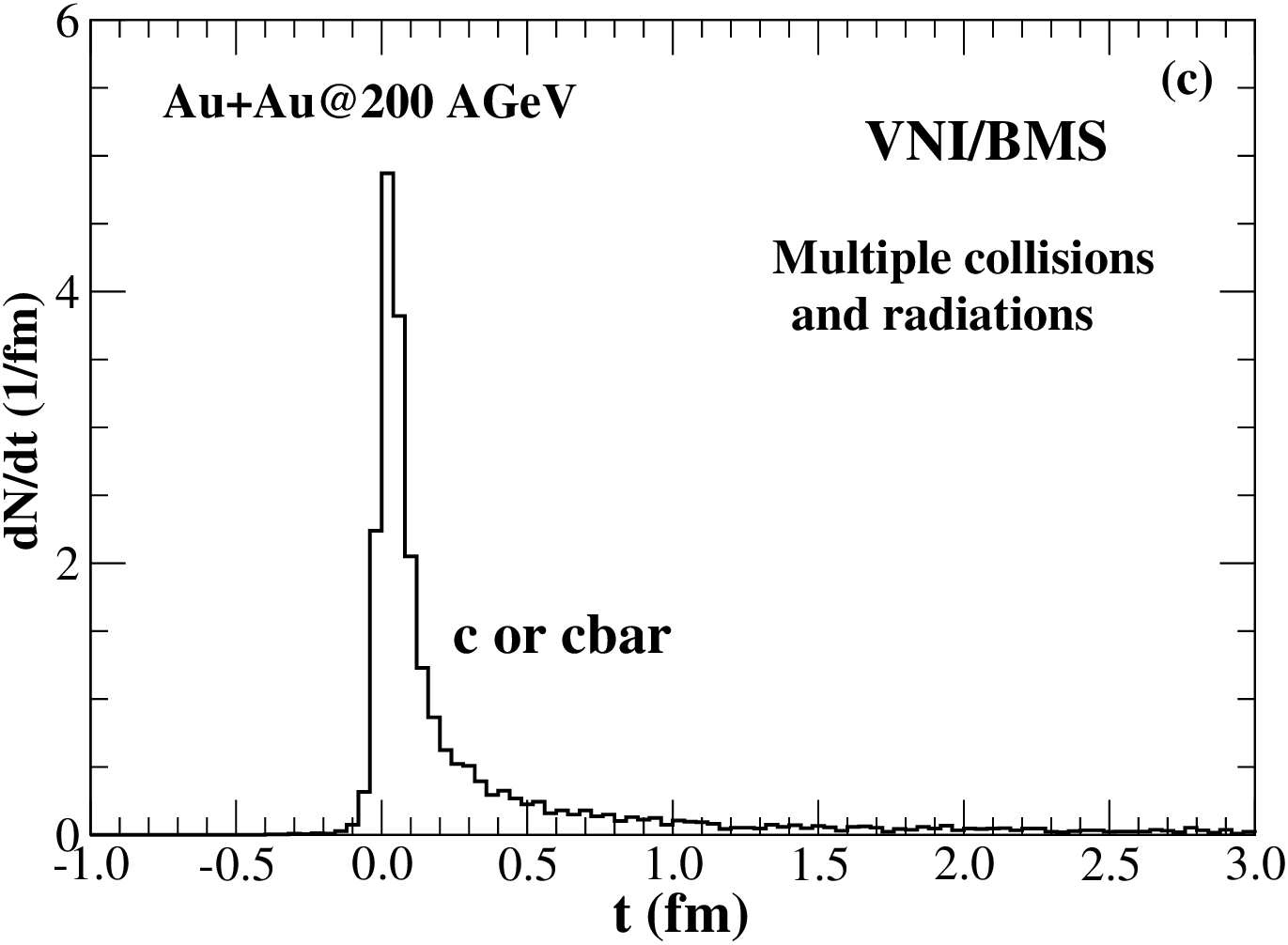}}
\caption{The distribution of production times of
charm (or anti-charm) quarks for the cases involving only primary-primary collisions (a), only multiple collisions (b) and multiple collisions and radiations of gluons (c). Note that $t=0$ is defined to occur when the two nuclei are in full overlap.}
\label{dndt}
\end{figure}

Now we proceed to results with  the proper accounting of multiple scatterings 
along with the 
changes in the flavour and momenta of the produced partons,
before they scatter again. These studies, we believe,  will help us identify the effects
of multiple scatterings already at the level of p+p collisions.
 
Looking at Table~\ref{table1}, 
we find there is an increase in the
number of collisions by about 20\% compared to the eikonal approximation.  
Compared to the primary-primary collision calculation, the increase is about 60\%. 
While the increase in the number of collisions for the case of 
multiple scatterings over and above that of the case of primary-primary
collisions is 
expected, the increase in the number of collisions above that for
the eikonal approximation comes as a surprise. This may be due to interactions that impart a 
large momentum to partons that previously would have been below the momentum cutoff for an interaction now acquiring sufficient transverse momentum for subsequent interactions.

The rapidity integrated $p_T$ and the $p_T$ integrated
rapidity spectra are shown in Fig.~\ref{pp_eik_mult}.
We see that a proper accounting of the changing momenta and the flavours of the
partons following a hard scattering enhances the production of charm quarks
at moderate transverse momenta. This effect will undoubtedly be even more prominent in
 $Au+Au$ collisions, where the probability of
multiple scatterings increases considerably. However there are already indications of non-uniform
variation of the factor by which the two calculations differ as a function of $p_T$ 
and $y$, presumably due to the larger density of interacting partons at midrapidity.

\begin{figure}
\centerline{\includegraphics*[width=8.0 cm]{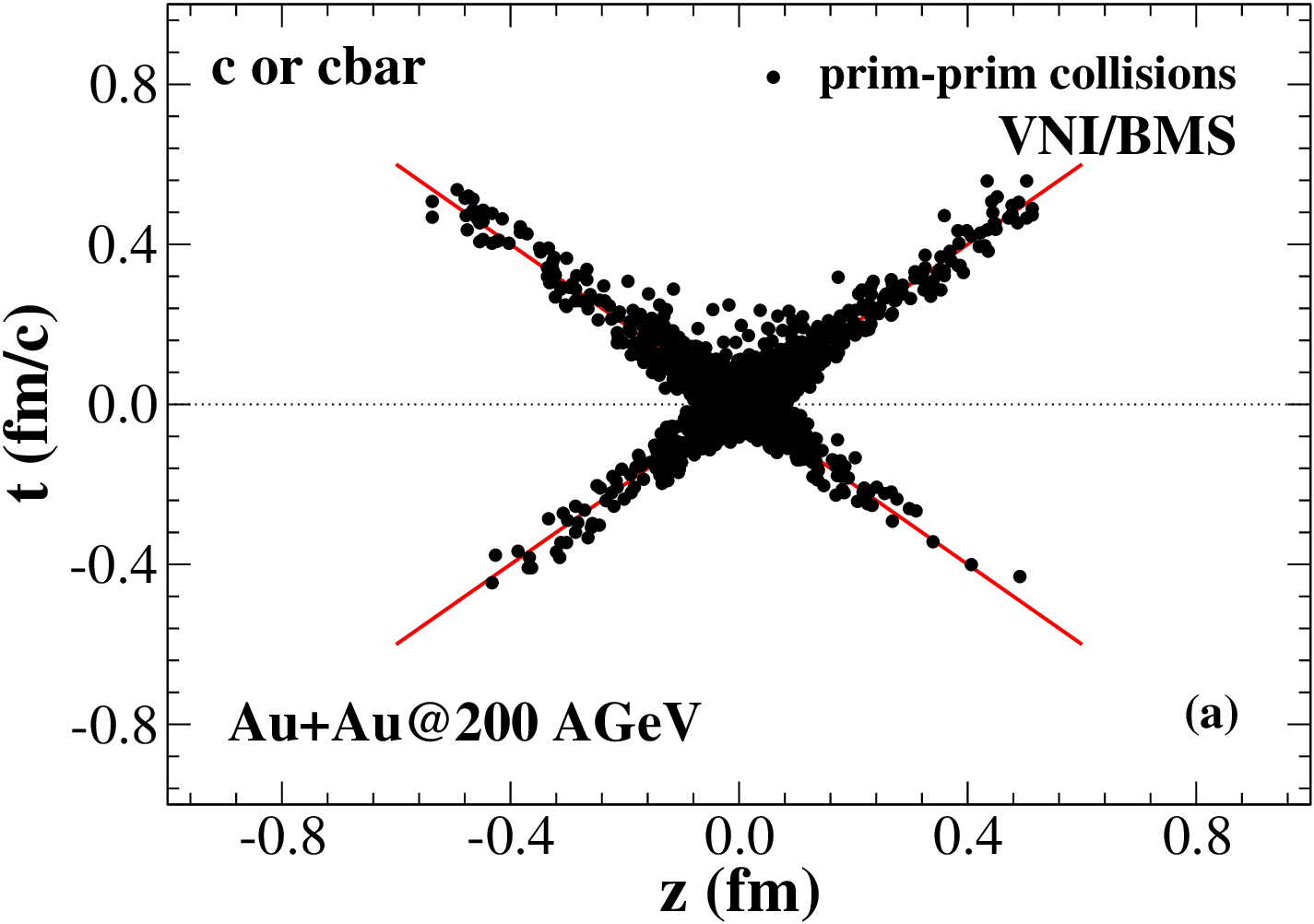}}
\centerline{\includegraphics*[width=8.0 cm]{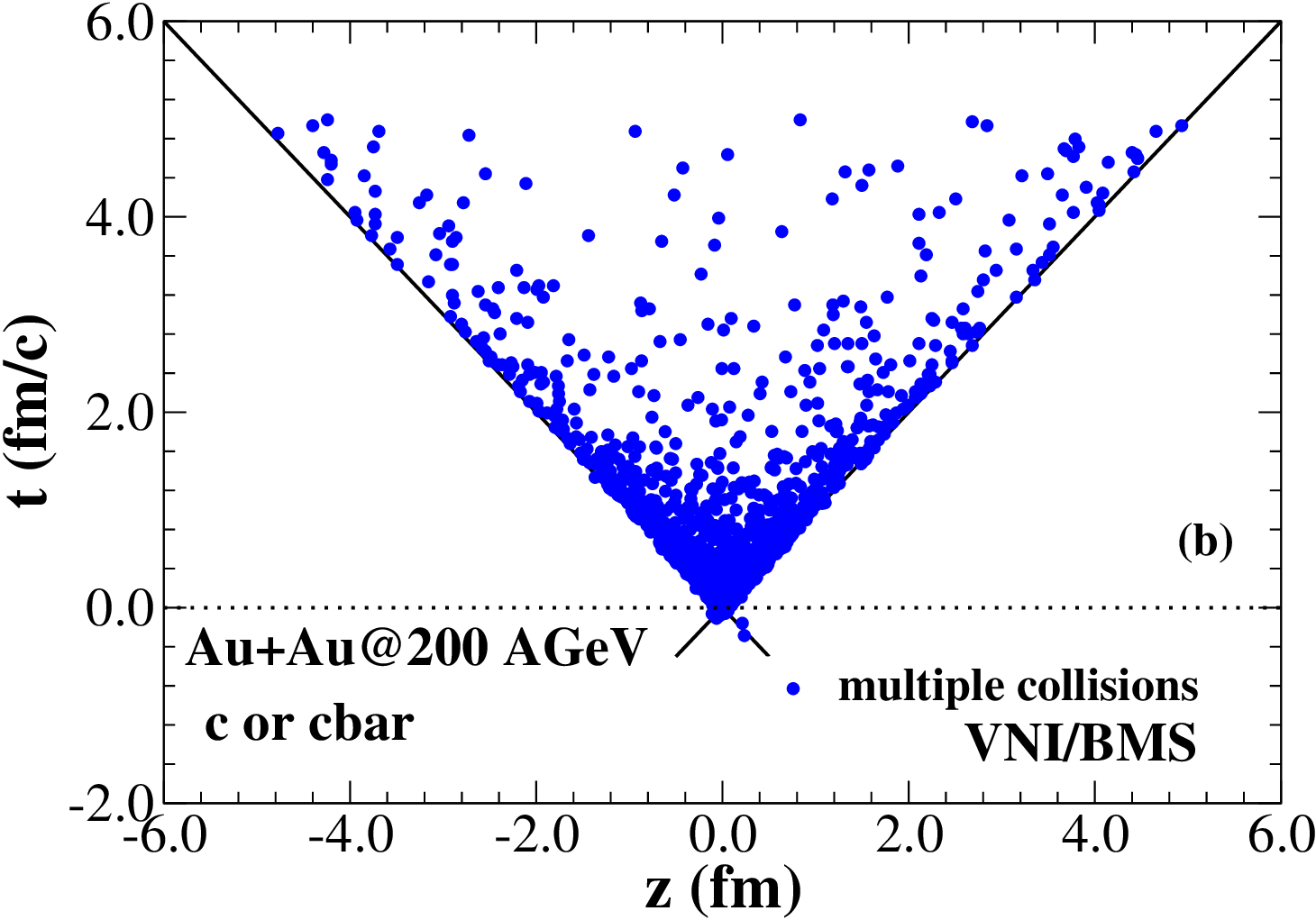}}
\centerline{\includegraphics*[width=8.0 cm]{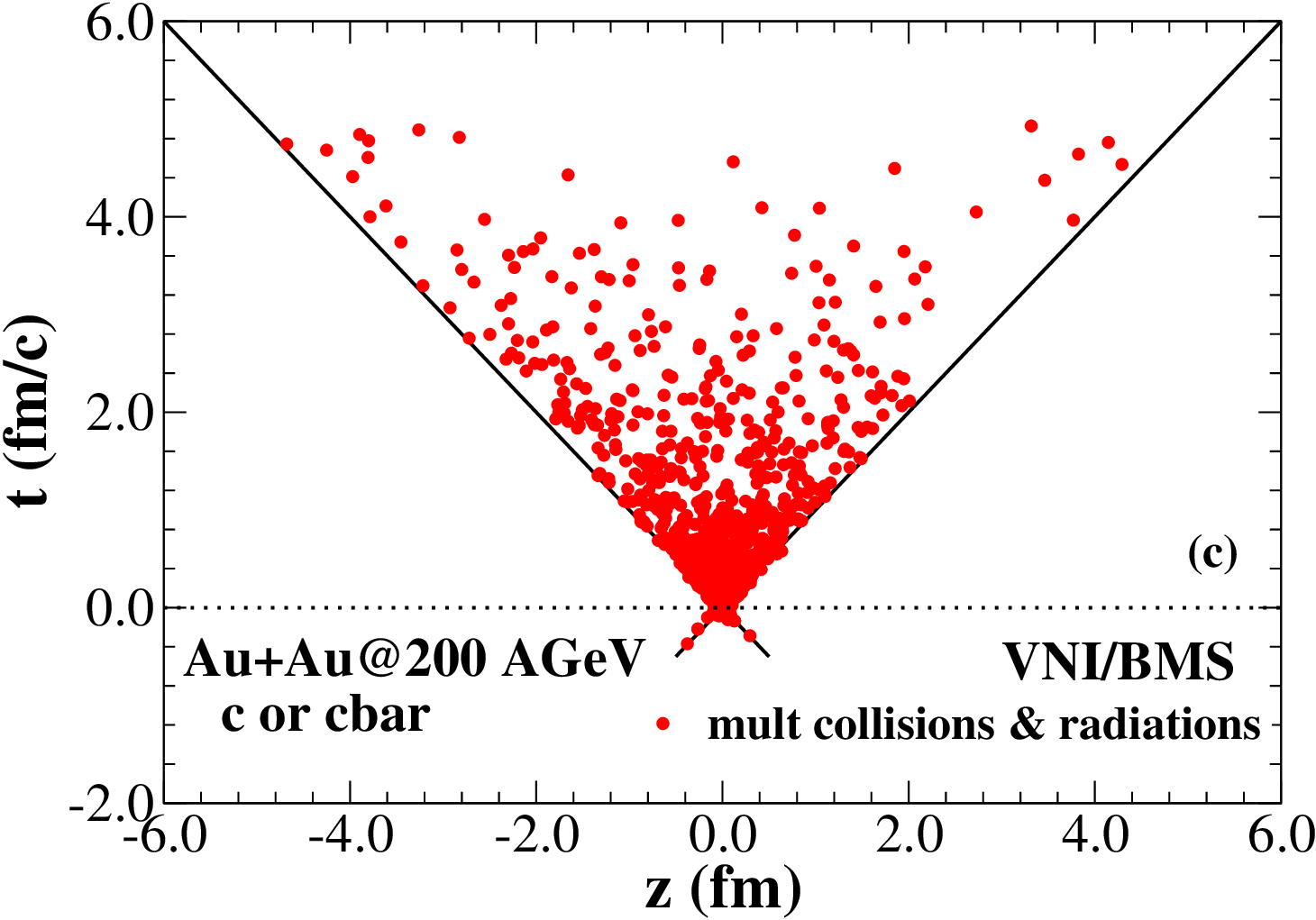}}
\caption{(Colour online)The $z$ and $t$ co-ordinates of production vertices of
charm (or anti-charm) quarks for the cases involving only primary-primary collisions (a), only multiple collisions (b) and multiple collisions and radiations of gluons (c). Note the change of scale along
the $z$ axis for the primary-primary case.}
\label{aa_tz}
\end{figure}
\begin{figure}
\centerline{\includegraphics*[width=8.0 cm]{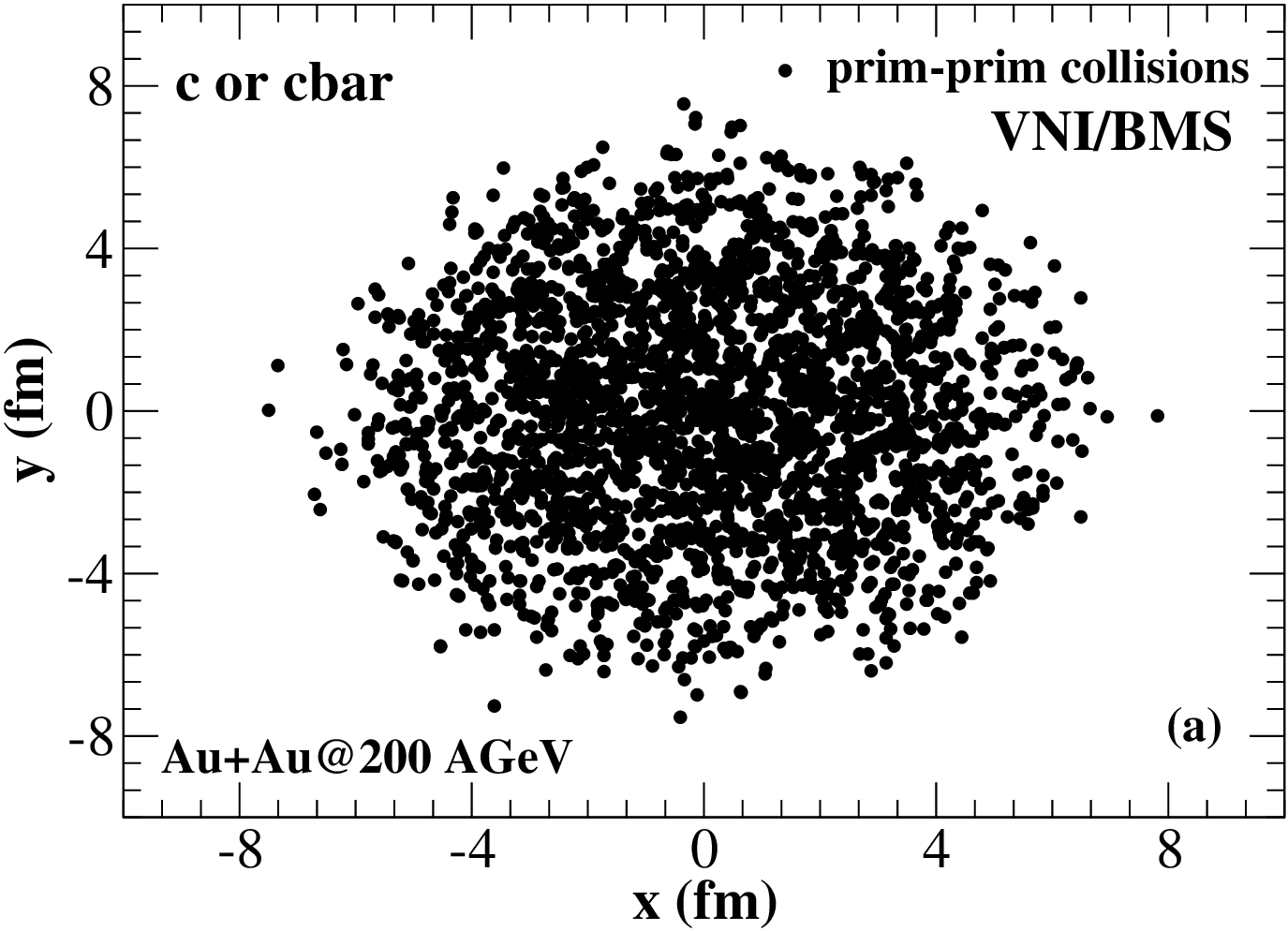}}
\centerline{\includegraphics*[width=8.0 cm]{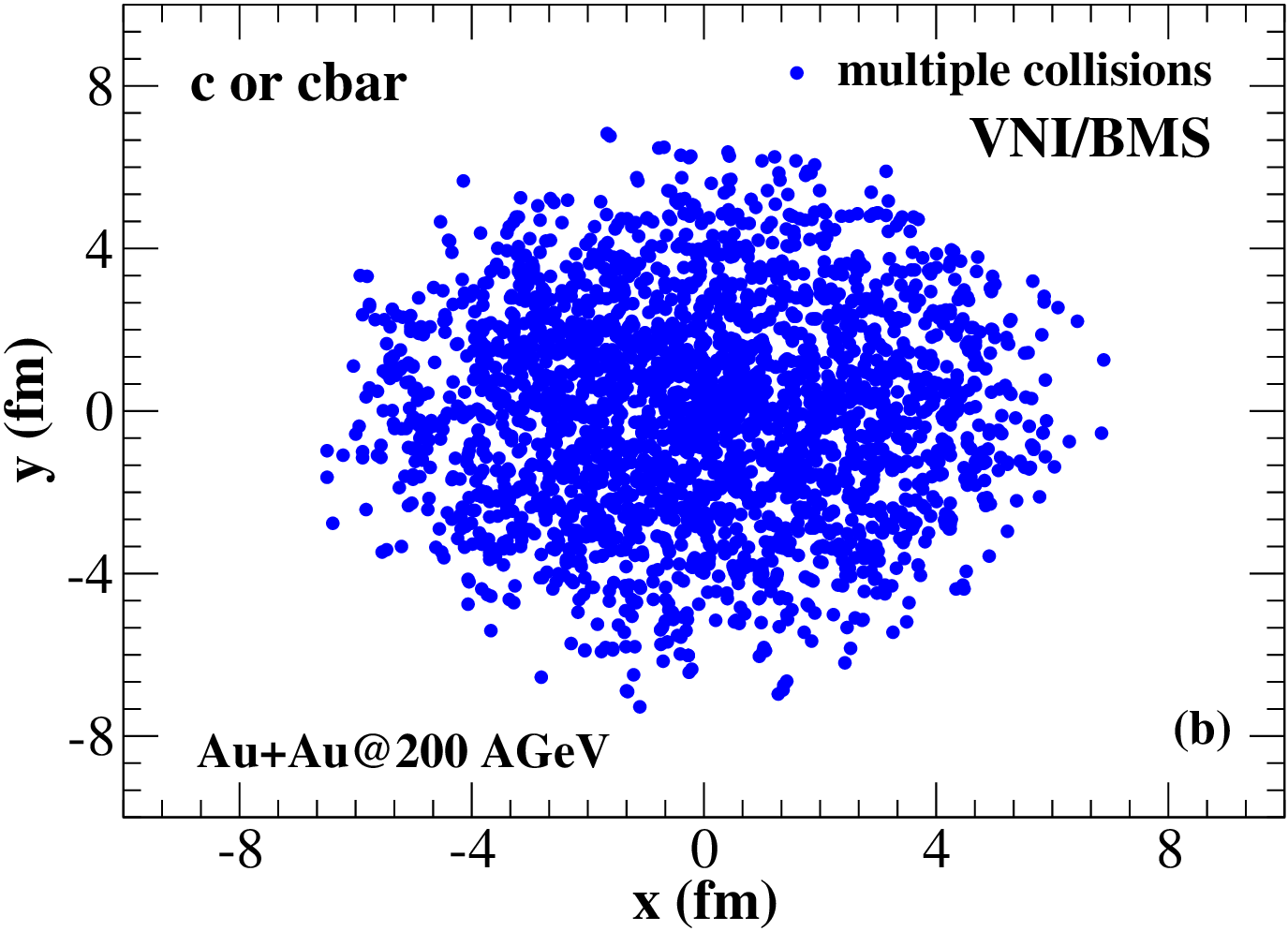}}
\centerline{\includegraphics*[width=8.0 cm]{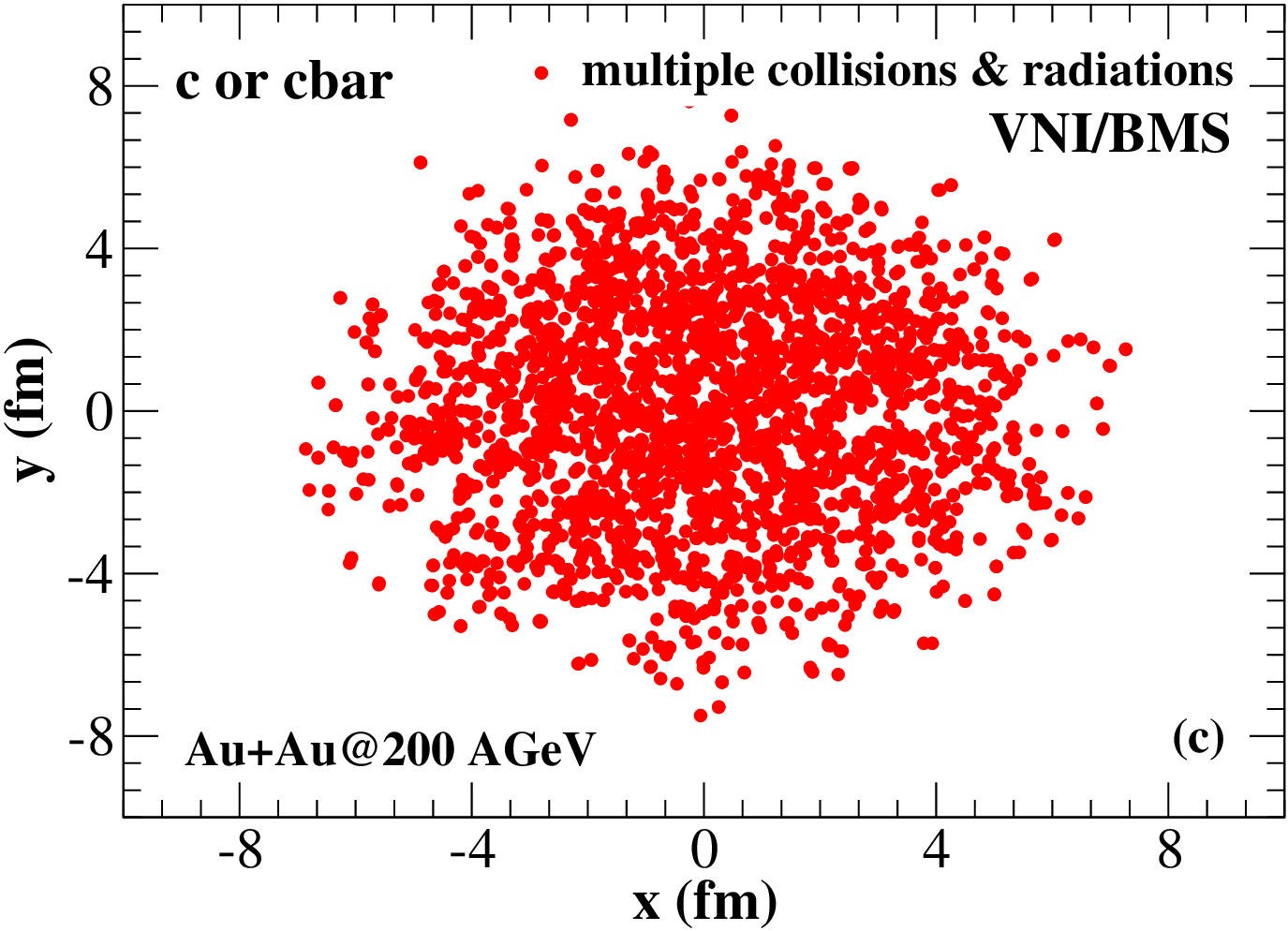}}
\caption{(Colour on-line) The $x$ and $y$ co-ordinates of production vertices of
charm (or anti-charm) quarks for the cases involving only primary-primary collisions (a), only multiple collisions (b) and multiple collisions and radiations of gluons (c).}
\label{aa_xy}
\end{figure}

\begin{figure}
\centerline{\includegraphics*[width=8.0 cm]{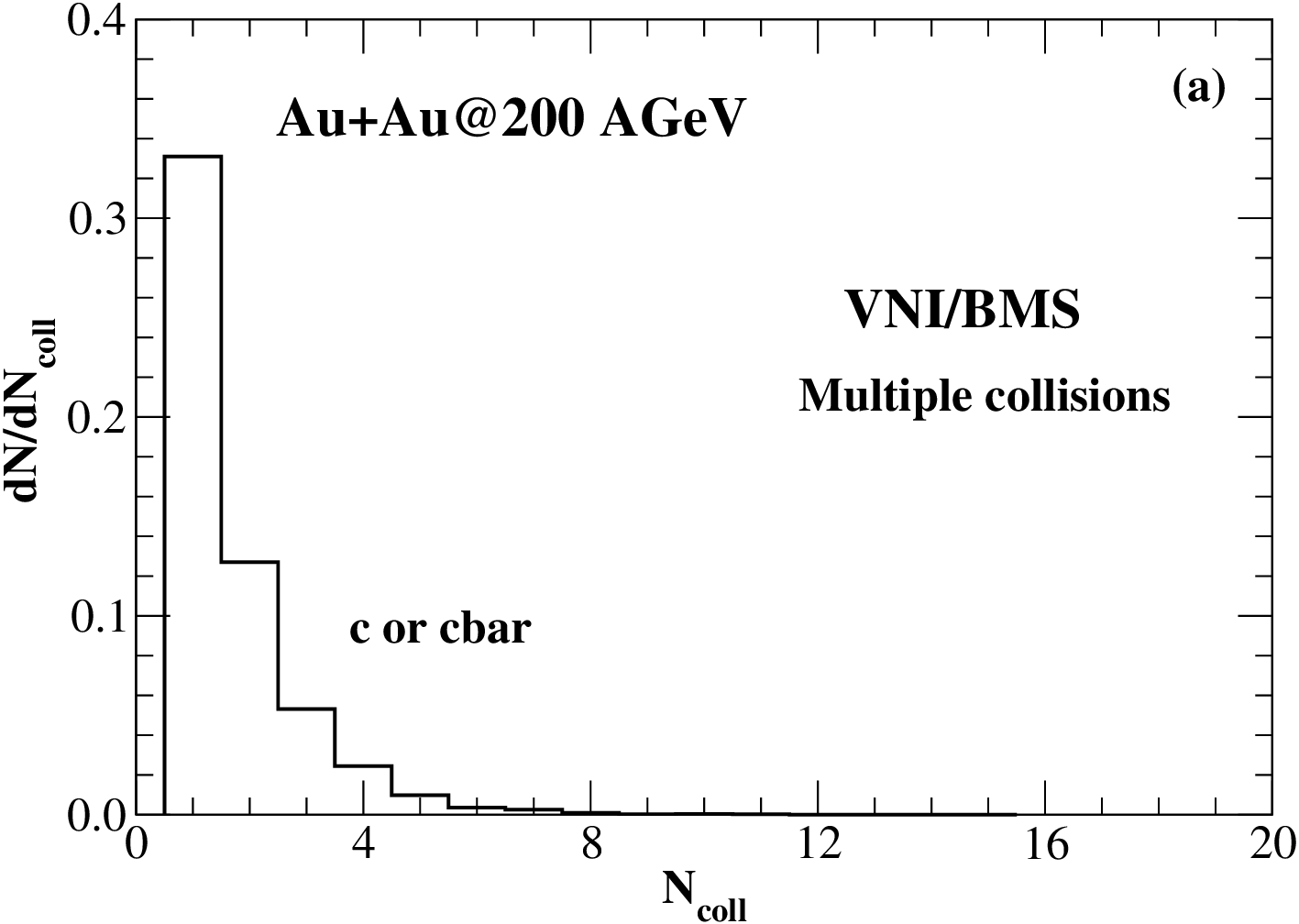}}
\centerline{\includegraphics*[width=8.0 cm]{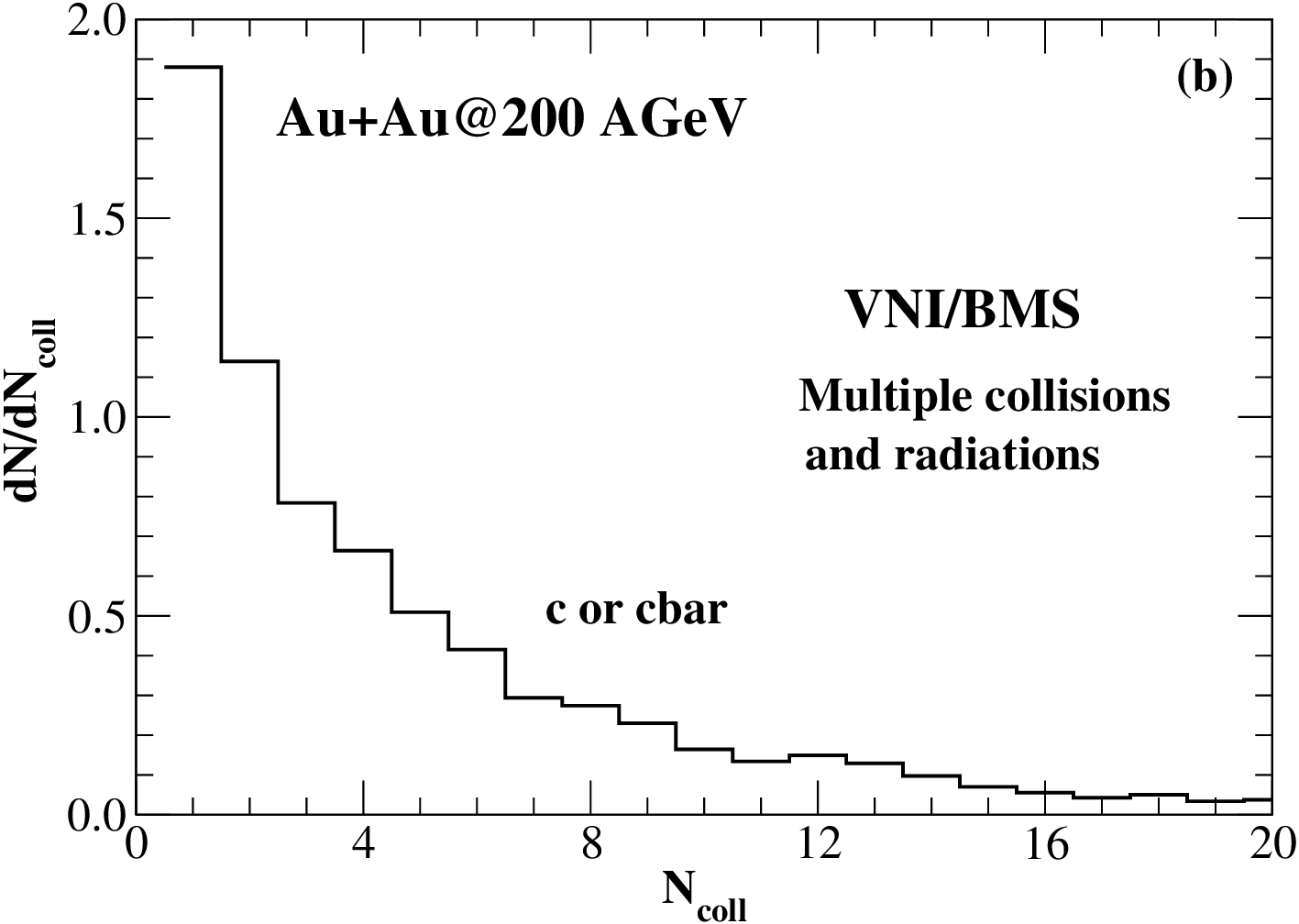}}
\caption{ (a) The distribution of number of collisions encountered by charm quarks when only multiple scattering among partons is permitted and (b) when both multiple scattering and radiations are permitted .}
\label{aa_coll}
\end{figure}

\begin{figure}
\centerline{\includegraphics*[width=8.0 cm]{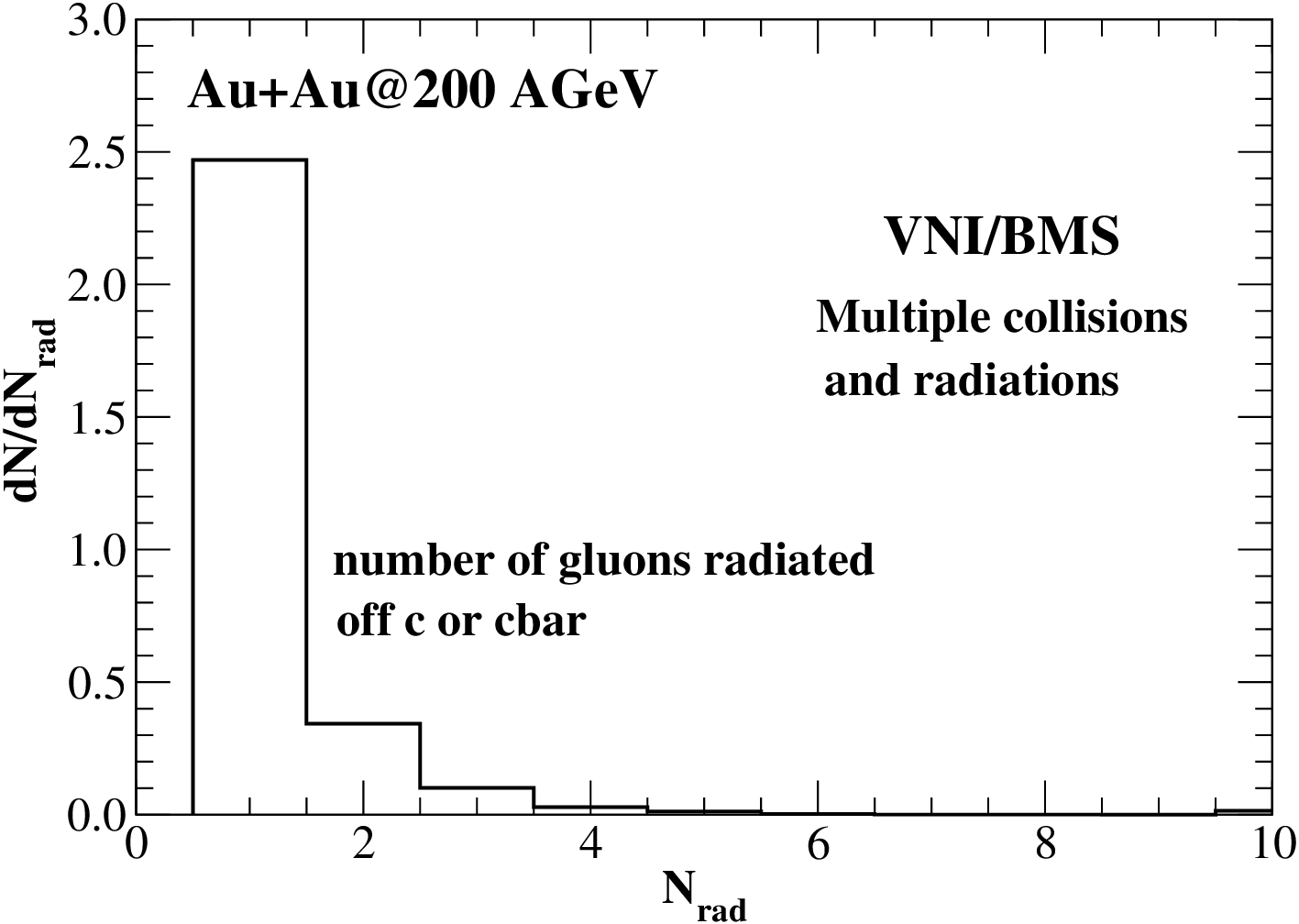}}
\caption{ The distribution of number of gluons radiated off charm quarks following a semi-hard scattering.}
\label{aa_rad}
\end{figure}

Finally, we include time-like branchings of partons in our calculation, which
leads to a rapid increase in the number of partons at midrapidity and subsequent multiple
scatterings. As expected, we observe  a considerable increase in the number of hard collisions, fueled
by the time-like branchings which leads  (mainly) to gluon multiplication.
Once again we notice that the increase in $gg \rightarrow q\bar{q}$ 
is reflected in the increase of the production of charm quarks.
This close correspondence may be due to fact that the medium generated in
$pp$ collisions is not yet dense and long lived enough to have several scatterings
by partons which can reduce their momenta such that further collisions do not
produce heavy quarks. The momenta of charm quarks at medium and large $p_T$ are also 
greatly affected (see Fig.~\ref{pp_full_mult}) while a comparison of the rapidity
distribution suggests increased interactions at central rapidities (Fig.~\ref{pp_full_mult}). 

\begin{figure}
\centerline{\includegraphics*[width=8.0 cm]{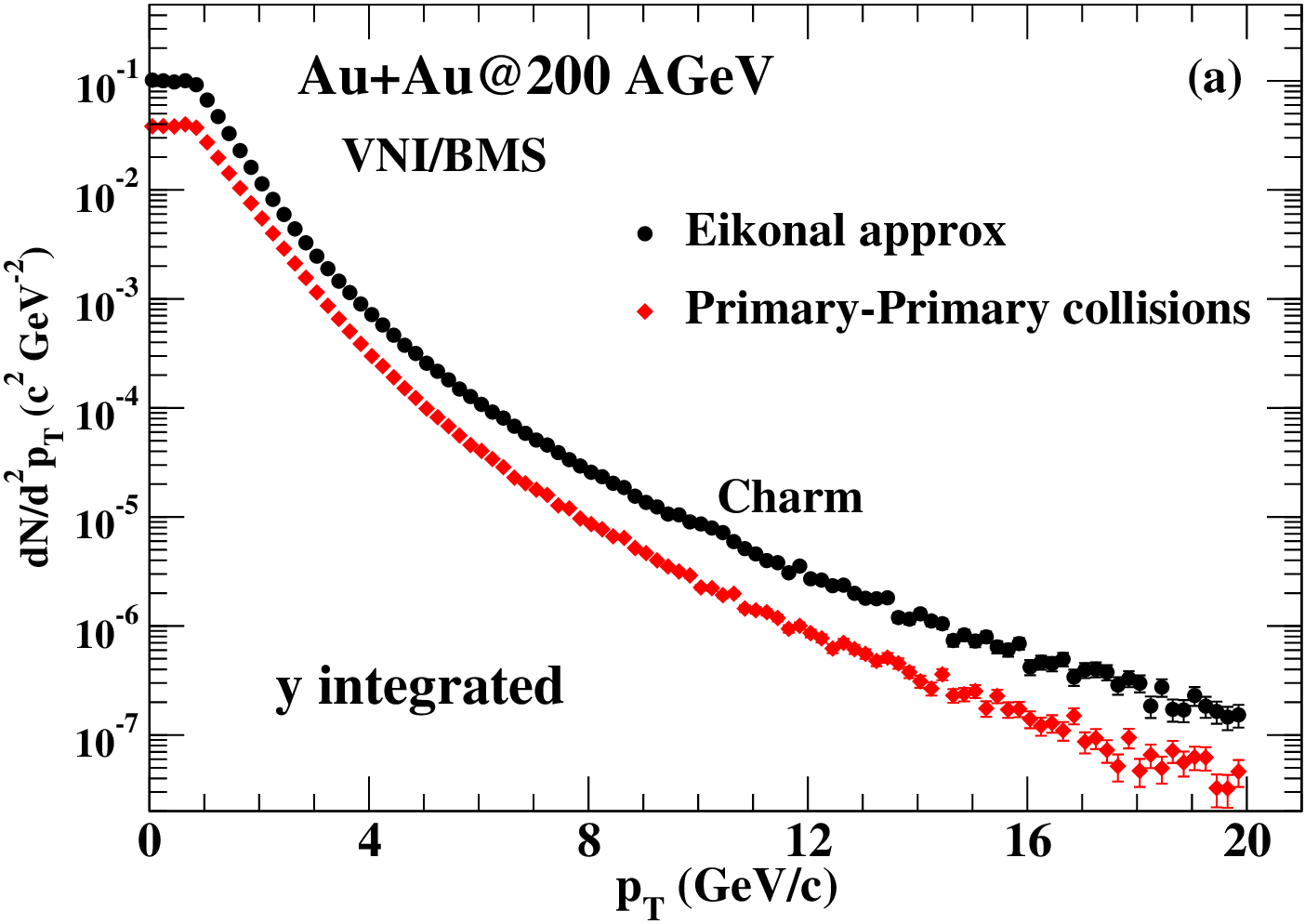}}
\centerline{\includegraphics*[width=8.0 cm]{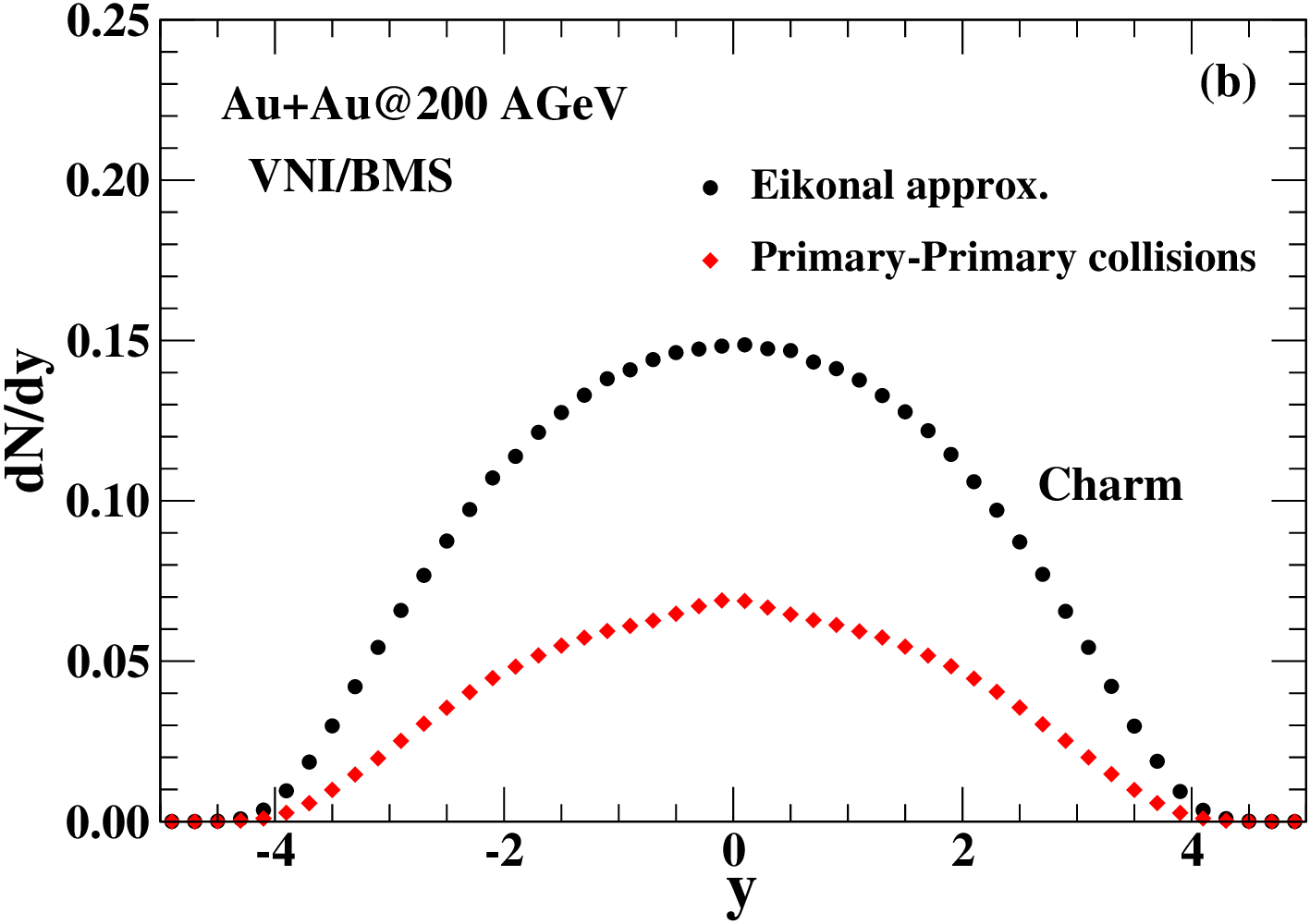}}
\caption{(colour online) (a) The $y$ integrated $p_T$ spectra  of charm quarks
using eikonal approximation and only primary-primary collisions for $Au+Au$ system at
$\sqrt{s_\textrm{NN}}$ = 200 GeV.
 The lower panel (b) gives the $p_T$ integrated rapidity
spectra for the same cases.}
\label{aa_eik_prim}
\end{figure}

\begin{figure}
\centerline{\includegraphics*[width=8.0 cm]{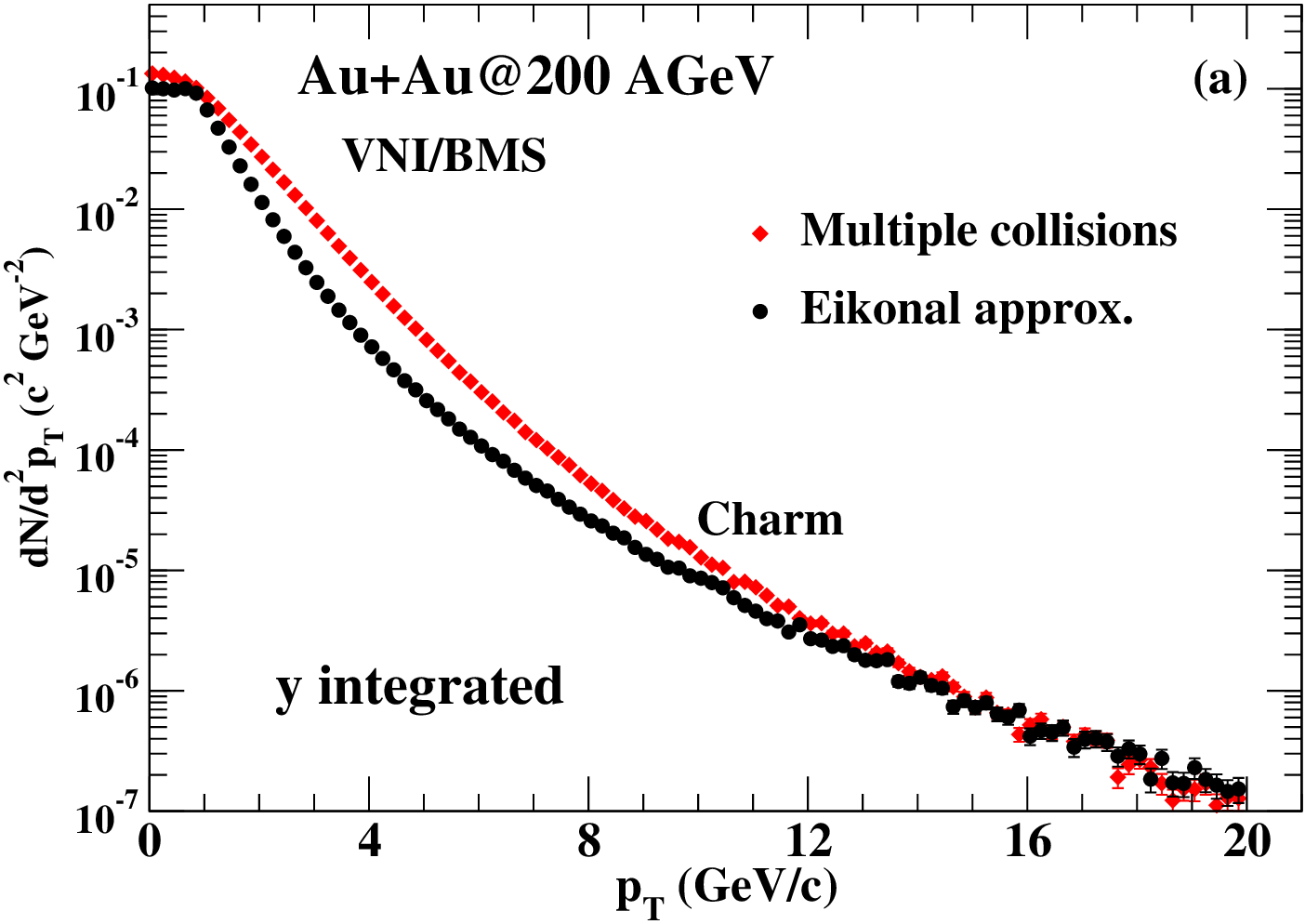}}
\centerline{\includegraphics*[width=8.0 cm]{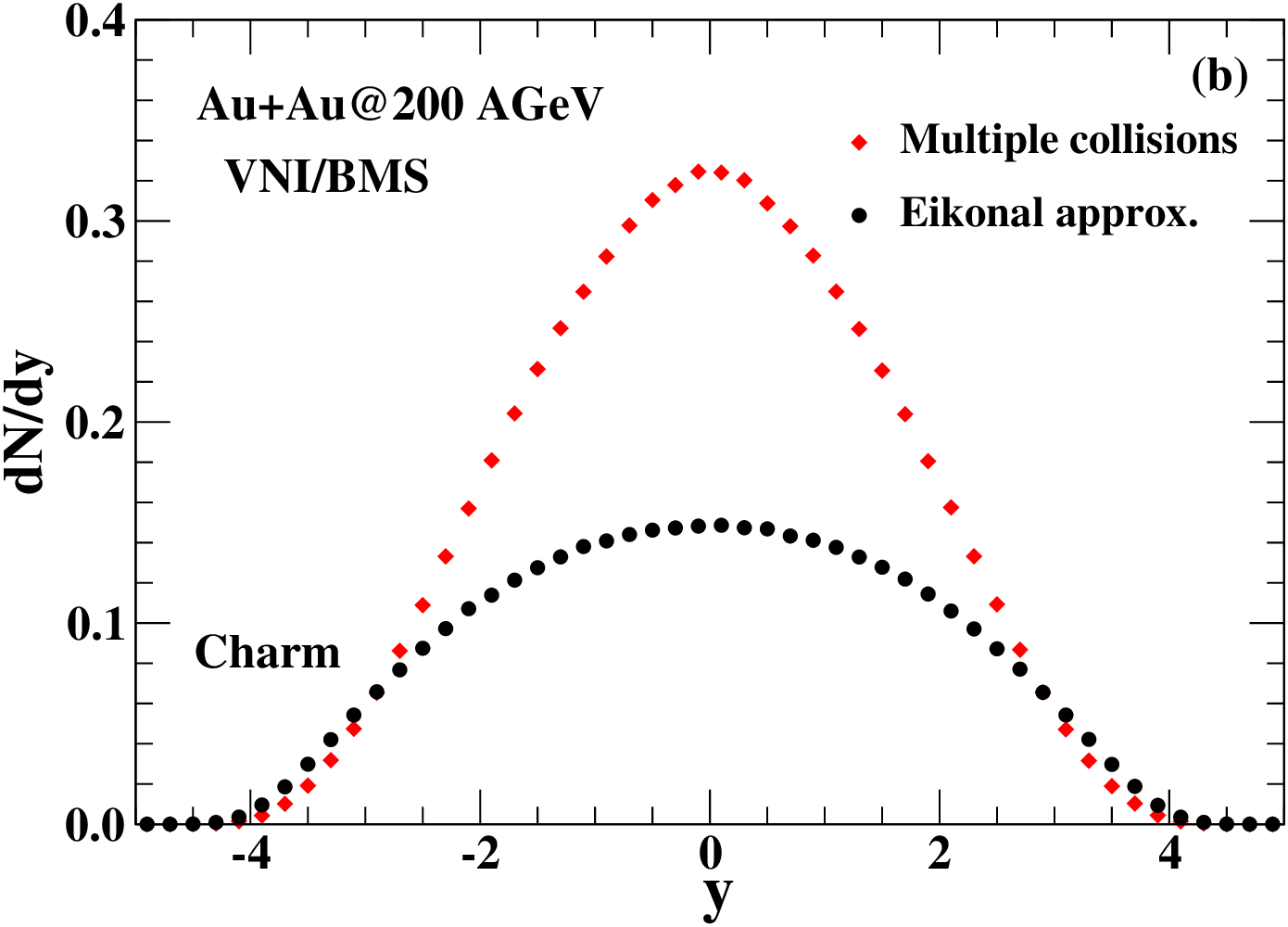}}
\caption{(colour online) (a) The $y$ integrated $p_T$ spectra of charm quarks
using eikonal approximation and multiple collisions for $Au+Au$ system at
$\sqrt{s_\textrm{NN}}$ = 200 GeV in parton cascade model.
 The lower panel  (b) gives the $p_T$ integrated rapidity
spectra for the same cases.}
\label{aa_eik_mult}
\end{figure}

  Even though not substantial, we also observe
perceptible change in the momentum distribution obtained in eikonal approximation
compared to the case for proper multiple collisions, as these actually evolve. This
would imply a slight correction to the standard calculations for the proton-proton baseline for the medium modification of particle production if 
estimated within the eikonal approximation. However, one would require very high precision data
to quantitatively investigate this effect. 

Charm production in $pp$ collisions has been measured by STAR experiment~\cite{Ye:2014ska} by measuring $D^0$ production and
other $D$ mesons. We can compare our results for
the distribution of charm quarks by accounting for the fragmentation ratio ($\approx 0.565$) for the $c \rightarrow D^0$ fragmentation as also in the
above reference for a comparison results from with {\tt PYTHIA} (tuned) and FONLL calculations. This amounts to assuming that the fragmentation function varies as $\delta (1-z)$ (see later). Our results for the calculations incorporating multiple scattering and radiations are shown in Fig.~\ref{pp_data}. In view of the $p_T^\text{cut-off}$ and $\mu_0$ used to regularize the
matrix elements and fragmentation functions used in our work we show the results for our calculations only for $p_T \geq $ 2 GeV.
A fair agreement is seen without any attempt to adjust any parameters.

\subsection{$Au+Au$ collisions at $\sqrt{s_{NN}}=$ 200 GeV}

When and where are most of the charm quarks produced in a nucleus-nucleus collision? How often do
they undergo collisions? How often do they radiate gluons? How does the charm production depend
on gluon multiplication? These and many other interesting details of charm production 
are easily discernible in the space-time description of the cascading partons in our calculations.
We proceed to address these questions in the following sections.



\subsubsection{Dynamics of charm production in cascading partons}

We start by calculating the spatial  distribution of the production vertices of the
charm quarks along the beam axis (Fig.~\ref{dndz}) for the different interaction scenarios introduced in the 
previous section.
We see that for all cases studied (primary interactions only, secondary rescattering and rescattering plus radiation) the production of charm quarks is symmetrically concentrated around the region of complete overlap ($z$=0). We further note that charm-quark production is mostly limited to the zone around $\pm $ 0.5 fm along the beam axis, due to the Lorentz-contraction of the colliding nuclei. 
 

Focusing on the case where  multiple collisions along  with the fragmentation of partons is permitted, we note that the number of produced charm quarks is significantly larger than in the other cases. However, the fraction of charm quarks produced at later separations is now much smaller than that for the case when only multiple scatterings are permitted (see Fig.~\ref{dndz}). Multiple collisions occur predominantly in the region of complete overlap. The colliding partons radiate gluons and rapidly lose energy, while the parton clouds continue to propagate through each other. However by the time the gluons undergo third or fourth interaction their energy has dropped significantly and subsequent interactions may not have enough energy to produce heavy quarks.  On the other hand, if the partons were not allowed to radiate gluons as in the case of multiple scatterings without radiations, even the later
collisions have enough energy to produce heavy quarks.  This scenario is further reinforced by the
observations of Fig.~\ref{dndt}, which shows the distribution of the production times of 
charm quarks.

Charm production generally commences by the time the nuclei start touching other, increases rapidly as they interpenetrate and  peaks when they are on top of each other and falls off as the nuclei disengage. A similar trend is seen in the behavior of the $z$-distribution of the production vertices.
When the multiple scatterings are permitted (without radiations), the
production of charm quarks continues (albeit at a very small rate) till almost 3 fm/$c$ after the complete over-lap of the nuclei, though most of them are produced by about 0.5 -- 1.0 fm/$c$.
For the last case, when the radiation of gluons is permitted, a rapid multiplication of partons and a considerable increase in multiple scatterings is observed. As a consequence, there is a large increase in production of charm quarks as well. However, this additional production is now mostly limited to early times, say about 0.3--0.5 fm/$c$. 
We attribute this restriction to early times to the energy lost by partons due to radiation of gluons following multiple scatterings: collisions at later times involve energies which are below the threshold for production of charm.
We emphasize however that even though the production of charm quarks is essentially over by about 0.5 fm/$c$ the collisions
(and radiations) will continue till about 3--4 fm/$c$ and thus charm quarks (and other high energy quarks or gluons) will continue to
lose energy due to semihard processes\cite{Bass:2002fh}.

In Fig~\ref{aa_tz} we show the actual distribution of $t$ and $z$ positions of the production vertices (for about 2700 charm quarks) for the three cases discussed above.
We notice that the $t$ and $z$ co-ordinates of the production vertices for the primary-primary collision
case are located close to the world lines describing the motion of the centers of the two nuclei and are limited to the time of overlap. For the case of multiple collisions only, the production continues to late times and fills up larger
distances from the point of complete overlap.  For the case when both multiple scatterings and radiation (of gluons) are permitted, we have 
a concentration of production vertices around the time of overlap, as at later times the partons do not have enough energy to cross the threshold for the production of charm-anticharm pairs.
The corresponding $x$ and $y$ co-ordinates for the three cases are shown in Fig.~\ref{aa_xy}. We have verified that these distributions closely
follow the distribution of binary collisions $n_\textrm{BC}(x,y)$ obtained using a Glauber model~\cite{Kolb:2003dz}.

How often do the charm quarks scatter after production and how often do they radiate gluons? The distribution of the number of scatterings suffered by charm quarks for the two cases that allow rescattering after production is given in Fig.~\ref{aa_coll}. The produced charm quarks, being secondary particles, will not scatter when only primary-primary parton interactions are considered. We also show the distribution of the number of gluon splittings by the charm quarks (Fig.~\ref{aa_rad}) for the case when we consider both multiple scattering and multiplication of gluons due to their radiation off scattered partons.

When only multiple scatterings are permitted, charm quarks will rarely scatter more than 3--4 times. We attribute this limitation to the number of partons remaining limited and thus the possibility of an interaction decreasing rapidly as the
parton clouds disengage.  However, when the fragmentation of final state partons 
produced in a hard scattering is allowed, the number of gluons increases rapidly and leads to a significantly larger number of scatterings and a significant production of gluons due to radiation. Now we find that charm quarks may scatter up to 20 times, with 8--10 scatterings not being too uncommon. Despite the restriction of the PCM to interactions above a $p_T$ cut-off, this large number of semi-hard scatterings has the potential to start propelling charm quarks towards thermalization!

Finally in Fig.~\ref{aa_rad} we provide results for the occurrences of the splitting of gluons off charm quarks following a semi-hard scattering (note that we only count the number of initial splittings, not the total number of gluons that may be emitted in each of these processes).
While one splitting is most frequent, we do observe some occurrences of multiple splittings. 

%
%

Table~\ref{table2} provides a quantitative summary of the above analysis.
For example, we find that total number of semi-hard scatterings for all
partons increases by a factor of 2.2 when 
we perform calculations in the eikonal approximation, a factor of about 3.5 when we perform the calculation using multiple scatterings and by a factor of about 19 when we perform calculations 
allowing for multiple scatterings and multiplication of partons due to time-like branchings or fragmentations --  in comparison to the calculation involving only primary-primary interactions. The number of collisions for the leading process for charm production included in the counting of, $gg \rightarrow q\bar{q}$, rises by a factors of 2.2, 4.3 and 44 respectively.

The production of charm quarks shows increases by factors of 2.5, 4, and 15.2 respectively. We note that while the rise in charm production in going from primary-primary
scatterings to the eikonal approximation to multiple
scatterings closely follows the trend for the rise in the $gg \rightarrow q\bar{q}$ process,
the rise in charm production when we consider multiple scatterings with parton multiplications
is much less (only a factor of about 15) compared to the rise in
such processes (about a factor of 44) for light parton interactions. As discussed previously, we attribute this reduced increase to gluon multiplication leading to an ensemble of less energetic partons so that
despite the increase in the number of multiple collisions that partons suffer, they may not be energetic enough for the production of charm in later interactions.

\begin{figure}
\centerline{\includegraphics*[width=8.0 cm]{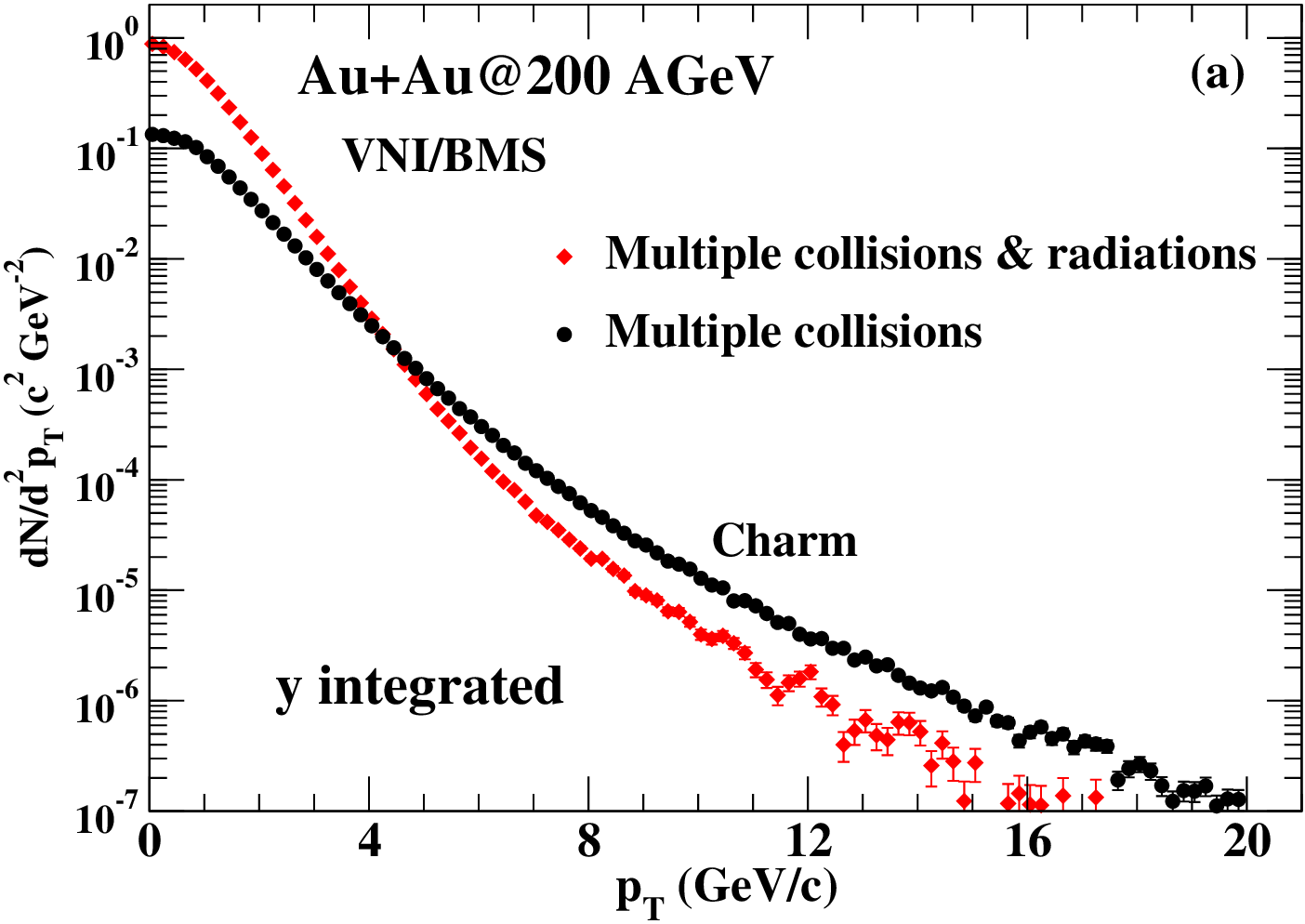}}
\centerline{\includegraphics*[width=8.0 cm]{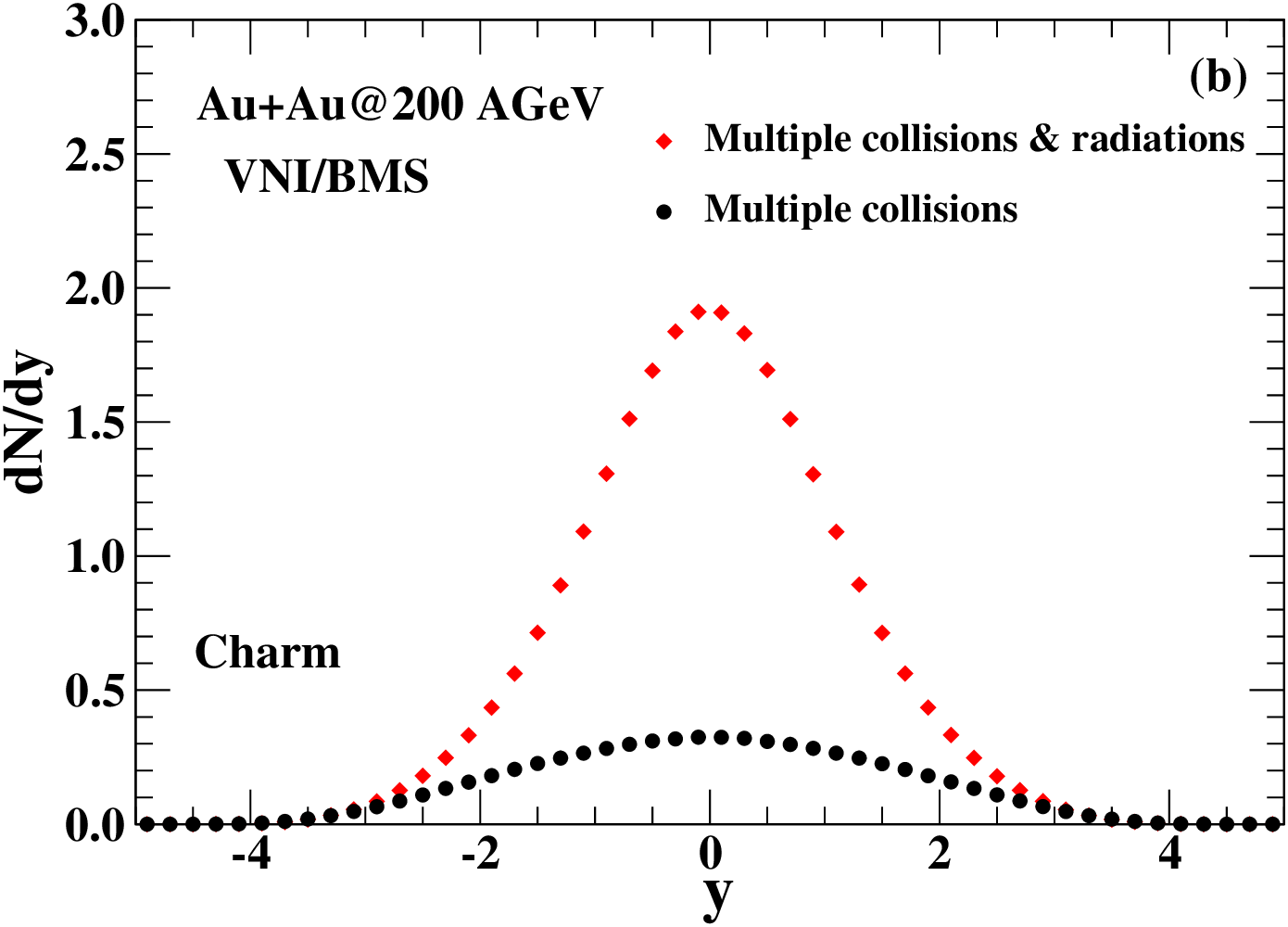}}
\caption{(colour online) (a) The $y$ integrated $p_T$ spectra of charm quarks
using multiple collisions and multiple collisions and fragmentations for $Au+Au$ system at
$\sqrt{s_\textrm{NN}}$ = 200 GeV in parton cascade model.
 The lower panel (b) gives the $p_T$ integrated rapidity
spectra for the same cases.}
\label{aa_mult_full}
\end{figure}

\begin{figure}
\centerline{\includegraphics*[width=8.0 cm]{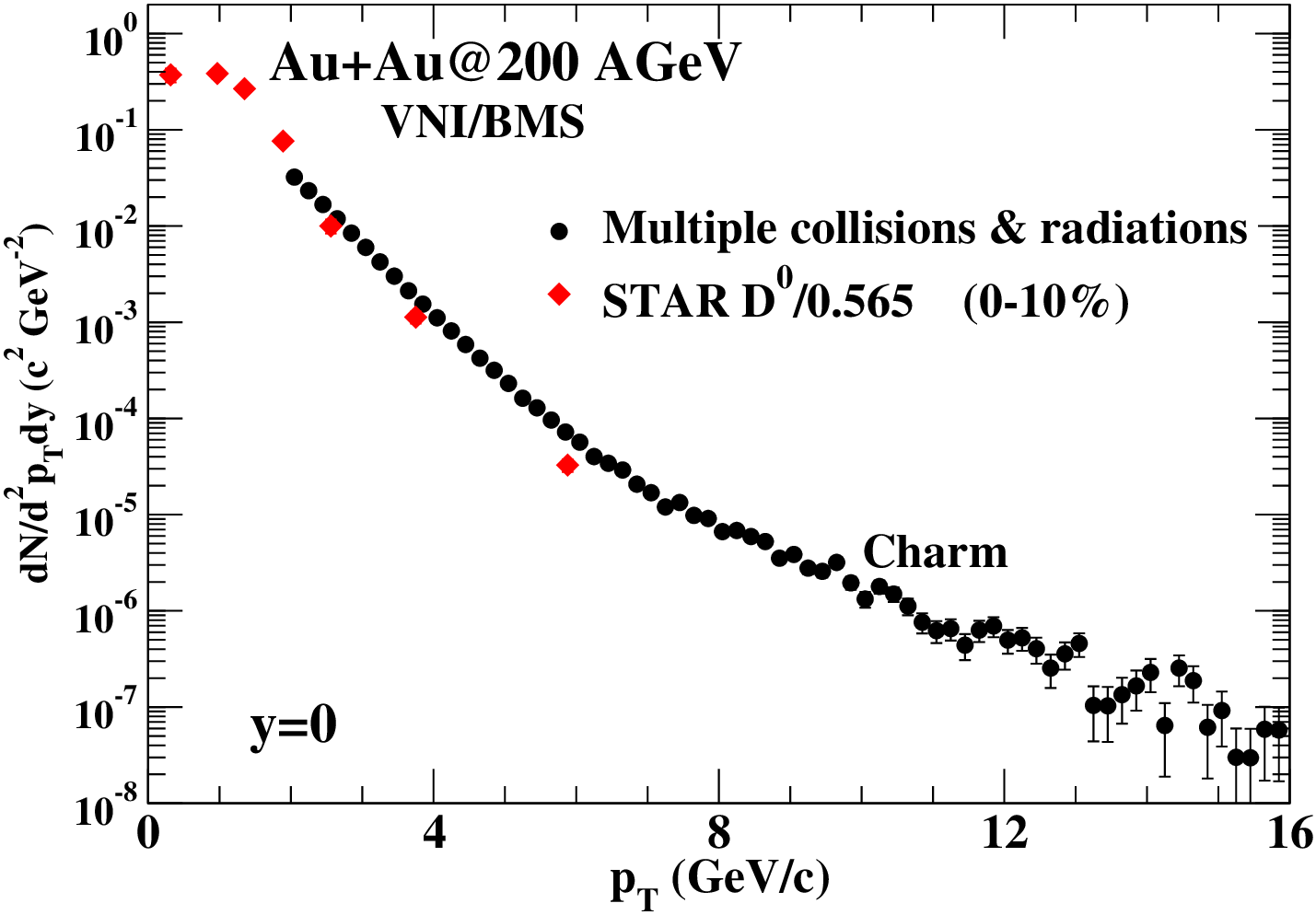}}
\caption{(colour online) The $p_T$ spectrum of charm quarks at central rapidity along with
the spectrum of $D$-mesons obtained with STAR Collaboration~\cite{Adamczyk:2014uip} for the most central collisions for
$Au+Au$ system at
$\sqrt{s_\textrm{NN}}$ = 200 GeV. The parton cascade calculation includes multiple scatterings as well as radiations of gluons off final state partons in semi-hard
scatterings.}
\label{aa_full_pt_y0}
\end{figure}

\subsubsection{Spectra of charm quarks in $Au+Au$ collision at $\sqrt{s_\textrm{NN}}$ = 200 GeV}

In this section we proceed to discuss our results for spectra of charm quarks produced in central collisions for $Au+Au$ at 
$\sqrt{s_\textrm{NN}}$ = 200 GeV, again using the same set of interaction models as in the previous section.

As a first step we compare our results for the implementations of the eikonal approximation
with those for primary-primary collisions (Fig.~\ref{aa_eik_prim}). We find that $p_T$
spectra are similar but not identical in shape and differ by factors of 2--5 varying with $p_T$
and also $y$.

\begin{figure}
\centerline{\includegraphics*[width=8.0 cm]{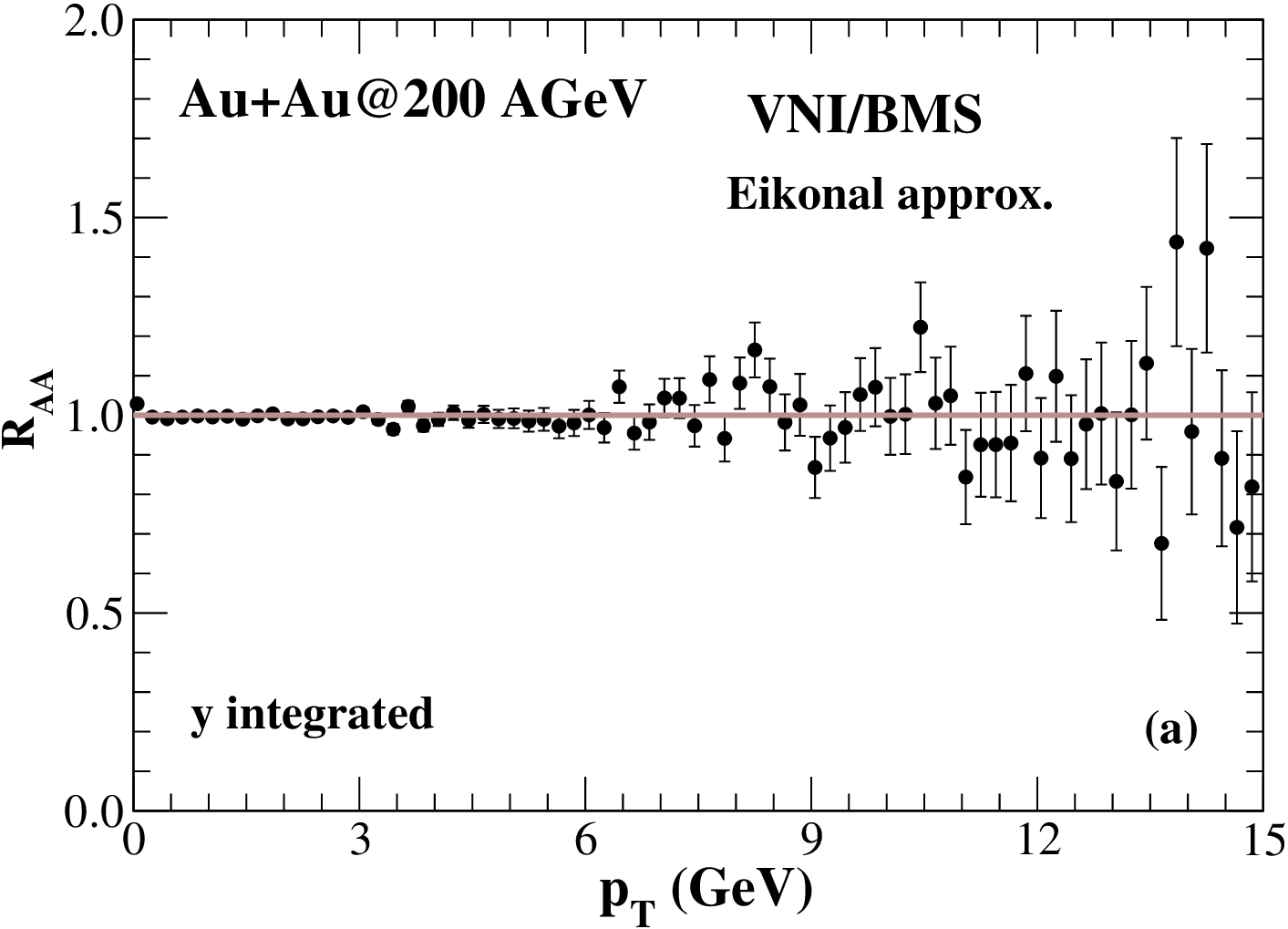}}
\centerline{\includegraphics*[width=8.0 cm]{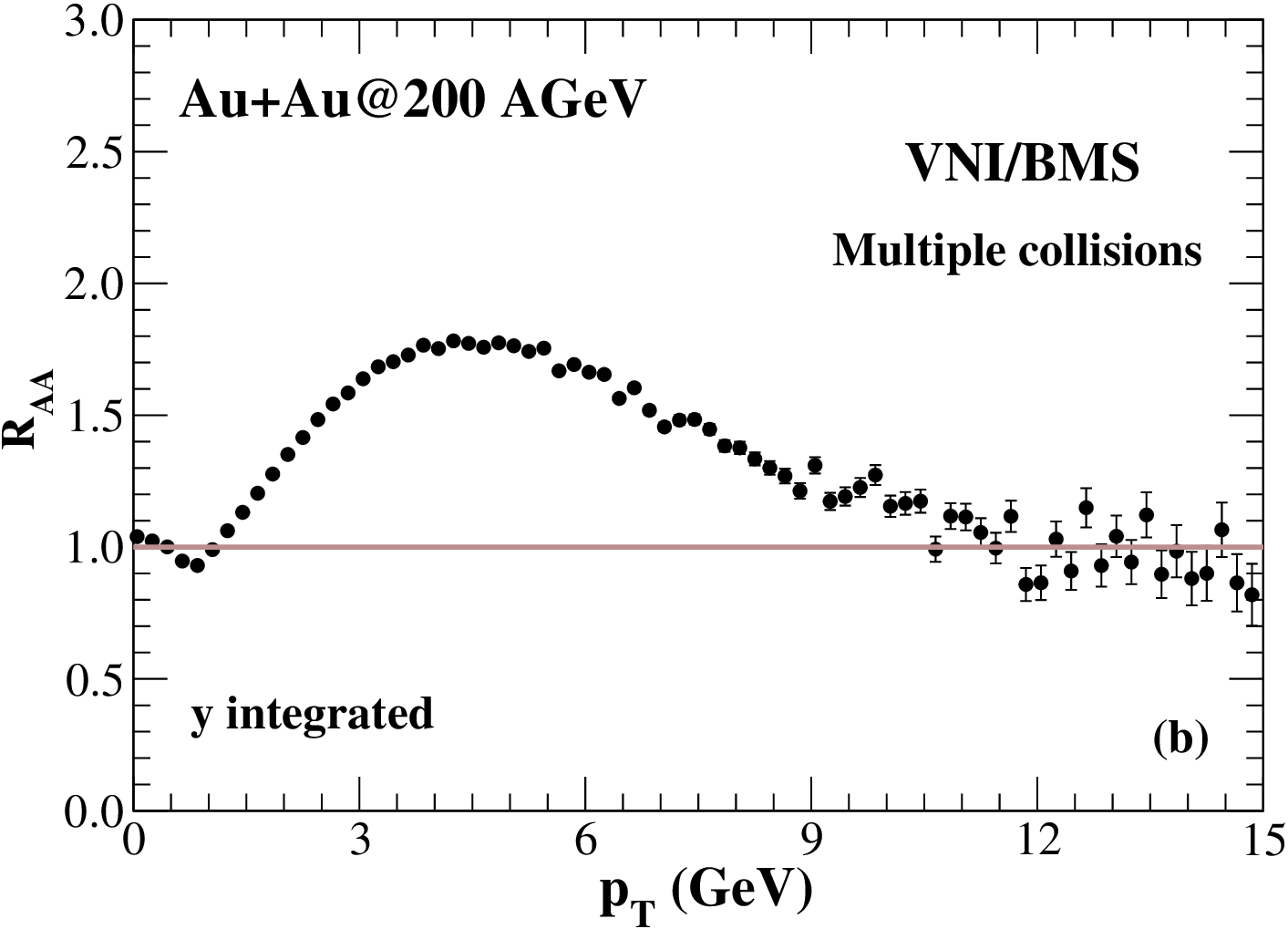}}
\centerline{\includegraphics*[width=8.0 cm]{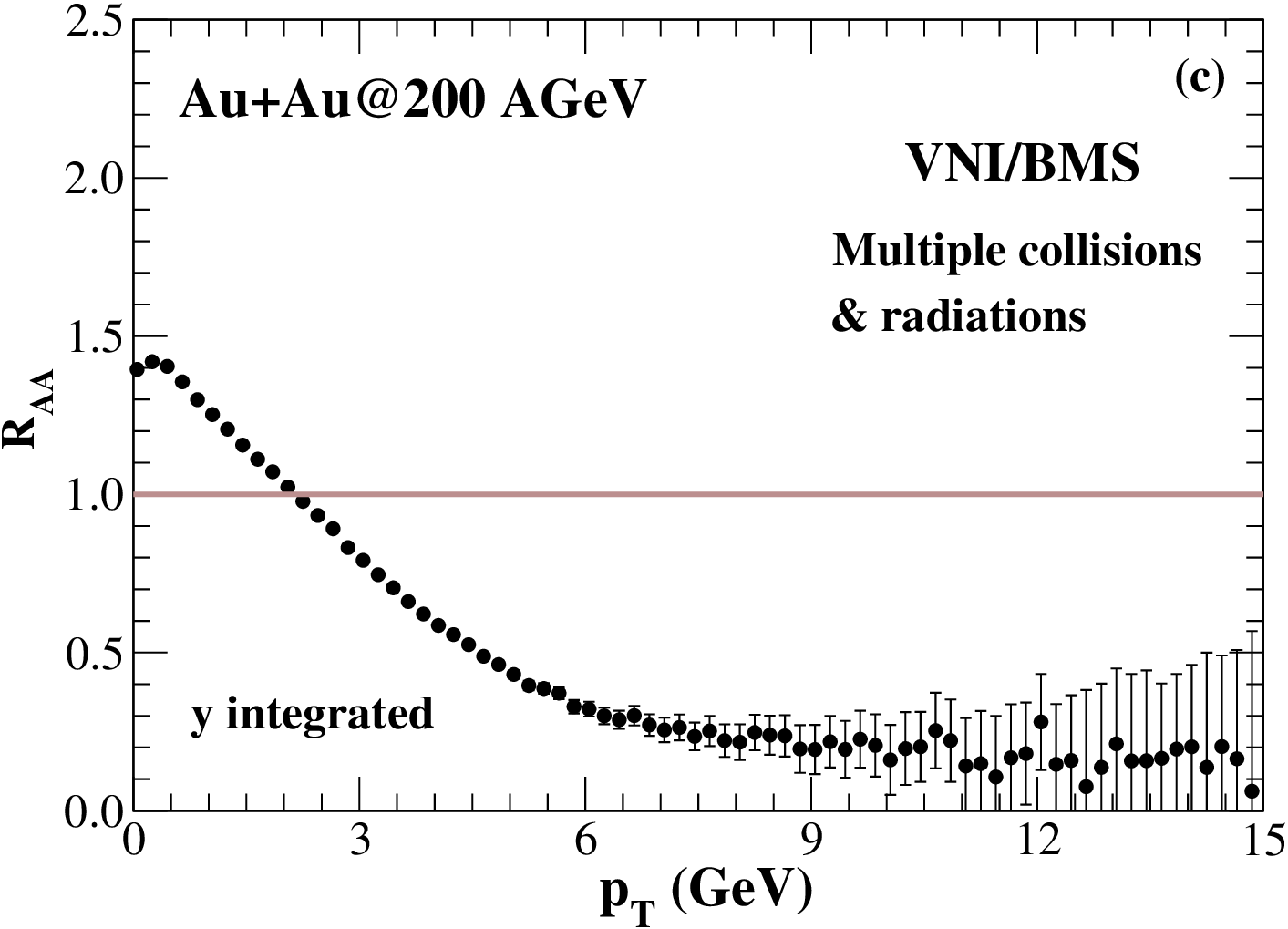}}
\caption{(colour online) rapidity integrated nuclear modification factor for eikonal approximation (a)
multiple collisions (b) and multiple collisions along with the parton multiplications (c) for
central collisions for $Au+Au$ system at $\sqrt{s_\textrm{NN}}$ = 200 GeV.} 
\label{raa}
\end{figure}
\begin{figure}
\centerline{\includegraphics*[width=8.0 cm]{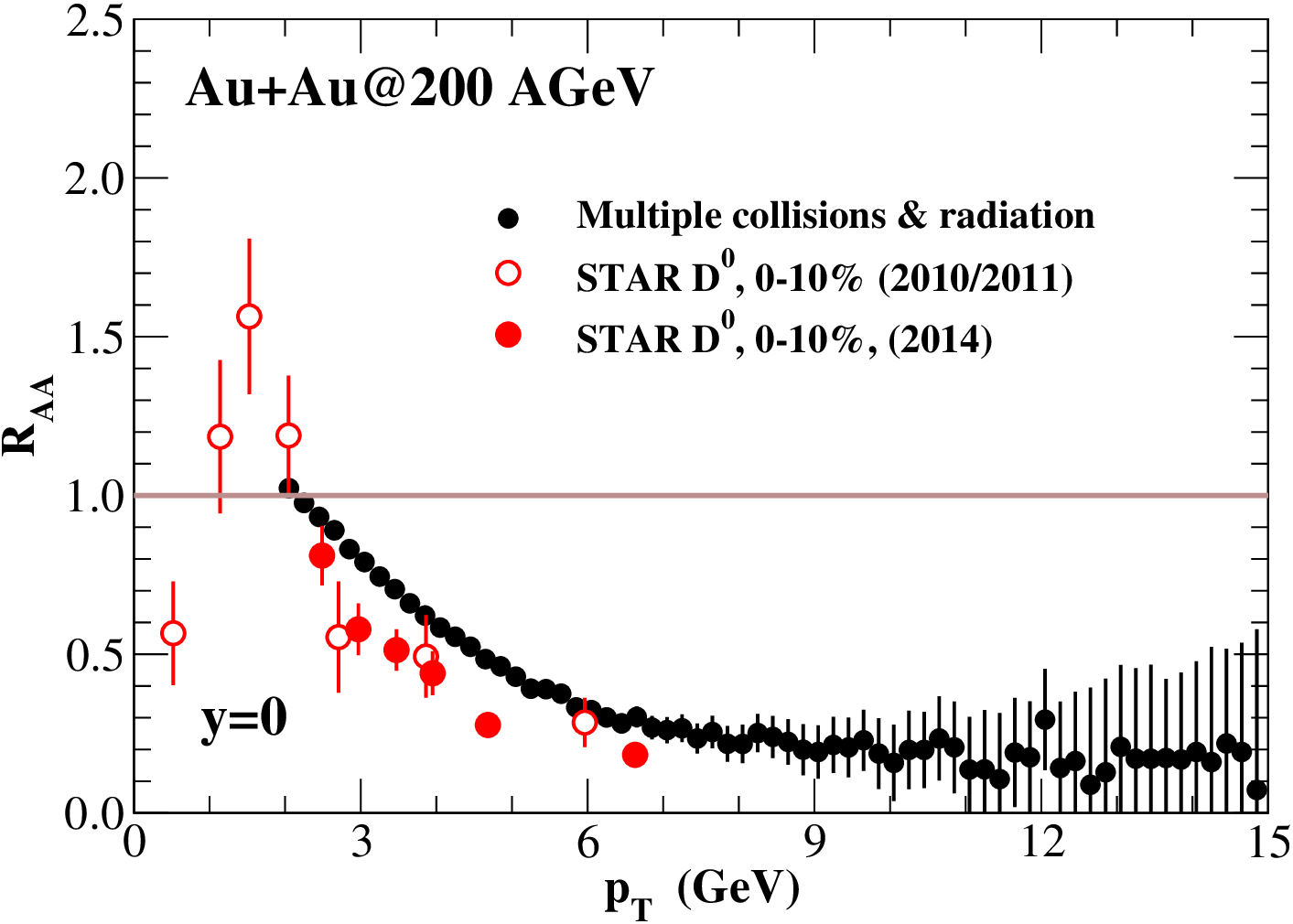}}
\caption{(colour online) Nuclear modification factor for central rapidity in parton cascade model 
using multiple collisions along with the parton multiplications for
central collisions for $Au+Au$ system at $\sqrt{s_\textrm{NN}}$ = 200 GeV. The experimental data (with only statistical errors) are from 
STAR Collaboration~\cite{Adamczyk:2014uip, Xie:2017nal}.} 
\label{raa_pt}
\end{figure}

The results for a comparison of the multiple scattering to that
of the eikonal approximation are shown in (Fig.~\ref{aa_eik_mult}).
The $p_T$ spectra for both calculations are quite similar for large $p_T$. However, we observe considerable differences at 
intermediate transverse momenta: while the results for the eikonal approximation
mimic a power-law, those for the case of multiple scattering exhibit more of an exponential shape.
This is a consequence of 
multiple scatterings altering the parton momenta, which are properly accounted for in the parton cascade model. This observation
is also supported by estimates of the production of charm quarks due to multiple scatterings 
among partons produced in semi-hard (mini)-jets by several authors~\cite{Younus:2010sx, Levai:1994dx, Lin:1994xma}. We  note that the production of charm is
proportionately higher at central rapidities as compared to larger rapidities in the
case of multiple scattering.

Now we study the effect of gluon splitting on the charm spectra: for this purpose we compare results for the production of charm quarks  in the 
multiple scattering mode to that in which multiple
scatterings as well as fragmentation of partons is enabled (Fig.~\ref{aa_mult_full}). 
We observe a substantially
larger production of charm at lower $p_T$ and a suppressed production of charm at larger $p_T$.
The increase in the production of charm at lower $p_T$ is a consequence of the large
increase in the population of low momentum  partons due to gluon multiplication. In addition, high momentum charm quarks loose energy via gluon splitting and multiple scatterings with lower momentum partons.

The rapidity distribution of charm shows an enhanced production at central 
rapidities: while the overall increase in production of charm is by a factor of about four,
at central rapidity the rise is by a factor of six.


Comparing data on final state hadrons, such as D-mesons, to the PCM is inherently challenging, due to the limitation of the PCM to parton scattering above a momentum cut-off $p_T^{\rm min}$ as well as due to the systematic uncertainties inherent in the modeling of hadronization. A realistic hadronization model, accounting for bulk hadronization, parton recombination and hadronization via fragmentation is currently out of reach for this particular PCM implementation. Nevertheless, we shall attempt a series of schematic comparisons, by assuming a simple hadronization scheme for D-mesons, based solely on the fragmentation of c-quarks. Starting with a fragmentation function for the production of $D$ mesons
from charm quarks  as $\delta(1-z)$ where $z$ is the fraction of momentum of 
the charm quark carried by the $D$ meson, we can assume that the $p_T$ of the D-mesons will be very similar to the $p_T$ of the charm quarks themselves. This ansatz has widely been used in the literature
(see e.g., Ref.~\cite{Vogt:1993kn, Brodsky:1991dj}). Please note, however, that more accurate fragmentation functions would show a narrow
peak around the ratio of the masses of the charm quark and $D$ mesons $M_c/M_D$ 
(Ref.~\cite{Peterson:1982ak}).
With this in mind we compare the results of our calculations for the $p_T$-spectrum of charm
quarks at central rapidity, with the results for $D^0$ mesons obtained by the STAR Collaboration~\cite{Adamczyk:2014uip} for the most central collisions 
(Fig.~\ref{aa_full_pt_y0})
after adjusting it for the fragmentation ratio for the $c \rightarrow D^0$ fragmentation.
 We show our results only for $p_T>2$~GeV, since our calculations do not include hydrodynamic flow, which is known to modify the spectra at lower transverse momenta and we have used a $p_T^\text{cut-off}$ and $\mu_0$ to regularize the pQCD matrices
and $P_{a\rightarrow c}$ fragmentations.
We see that the shape for the calculated spectrum
is quite similar to the experimental result above a $p_T$ of 2-3 GeV, where fragmentation should start to significantly contribute to charm quark hadronization.
 We do add 
that while our calculations are for zero impact parameter, the experimental data are for 0--10\% central collisions
which have about a 10\% smaller number of collisions.

\subsubsection{Nuclear Modification Factor for Charm Production}

The nuclear modification factor is defined by:
\begin{equation}
R_\textrm{AA}=\frac{dN_\textrm{AA}/dp_T dy}{N_\text{coll} \times dN_\text{pp}/dp_T dy}
\end{equation}
where $N_\text{coll}$ is the number of binary nucleon-nucleon collisions for a given centrality.
$R_\textrm{AA}$ is a measure of modification of the production processes due to the presence of the medium produced
in the wake of the nucleus-nucleus collisions and has been used extensively to study the phenomenon
of jet-quenching.

As a first step we compare calculations for the rapidity integrated $R_\textrm{AA}$ for different interaction assumptions (Fig.~\ref{raa}).
Unsurprisingly, we find that the results for the eikonal approximation are consistent with one,
 as one would expect from these calculations.
The results for multiple scattering suggests a considerable increase in the production of charm quarks at intermediate
$p_T$ beyond that which one would have expected on the basis of a mere superposition of $pp$ collisions and is consistent with our findings discussed in previous sections.

The results for the full calculation, including multiple scatterings as well as parton fragmentations (mostly 
radiation of gluons) provide some interesting findings: in the low $p_T$ domain, we observe
an enhancement of charm production of more than 50\% beyond what
one would expect from mere superposition of $pp$ collisions. Noting that the $p_T$ of charm quarks can get strongly 
affected by hydrodynamic flow \cite{Cao:2014fna}, these result are expected to change
if a more realistic bulk medium evolution is included in the calculation. At large $p_T$ we see
a substantial suppression of charm production. Thus we find that while the inclusion of mere multiple scattering can lead to
increase in production of charm at intermediate $p_T$, the inclusion of gluon emission rapidly depletes the momenta of the partons. Please note that this significant energy-loss is due to vacuum radiation matrix elements, since the PCM does not contain any medium modified matrix elements. Please also note that the presence of hydrodynamic flow would significantly alter the shape of $R_{AA}$ at low transverse momenta.

In Fig~\ref{raa_pt} we show a comparison of the nuclear modification factor $R_\text{AA}$ for central collisions at central rapidity. 
Interestingly we find a significant suppression, even down to intermediate $p_T$, indicating that charm interactions in the early non-equilibrium phase of the reaction play a significant role. Clearly, the PCM calculation of $R_\text{AA}$ suppression is less than observed in the
data, i.e. it cannot account for the full suppression observed. Also, the radial flow 
bump at low $p_T$ is missing in the PCM calculation. Yet our results indicate that a comprehensive calculation of heavy flavour energy loss would need to account for both, the early time 
dynamics as described by the PCM as well as the later time dynamics as described by 
some form of hydro-kinetic transport \cite{Rapp:2009my,Gossiaux:2010yx,Cao:2014fna,Das:2015ana,Song:2015ykw}. 
Please note that the 
experimental data are for 0--10\% centrality while our calculations are
for impact parameter equal to zero.

\section{Summary}
We have studied the production and scattering of heavy quarks in parton cascade model 
and presented results for production of charm quarks in $pp$ and $AuAu$ collisions at $\sqrt{s_\textrm{NN}}$ = 200 GeV.
We have verified the PCM implementation by comparing 
an eikonal approximation implementation against an independent calculation using the formalism of minijets. The effect of charm-medium interactions is studied by successively opening up reaction channels,  starting with only primary-primary
collisions at first, then allowing for multiple scatterings and finally switching on  multiple scatterings along with parton splittings.
Interestingly, while the last named calculations lead to a very large increase in multiple collisions, 
the charm production does not increase by the same factor, mainly due to their production threshold as the initial light partons lose energy while radiating gluons.
The space-time distribution of the production vertices of charm quarks provide valuable 
insight into the production process. The transverse momentum spectrum obtained using the complete calculation is found to
have a shape similar to the experimental results, confirming the importance of radiative energy loss suffered by heavy quarks at intermediate
and large $p_T$. Results for different systems, energies and cetralities  would provide a very valuable
confirmation of our procedure and findings. 
The considerable suppression of production of charm quarks having large transverse momenta due to the multiple collisions and radiations seen in the present work suggests that a significant contribution to jet-quenching could also arise from the same mechanism. This is under investigation. 


\section*{Acknowledgments} 
DKS gratefully acknowledges the support by the Department of Atomic
Energy, India. This research was supported in part by the ExtreMe Matter Institute EMMI at the GSI Helmholtzzentrum für Schwerionenforschung, Darmstadt,
Germany. We thankfully acknowledge the High Performance Computing Facility of Variable 
Energy Cyclotron Centre Kolkata for all the help
during these calculations performed over several months. SAB acknowledges support by US Department of Energy grant DE-FG02-05ER41367.

\newpage

\onecolumngrid
\begin{table} 
\begin{center}
\begin{tabular}{l r r r r}
\hline
\multicolumn{5}{c} {The number of scatterings and time like branchings involving different sub-processes
in $pp$ collisions at $\sqrt{s_{NN}}$ = 200 GeV.} \\
\cline{1-5}
\hline
\hline
process & {\hspace {0.9 cm}} Only Primary-Primary & {\hspace {0.9 cm}} Eikonal & {\hspace {0.9 cm}}  Multiple & {\hspace {0.9 cm}} Multiple collisions \\

 & {\hspace {0.9 cm}}  collisions & {\hspace {0.9 cm}}  approximation & {\hspace {0.9 cm}}   collisions & {\hspace {0.9 cm}}  and radiation \\
\hline
\hline
q+q  $\rightarrow$ q+q    &   0.204  &  0.255   &   0.269   &    0.364 \\
q+$\bar q$ $\rightarrow$ q+$\bar q$  &    0.002  &  0.002    &  0.002   &    0.012 \\
q+$\bar q$ $\rightarrow$ g+g    &   0.006  &  0.008   &   0.008    &   0.032 \\
q+$\bar q$ $\rightarrow$ g+$\gamma$  &     0.0   &  0.0   &   0.0   &    0.0 \\
q+$\bar q$  $\rightarrow$ 2 $\gamma$  &   0.0  &  0.0   &   0.0    &   0.0  \\
 q+g   $\rightarrow$ q+g   &    1.062  &  1.447   &   1.617  &     2.874  \\
 q+g  $\rightarrow$ q+$\gamma$ &    0.0  &  0.001   &   0.001  &   0.004 \\
 g+g  $\rightarrow$  q+$\bar q$  &    0.008  &  0.011   &   0.014   &    0.052  \\
 g+g  $\rightarrow$ g+g    &   0.985  &  1.533   &   1.852    &   4.814  \\
\hline
\hline
total hard a+b $\rightarrow$ c+d &  2.267  &  3.257   &   3.764   &    8.152 \\
\hline
\hline

 q    $\rightarrow$ q+g   &  & & & 0.710 \\

 g    $\rightarrow$ g+g   & & & &   2.989 \\

 g    $\rightarrow$ q+$\bar q$ & & & & 0.255 \\

 q  $\rightarrow$  q+$\gamma$  & & & &  0.0 \\

\hline
\hline
 Total a $\rightarrow$ b+c  & & & &   3.954 \\
\hline
\hline
$N_c$ (=$\bar {N_c}$)/event & 4.98 $\times$ 10$^{-4}$ & 7.62 $\times$ 10$^{-4}$ & 9.08 $\times$ 10$^{-4}$ & 34.95 $\times$ 10$^{-4}$ \\

\hline
\hline
\end{tabular}
\caption{}
\label{table1} 
\end{center}
\end{table}
\twocolumngrid
\newpage
\onecolumngrid

\begin{table} 
\begin{center}
\begin{tabular}{l r r r r}
\hline
\multicolumn{5}{c} {The number of scatterings and time like branchings involving different sub-processes
in $Au+Au$ collisions at  200A GeV.} \\
\cline{1-5}
\hline
\hline
process & {\hspace {0.9 cm}} Only Primary-Primary & {\hspace {0.9 cm}} Eikonal & {\hspace {0.9 cm}}  Multiple & {\hspace {0.9 cm}} Multiple collisions \\

 & {\hspace {0.9 cm}}  collisions & {\hspace {0.9 cm}}  approximation & {\hspace {0.9 cm}}   collisions & {\hspace {0.9 cm}}  and radiation \\
\hline
\hline
q+q  $\rightarrow$ q+q    &  111.2  &  185.3 &  227.8  &   591.5 \\
q+$\bar q$ $\rightarrow$ q+$\bar q$  &  1.1  &  1.4   &  2.2  & 34.0 \\ 
q+$\bar q$ $\rightarrow$ g+g    &  3.7 &  5.4 &   7.6 & 86.7 \\ 
q+$\bar q$ $\rightarrow$ g+$\gamma$  &  0.0  &  0.1    & 0.0   &  0.8 \\   
q+$\bar q$  $\rightarrow$ 2 $\gamma$  &  0.0  & 0.0   &    0.0   &  0.0 \\
 q+g   $\rightarrow$ q+g   &   527.7  &     1056.9   & 1538.3   &  6370.0 \\
 q+g  $\rightarrow$ q+$\gamma$ &   0.3  &  0.4  & 0.7  & 11.8 \\
 g+g  $\rightarrow$  q+$\bar q$  &   3.7  & 8.0  & 16.3   &   164.8 \\
 g+g  $\rightarrow$ g+g    & 432.7 & 1109.8  &  2020.6  & 13304.0  \\
\hline
\hline
total hard a+b $\rightarrow$ c+d &  1080.6  & 2367.3  & 3813.4  & 20563.7  \\
\hline
\hline

 q    $\rightarrow$ q+g   &  & & & 787.0 \\

 g    $\rightarrow$ g+g   & & & &   3868.2 \\

 g    $\rightarrow$ q+$\bar q$ & & & & 345.5 \\

 q  $\rightarrow$  q+$\gamma$  & & & &  0.0 \\

\hline
\hline
 Total a $\rightarrow$ b+c  & & & &   5000.8 \\
\hline
\hline
$N_c$ (=$\bar {N_c}$)/event &  0.33  &   0.82  &   1.33  & 5.21 \\

\hline
\hline
\end{tabular}
\caption{}
\label{table2} 
\end{center}
\end{table}

\twocolumngrid 


\begin{thebibliography}{10}

\bibitem{Gyulassy:2004zy}
M.~Gyulassy and L.~McLerran,
\newblock Nucl. Phys. {\bf A750}, 30 (2005), nucl-th/0405013.

\bibitem{Muller:2006ee}
B.~Muller and J.~L. Nagle,
\newblock Ann. Rev. Nucl. Part. Sci. {\bf 56}, 93 (2006),
  arXiv:nucl-th/0602029.

\bibitem{Muller:2012zq}
B.~Muller, J.~Schukraft, and B.~Wyslouch,
\newblock Ann.Rev.Nucl.Part.Sci. {\bf 62}, 361 (2012), arXiv:1202.3233.

\bibitem{Wang:2016opj}
X.-N. Wang, editor,
\newblock {\em {Quark-Gluon Plasma 5}} (World Scientific, New Jersey, 2016).

\bibitem{Kolb:2003dz}
P.~F. Kolb and U.~W. Heinz,
\newblock (2003), arXiv:nucl-th/0305084.

\bibitem{Song:2010mg}
H.~Song, S.~A. Bass, U.~Heinz, T.~Hirano, and C.~Shen,
\newblock Phys.Rev.Lett. {\bf 106}, 192301 (2011), arXiv:1011.2783.

\bibitem{Alver:2007qw}
PHOBOS, B.~Alver {\em et~al.},
\newblock Phys. Rev. Lett. {\bf 104}, 142301 (2010), arXiv:nucl-ex/0702036.

\bibitem{Alver:2008zza}
B.~Alver {\em et~al.},
\newblock Phys. Rev. {\bf C77}, 014906 (2008), arXiv:0711.3724.

\bibitem{Alver:2010gr}
B.~Alver and G.~Roland,
\newblock Phys. Rev. {\bf C81}, 054905 (2010), arXiv:1003.0194,
\newblock [Erratum: Phys. Rev.C82,039903(2010)].

\bibitem{Fries:2003vb}
R.~J. Fries, B.~Muller, C.~Nonaka, and S.~A. Bass,
\newblock Phys. Rev. Lett. {\bf 90}, 202303 (2003), nucl-th/0301087.

\bibitem{Greco:2003xt}
V.~Greco, C.~M. Ko, and P.~Levai,
\newblock Phys. Rev. Lett. {\bf 90}, 202302 (2003), nucl-th/0301093.

\bibitem{Svetitsky:1987gq}
B.~Svetitsky,
\newblock Phys.Rev. {\bf D37}, 2484 (1988).

\bibitem{GolamMustafa:1997id}
M.~Golam~Mustafa, D.~Pal, and D.~Kumar~Srivastava,
\newblock Phys.Rev. {\bf C57}, 889 (1998), arXiv:nucl-th/9706001.

\bibitem{Mustafa:1997pm}
M.~G. Mustafa, D.~Pal, D.~K. Srivastava, and M.~Thoma,
\newblock Phys. Lett. {\bf B428}, 234 (1998), arXiv:nucl-th/9711059.

\bibitem{Gyulassy:2000fs}
M.~Gyulassy, P.~Levai, and I.~Vitev,
\newblock Phys. Rev. Lett. {\bf 85}, 5535 (2000), nucl-th/0005032.

\bibitem{Djordjevic:2003zk}
M.~Djordjevic and M.~Gyulassy,
\newblock Nucl. Phys. {\bf A733}, 265 (2004), arXiv:nucl-th/0310076.

\bibitem{Moore:2004tg}
G.~D. Moore and D.~Teaney,
\newblock Phys. Rev. {\bf C71}, 064904 (2005), hep-ph/0412346.

\bibitem{Peigne:2008nd}
S.~Peigne and A.~Peshier,
\newblock Phys. Rev. {\bf D77}, 114017 (2008), arXiv:0802.4364.

\bibitem{Das:2010tj}
S.~K. Das, J.-e. Alam, and P.~Mohanty,
\newblock Phys. Rev. {\bf C82}, 014908 (2010), arXiv:1003.5508.

\bibitem{Gossiaux:2010yx}
P.~B. Gossiaux, J.~Aichelin, T.~Gousset, and V.~Guiho,
\newblock J. Phys. {\bf G37}, 094019 (2010), arXiv:1001.4166.

\bibitem{Abir:2012pu}
R.~Abir, U.~Jamil, M.~G. Mustafa, and D.~K. Srivastava,
\newblock Phys. Lett. {\bf B715}, 183 (2012), arXiv:1203.5221.

\bibitem{Das:2015ana}
S.~K. Das, F.~Scardina, S.~Plumari, and V.~Greco,
\newblock Phys. Lett. {\bf B747}, 260 (2015), arXiv:1502.03757.

\bibitem{Younus:2010sx}
M.~Younus and D.~K. Srivastava,
\newblock J. Phys. {\bf G37}, 115006 (2010), arXiv:1008.1120.

\bibitem{Cao:2013ita}
S.~Cao, G.-Y. Qin, and S.~A. Bass,
\newblock Phys.Rev. {\bf C88}, 044907 (2013), arXiv:1308.0617.

\bibitem{vanHees:2007me}
H.~van Hees, M.~Mannarelli, V.~Greco, and R.~Rapp,
\newblock Phys. Rev. Lett. {\bf 100}, 192301 (2008), arXiv:0709.2884.

\bibitem{Rapp:2009my}
R.~Rapp and H.~van Hees,
\newblock {Heavy Quarks in the Quark-Gluon Plasma},
\newblock in {\em {Quark-gluon plasma 4}}, pp. 111--206, 2010, arXiv:0903.1096.

\bibitem{Andronic:2015wma}
A.~Andronic {\em et~al.},
\newblock Eur. Phys. J. {\bf C76}, 107 (2016), arXiv:1506.03981.

\bibitem{Xie:2016iwq}
STAR, G.~Xie,
\newblock Nucl. Phys. {\bf A956}, 473 (2016), arXiv:1601.00695.



\bibitem{Cassing:2000vx} 
  W.~Cassing, E.~L.~Bratkovskaya and A.~Sibirtsev,
  Nucl.\ Phys.\ A {\bf 691}, 753 (2001)
  [nucl-th/0010071].

\bibitem{Bratkovskaya:2003ux} 
  E.~L.~Bratkovskaya, W.~Cassing and H.~Stoecker,
  Phys.\ Rev.\ C {\bf 67}, 054905 (2003)
  doi:10.1103/PhysRevC.67.054905
  [nucl-th/0301083].
 
\bibitem{Bratkovskaya:2004ec} 
  E.~L.~Bratkovskaya, W.~Cassing, H.~Stoecker and N.~Xu,
  Phys.\ Rev.\ C {\bf 71}, 044901 (2005)
  doi:10.1103/PhysRevC.71.044901
  [nucl-th/0409047].

\bibitem{Linnyk:2008uf} 
  O.~Linnyk, E.~L.~Bratkovskaya and W.~Cassing,
  Nucl.\ Phys.\ A {\bf 807}, 79 (2008)
  doi:10.1016/j.nuclphysa.2008.03.016
  [arXiv:0801.4282 [nucl-th]].

\bibitem{Linnyk:2008hp} 
  O.~Linnyk, E.~L.~Bratkovskaya and W.~Cassing,
  Int.\ J.\ Mod.\ Phys.\ E {\bf 17}, 1367 (2008)
  doi:10.1142/S0218301308010507
  [arXiv:0808.1504 [nucl-th]].

\bibitem{Song:2015sfa} 
  T.~Song, H.~Berrehrah, D.~Cabrera, J.~M.~Torres-Rincon, L.~Tolos, W.~Cassing and E.~Bratkovskaya,
  Phys.\ Rev.\ C {\bf 92}, no. 1, 014910 (2015)
  doi:10.1103/PhysRevC.92.014910
  [arXiv:1503.03039 [nucl-th]].


\bibitem{Song:2015ykw} 
  T.~Song, H.~Berrehrah, D.~Cabrera, W.~Cassing and E.~Bratkovskaya,
  Phys.\ Rev.\ C {\bf 93}, no. 3, 034906 (2016)
  doi:10.1103/PhysRevC.93.034906
  [arXiv:1512.00891 [nucl-th]].


\bibitem{Levai:1994dx}
P.~Levai, B.~Muller, and X.-N. Wang,
\newblock Phys. Rev. {\bf C51}, 3326 (1995), arXiv:hep-ph/9412352.

\bibitem{Lin:1994xma}
Z.-w. Lin and M.~Gyulassy,
\newblock Phys. Rev. {\bf C51}, 2177 (1995), arXiv:nucl-th/9409007,
\newblock [Erratum: Phys. Rev.C52,440(1995)].

\bibitem{Mangano:1991jk}
M.~L. Mangano, P.~Nason, and G.~Ridolfi,
\newblock Nucl. Phys. {\bf B373}, 295 (1992).

\bibitem{Frixione:1997ma}
S.~Frixione, M.~L. Mangano, P.~Nason, and G.~Ridolfi,
\newblock Adv. Ser. Direct. High Energy Phys. {\bf 15}, 609 (1998),
  arXiv:hep-ph/9702287.

\bibitem{Younus:2011mn}
M.~Younus, U.~Jamil, and D.~K. Srivastava,
\newblock J. Phys. {\bf 39}, 025001 (2012), arXiv:1108.0855.

\bibitem{He:2011qa}
M.~He, R.~J. Fries, and R.~Rapp,
\newblock Phys.Rev. {\bf C86}, 014903 (2012), arXiv:1106.6006.

\bibitem{Gossiaux:2011ea}
P.~B. Gossiaux {\em et~al.},
\newblock arXiv:1102.1114.

\bibitem{Uphoff:2012gb}
J.~Uphoff, O.~Fochler, Z.~Xu, and C.~Greiner,
\newblock Phys. Lett. {\bf B717}, 430 (2012), arXiv:1205.4945.

\bibitem{Scardina:2014lda}
F.~Scardina, S.~K. Das, S.~Plumari, D.~Perricone, and V.~Greco,
\newblock J. Phys. Conf. Ser. {\bf 535}, 012019 (2014).

\bibitem{Uphoff:2014hza}
J.~Uphoff, O.~Fochler, Z.~Xu, and C.~Greiner,
\newblock J. Phys. {\bf G42}, 115106 (2015), arXiv:1408.2964.

\bibitem{Geiger:1991nj}
K.~Geiger and B.~Muller,
\newblock Nucl. Phys. {\bf B369}, 600 (1992).

\bibitem{Geiger:1992si}
K.~Geiger,
\newblock Phys. Rev. {\bf D46}, 4965 (1992).

\bibitem{Geiger:1992ac}
K.~Geiger,
\newblock Phys. Rev. {\bf D46}, 4986 (1992).

\bibitem{Geiger:1993py}
K.~Geiger,
\newblock Phys. Rev. {\bf D48}, 4129 (1993).

\bibitem{Geiger:1993ix}
K.~Geiger and J.~I. Kapusta,
\newblock Phys. Rev. {\bf D47}, 4905 (1993).

\bibitem{Geiger:1994he}
K.~Geiger,
\newblock Phys. Rept. {\bf 258}, 237 (1995).

\bibitem{Bass:2002fh}
S.~A. Bass, B.~Muller, and D.~K. Srivastava,
\newblock Phys. Lett. {\bf B551}, 277 (2003), nucl-th/0207042.

\bibitem{Younus:2013rja}
M.~Younus, C.~E. Coleman-Smith, S.~A. Bass, and D.~K. Srivastava,
\newblock Phys. Rev. {\bf C91}, 024912 (2015), arXiv:1309.1276.

\bibitem{Xu:2004mz}
Z.~Xu and C.~Greiner,
\newblock Phys. Rev. {\bf C71}, 064901 (2005), hep-ph/0406278.

\bibitem{Cao:2011et}
S.~Cao and S.~A. Bass,
\newblock Phys. Rev. {\bf C84}, 064902 (2011), arXiv:1108.5101.

\bibitem{Combridge:1978kx}
B.~Combridge,
\newblock Nucl.Phys. {\bf B151}, 429 (1979).

\bibitem{Collins:1985gm}
J.~C. Collins, D.~E. Soper, and G.~F. Sterman,
\newblock Nucl. Phys. {\bf B263}, 37 (1986).

\bibitem{Kang:2016ofv} 
  Z.~B.~Kang, F.~Ringer and I.~Vitev,
  JHEP {\bf 1703}, 146 (2017)
  doi:10.1007/JHEP03(2017)146
  [arXiv:1610.02043 [hep-ph]].

\bibitem{Sjostrand:1985xi}
T.~Sjostrand,
\newblock Phys. Lett. {\bf B157}, 321 (1985).

\bibitem{Altarelli:1977zs}
G.~Altarelli and G.~Parisi,
\newblock Nucl. Phys. {\bf B126}, 298 (1977).

\bibitem{Eichten:1984eu}
E.~Eichten, I.~Hinchliffe, K.~D. Lane, and C.~Quigg,
\newblock Rev. Mod. Phys. {\bf 56}, 579 (1984),
\newblock [Addendum: Rev. Mod. Phys.58,1065(1986)].


\bibitem{Ye:2014ska} 
  Z.~Ye [STAR Collaboration],
  Nucl.\ Phys.\ A {\bf 932}, 45 (2014).
  doi:10.1016/j.nuclphysa.2014.09.097

\bibitem{Adamczyk:2014uip}
STAR, L.~Adamczyk {\em et~al.},
\newblock Phys. Rev. Lett. {\bf 113}, 142301 (2014), arXiv:1404.6185.

\bibitem{Xie:2017nal}
STAR, G.~Xie,
\newblock (2017), arXiv:1701.01878.

\bibitem{Vogt:1993kn}
R.~Vogt, B.~V. Jacak, P.~L. McGaughey, and P.~V. Ruuskanen,
\newblock Phys. Rev. {\bf D49}, 3345 (1994), arXiv:hep-ph/9309213.

\bibitem{Brodsky:1991dj}
S.~J. Brodsky, P.~Hoyer, A.~H. Mueller, and W.-K. Tang,
\newblock Nucl. Phys. {\bf B369}, 519 (1992).

\bibitem{Peterson:1982ak}
C.~Peterson, D.~Schlatter, I.~Schmitt, and P.~M. Zerwas,
\newblock Phys. Rev. {\bf D27}, 105 (1983).

\bibitem{Cao:2014fna}
S.~Cao, Y.~Huang, G.-Y. Qin, and S.~A. Bass,
\newblock J. Phys. {\bf G42}, 125104 (2015), arXiv:1404.3139.

\end{thebibliography}

\end{document}